\documentclass[a4paper,11pt]{article}
\usepackage{amssymb}
\usepackage{amsfonts}
\usepackage{graphicx}
\usepackage{amsmath}
\usepackage{color}

\DeclareFontFamily{U}{rsf}{}
\DeclareFontShape{U}{rsf}{m}{n}{
  <5> <6> rsfs5 <7> <8> <9> rsfs7 <10-> rsfs10}{}
\DeclareMathAlphabet\Scr{U}{rsf}{m}{n} \makeatletter
\@addtoreset{equation}{section} \makeatother

\def\be{\begin{equation}}
\def\ee{\end{equation}}
\def\ba{\begin{array}}
\def\ea{\end{array}}

\newcommand{\bea}{\begin{eqnarray}}
\newcommand{\eea}{\end{eqnarray}}

\begin{document}

\begin{titlepage}

\vskip 4.0 cm
\begin{center}  {\Huge{\bf Black Hole Attractors} \\\vskip 0.2 cm {\bf and $U(1)$ Fayet-Iliopoulos Gaugings:}}\\\vskip 0.2 cm {\huge{\bf Analysis and Classification}}

\vskip 3.5 cm

{\Large{\bf Davide Astesiano$^{1,2}$}, {\bf Sergio L. Cacciatori$^{1,2}$}, {\bf Alessio Marrani$^{3}$}}

\vskip 1.5 cm

$^1${\sl Dipartimento di Scienza ed Alta Tecnologia,\\ Universit\`a dell'Insubria, Via Valleggio 11, I-22100
Como, Italy\\
\texttt{dastesiano@uninsubria.it}\\ \texttt{sergio.cacciatori@uninsubria.it}}\\

\vskip 0.5
cm

$^2${\sl INFN, sezione di Milano,\\Via Celoria 16, I-20133, Milano, Italy}\\

\vskip 0.5 cm

$^3${\sl Centro Studi e Ricerche Enrico Fermi,\\Via Panisperna 89A,
I-00184, Roma, Italy\\ \texttt{jazzphyzz@gmail.com}}


 \vskip 4.0 cm

\begin{abstract}
We classify the critical points of the effective black hole potential which governs the attractor mechanism taking place at the horizon of static dyonic extremal black holes in $\mathcal{N}=2$, $D=4$ Maxwell-Einstein supergravity with $U(1)$ Fayet-Iliopoulos
gaugings. We use a manifestly symplectic covariant formalism, and we consider both spherical and hyperbolic horizons, recognizing the relevant sub-classes to which some representative examples belong. We also exploit projective special K\"{a}hler geometry
of vector multiplets scalar manifolds, the $U$-duality-invariant quartic structure (and 2-polarizations thereof) in order to retrieve and generalize various expressions of the entropy of asymptotically AdS$_{4}$ BPS black holes, in the cases in which the scalar
manifolds are symmetric spaces. Finally, we present a novel static extremal black hole solution to the $STU$ model, in which the dilaton interpolates between an hyperbolic near-horizon geometry and AdS$_{4}$ at infinity.

 \end{abstract}
\vspace{24pt}

\end{center}

\end{titlepage}

\newpage \tableofcontents \newpage


\section{\label{Intro}Introduction}

The relevance of black holes (BHs) in AdS spaces is known for several
reasons. One is of course the AdS/CFT correspondence and its several
applications, for instance to condensed matter physics (see e.g. \cite%
{Hartnoll}), Fermi liquids physics \cite{Iizuka}, and superconductivity \cite%
{Hartnoll2}, to name a few. In such frameworks, the coupling to
electromagnetic charges and scalar fields is of utmost importance, at least
in order to deal with for realistic physical models; as a consequence,
gauged supergravity models including Abelian gauge fields and coupled to
non-linear sigma models quite naturally acquire a key role. On the other
hand, BPS solutions provide examples in which supersymmetric conformal field
theories are defined on curved backgrounds, the conformal boundaries.
However, non-BPS as well as non-extremal BH solutions turn out to have
intriguing applications within the holographic paradigm, such as, for
example, to finite temperature condensed matter systems. Another important
and more recently established framework is the Kerr/CFT correspondence,
which offer valuable insights into the microscopic description and
computation of BH entropy (cf. e.g. \cite{Guica}, \cite{Benini}).

The structure of single-center extremal BPS black holes in $\mathcal{N}=2$, $%
D=4$ ungauged supergravity is well known: they are asymptotically flat
solutions to Maxwell-Einstein equations, preserving eight supersymmetries at
spatial infinity (at which, due to the absence of a gauge potential, scalar
fields are unfixed moduli), then breaking all supersymmetry when radially
flowing towards the event horizon, and finally restoring half of the
supersymmetries when the scalar fields, regardless of their asymptotical
values, are attracted to fixed values, purely dependent on the conserved
electric and magnetic BH charges, at the spherically symmetric horizon. This
is the celebrated attractor mechanism \cite{AM1,AM2}. In gauged
supergravity, the physical scenarios open up to a wide range of
possibilities, one of which will be the object of the present investigation.
Recent years witnessed unanticipated progress in finding BPS, as well as
non-BPS and non-extremal, thermal BH solutions in generally matter coupled $%
\mathcal{N}=2$ gauged supergravity in $D=4$ space-time dimensions; see for
instance \cite{Cacciatori:2020kso}-\nocite%
{Astesiano,CK,Gnecchi,Klemm,Chow-1,Chow-2,Chow-3,Chong,Hristov-1,H-2,H-3,Katmadas,H-5,H-6,Katmadas-Tomasiello}%
\cite{Goulart}. In presence of gauging (of the isometries of the scalar
manifolds), the supersymmetric BH solutions may be asymptotically AdS$_{4}$
and preserve all eight supersymmetries, with the scalars being fixed at
critical points (at least local minima) of the gauge potential itself. In
this framework, the near-horizon geometry of extremal BHs is no more the
Bertotti-Robinson conformally-flat AdS$_{2}\times S^{2}$ geometry, but
rather generalizes to spherical, hyperbolic and also flat configurations
(with generally non-vanishing Weyl tensor).

Whereas the aforementioned attractor mechanism and its exploitation in terms
of critical dynamics of an effective black hole potential \cite{AM2} is well
studied in the ungauged theory, a systematic study of the attractor
mechanism and of the corresponding (generalized) effective BH potential is
still missing in gauged supergravity, notwithstanding the wealth of
possibilities of the gauged scenario. Investigation of attractor flows in
presence of gauging started to be carried out in the BPS case in \cite{DG}
and \cite{Kachru:2011ps}, as well as in \cite{BFMY-FI} and \cite{Chimento}
in the effective black hole potential formalism (in \cite{Chimento} the
coupling to hypermultiplets was considered, too).

\cite{BFMY-FI}, \cite{Chimento} and \cite{KPR} provided the construction of
an effective BH potential $V_{\text{eff}}$ which depends on the gauge
potential $V$, and moreover generalizes the BH potential $V_{BH}$ of the
ungauged case (to which it reduces in the limit of zero gauging). At the
(unique) event horizon of extremal BHs, the (at least local minima) critical
points of $V_{\text{eff}}$ govern the attractor mechanism;
despite scalar fields are not generally all stabilized at the BH horizon
(and thus a moduli space of \textquotedblleft flat" directions is present),
the non-negative value of $V_{\text{eff}}$ at the BH horizon provides the
Bekenstein-Hawking BH entropy\footnote{ In general, proper
extremal BH attractors are defined by (\textit{at least} local) minima of
the effective BH potential, both in the ungauged and gauged theory. For what
concerns the ungauged case, in the symmetric cosets of special geometry, all
critical points of $V_{BH}$ are characterized by an Hessian matrix with
strictly non-negative eigenvalues (with vanishing eigenvalues corresponding
to \textquotedblleft flat" directions of $V_{BH}$ itself) \cite%
{Moduli-Spaces}. In the gauged framework under consideration, we are
assuming the same to hold for the critical points of $V_{\text{eff}}$;
indeed, the zero Hessian eigenvalues seems to be ubiquitous also in presence
of gauging (see e.g. \cite{CK}). We leave a detailed analysis of the Hessian
modes at the critical points of $V_{\text{eff}}$ for further future work.}
(in units of $\pi $):
\begin{equation}
\frac{S}{\pi }=\left. V_{\text{eff}}\right\vert _{\partial V_{\text{eff}}=0}.
\end{equation}

The present paper concerns the classification of the critical points of $V_{%
\text{eff}}$ and the study of the corresponding properties of the extremal
BH solutions, in $\mathcal{N}=2$ $D=4$ supergravity coupled to vector
multiplets in presence of a (generally dyonic) $U(1)$ Fayet-Iliopoulos (FI)
gauging. By denoting the number of vector multiplets with $n_{v}$, and
developing on the findings of \cite{KPR}, we will exploit a manifestly
symplectic, $Sp(2n_{v}+2,\mathbb{R})$-covariant formalism. Furthermore, we
will use structural identities of the special K\"{a}hler geometry of vector
multiplets' scalar manifolds in order to completely classify all the
possible extremal BH solutions with spherical or hyperbolic near-geometries.
As it will be evident from the treatment below, our analysis encompasses
both BPS and non-BPS configurations, and we will provide detailed analysis
of BPS sub-sectors throughout.

Upon extremizing $V_{\text{eff}}$, two main classes of critical points arise
out; namely :

\begin{description}
\item[\textbf{Class I}] , corresponding to critical points of $V_{\text{eff}%
} $ which are also critical points of both $V_{BH}$ and $V$ (all placed at
the horizon):
\begin{equation}
\left.
\begin{array}{r}
\partial _{i}V_{BH}=0; \\
\partial _{i}V=0;%
\end{array}
\right\} \Rightarrow \partial _{i}V_{\text{eff}}=0,~\forall i.
\end{equation}

\item[\textbf{Class II}] , corresponding to critical points of $V_{\text{eff}%
}$ which are \textit{not} critical points of $V_{BH}$ \textit{nor} of $V$,
with the gradients of $V_{BH}$ and of $V$ being proportional:
\begin{equation}
\partial _{i}V_{BH}=\frac{\left( 2V_{BH}V+\kappa \sqrt{1-4V_{BH}V}-1\right)}{%
2V^{2}}\partial _{i}V,~\forall i.
\end{equation}
\end{description}

For both classes, at least for \textit{symmetric} scalar manifolds'
geometries, the BH entropy can be related to suitable (2-)polarizations of a
quartic structure, invariant under $U$-duality\footnote{%
Here, $U$-duality is referred to as the \textquotedblleft
continuous\textquotedblright\ symmetries of \cite{CJ-1}; their discrete
versions are the $U$-duality non-perturbative string theory symmetries
introduced by Hull and Townsend \cite{HT-1}, which occur when the
Dirac-Schwinger-Zwanzinger quantization condition is enforced.}, and
primitive in all cases but for minimal coupling of the vector multiplets
\cite{Luciani, FKM-minimal}. Thus, our analysis provides an
extension of the analysis of \cite{Halmagyi:2013qoa} (subsequently developed in \cite%
{H-4}), in which algebraic BPS equations supported by generally dyonic
charge configuration, and with a cubic prepotential function, were solved.
Thence, we will recognize some examples among the currently known solutions
as belonging to a corresponding sub-class of the aforementioned two main
classes of critical points of $V_{\text{eff}}$. It should also be remarked
here that an interesting outcome is provided by the explicit construction of
a novel static extremal BH solution in $U(1)$ FI gauged supergravity,
supported by both non-BPS and BPS charge configurations.\bigskip

All in all, the general structure of this paper splits up into three main
parts:

\begin{enumerate}
\item In the first part (Secs. \ref{Ids}-\ref{BPSBPS}), we will exploit
special K\"{a}hler geometry and the 2-polarizations of the quartic invariant
structure in symmetric special cosets, in order to retrieve, and further
generalize in various ways, some known results on the entropy of extremal BH
attractors.

\item In the second part (Secs. \ref{Sec-Eff}-\ref{Class}), we consider the
effective BH potential $V_{\text{eff}}$ introduced in \cite{BFMY-FI,
Chimento, KMPR} and provide a complete classification of its critical
points, pointing out the existence of various (yet undiscovered) BPS
sub-sectors.

\item In the third part (Sec. \ref{Ex}), we will provide some examples of
known solutions, and determine their placement in the classification given
in the second part. Furthermore, we will also present a novel static
extremal BH solution to the $STU$ model, in which the dilaton interpolates
between an hyperbolic near-horizon geometry and AdS$_{4}$ at infinity.
\end{enumerate}

Some final remarks and four Appendices conclude the paper.


\section{\label{Ids}Identities and fluxes in projective special geometry}

We start by introducing the symplectic vectors $\mathcal{Q}$ and $\mathcal{G}
$ of electric-magnetic black hole fluxes resp. $U(1)$ Fayet-Iliopoulos (FI)
gaugings of $\mathcal{N}=2$, $D=4$ Maxwell-Einstein supergravity \label{CK,
DG, KPR}, which in the so-called 4D/5D special coordinates' symplectic frame
can be written as
\begin{eqnarray}
\mathcal{Q} &:&=\left( p^{0},p^{i},q_{0},q_{i}\right) ^{T};  \label{Q} \\
\mathcal{G} &:&=\left( g^{0},g^{i},g_{0},g_{i}\right) ^{T},  \label{G}
\end{eqnarray}
where the naught index pertains to the graviphoton, and $i=1,...,n$, with $n$
denoting the number of vector multiplets\footnote{%
In order to compare our results to Halmagy's treatment \cite{H1}, we here
only deal with $U(1)$ FI gauging (namely, only vector multiplets). After
\cite{KPR}, it is however possible to straightforwardly include also
hypermultiplets' Abelian gaugings, by simply replacing $\mathcal{G}$ with $%
\mathcal{P}:=\mathcal{P}^{x}Q^{x}$ . In this case, no assumptions on the
geometry of the hypermultiplets' scalar manifold are needed. We leave the
detailed treatment of such a framework to future investigation.}. Moreover, $%
\kappa $ is related to the constant scalar curvature $R=2\kappa $ of the
(unique) event horizon of the static extremal black hole solution under
consideration. In the following treatment we will consider $\kappa =1$
(spherical) or $\kappa =-1$ (hyperbolic) near-horizon geometry\footnote{%
The case $\kappa =0$, corresponding to extremal black holes with flat
horizon, deserves a separate treatment, which we leave for future
investigation.}.

The following identities holds in the projective special K\"{a}hler geometry
of the vector multiplets' scalar manifold\footnote{%
We will henceforth denote the imaginary unit as $\mathbf{i}$.} $M_{v}$ (with
$\dim _{\mathbb{C}}M_{v}=n$; cf. e.g. \cite{K-rev, AoB, KMPR}, and Refs.
therein):
\begin{eqnarray}
\mathcal{Q} &=&\mathbf{i}\overline{\mathcal{Z}}\mathcal{V}-\mathbf{i}
\mathcal{Z}\overline{\mathcal{V}}+\mathbf{i}\mathcal{Z}^{\bar{\imath}}
\overline{\mathcal{V}}_{\bar{\imath}}-\mathbf{i}\overline{\mathcal{Z}}^{i}
\mathcal{V}_{i};  \label{id-1} \\
\mathcal{G} &=&\mathbf{i}\overline{\mathcal{L}}\mathcal{V}-\mathbf{i}
\mathcal{L}\overline{\mathcal{V}}+\mathbf{i}\mathcal{L}^{\bar{\imath}}
\overline{\mathcal{V}}_{\bar{\imath}}-\mathbf{i}\overline{\mathcal{L}}^{i}
\mathcal{V}_{i},  \label{id-2}
\end{eqnarray}
where
\begin{eqnarray}
\mathcal{Z} &:&=\left\langle \mathcal{Q},\mathcal{V}\right\rangle ,~\mathcal{%
Z}_{i}\equiv D_{i}\mathcal{Z}:=\left\langle \mathcal{Q},\mathcal{V}%
_{i}\right\rangle ,  \label{Z} \\
\mathcal{L} &:&=\left\langle \mathcal{G},\mathcal{V}\right\rangle ,~\mathcal{%
L}_{i}\equiv D_{i}\mathcal{L}:=\left\langle \mathcal{G},\mathcal{V}%
_{i}\right\rangle ,  \label{L}
\end{eqnarray}
with $\left\langle \cdot ,\cdot \right\rangle $ denoting the symplectic
product defined in the flat symplectic bundle constructed over the special K%
\"{a}hler-Hodge manifold $M_{v}$. We adopt the notation of
\cite{K-rev, AoB, KMPR}; see also App. \ref{STU-Details}.

By using the results of \cite{CDF, M-Horizon, FMY-FD}, one can prove the
following \textquotedblleft two-centered" special K\"{a}hler identities:
\begin{eqnarray}
\frac{1}{2}\left\langle \mathcal{Q},\mathcal{G}\right\rangle &=&-\text{Im}%
\left( \mathcal{Z}\overline{\mathcal{L}}\right) +\text{Im}\left( \mathcal{Z}%
_{i}\overline{\mathcal{L}}^{i}\right) =-\text{Im}\left( \mathcal{Z}\overline{%
\mathcal{L}}-\mathcal{Z}_{i}\overline{\mathcal{L}}^{i}\right) ;  \label{111}
\\
-\frac{1}{2}\mathcal{Q}^{T}\mathcal{M}\left( \mathcal{N}\right) \mathcal{G}
&=&\text{Re}\left( \mathcal{Z}\overline{\mathcal{L}}\right) +\text{Re}\left(
\mathcal{Z}_{i}\overline{\mathcal{L}}^{i}\right) =\text{Re}\left( \mathcal{Z}%
\overline{\mathcal{L}}+\mathcal{Z}_{i}\overline{\mathcal{L}}^{i}\right) ;
\label{112} \\
\frac{1}{2}\mathcal{Q}^{T}\mathcal{M}\left( \mathcal{F}\right) \mathcal{G}
&=&-\text{Re}\left( \mathcal{Z}\overline{\mathcal{L}}\right) +\text{Re}%
\left( \mathcal{Z}_{i}\overline{\mathcal{L}}^{i}\right) =-\text{Re}\left(
\mathcal{Z}\overline{\mathcal{L}}-\mathcal{Z}_{i}\overline{\mathcal{L}}%
^{i}\right) ,  \label{113}
\end{eqnarray}%
where $\overline{\mathcal{L}}^{i}=g^{i\bar{\jmath}}\overline{\mathcal{L}}_{%
\bar{\jmath}}$, and $\mathcal{N}=\mathcal{N}_{\Lambda \Sigma }$ and $%
\mathcal{F}=F_{\Lambda \Sigma }$ respectively are the (complexified) kinetic
vector matrix and the Hessian matrix of the prepotential $F$.
The
symplectic, real, symmetric $\left( 2n+2\right) \times \left( 2n+2\right) $
matrix $\mathcal{M}\left( \mathcal{N}\right) $ is defined as\begin{equation}
\mathcal{M}\left( \mathcal{N}\right) =\left(
\begin{array}{cc}
\text{Im}\left( \mathcal{N}\right) +\text{Re}\left( \mathcal{N}\right) \text{Im}^{-1}\left( \mathcal{N}\right) \text{Re}\left( \mathcal{N}\right)  & \;-\text{Re}\left( \mathcal{N}\right) \text{Im}^{-1}\left( \mathcal{N}\right)
\\
-\text{Im}^{-1}\left( \mathcal{N}\right) \text{Re}\left( \mathcal{N}\right)
& \text{Im}^{-1}\left( \mathcal{N}\right)
\end{array}\right) ,\label{M(N)}
\end{equation}and $\mathcal{M}\left( \mathcal{F}\right) $ is defined the same way, with $\mathcal{N}_{\Lambda \Sigma }\rightarrow F_{\Lambda \Sigma }$. In terms of
the covariantly holomorphic sections $\mathcal{V}$ and of its covariant
derivatives $\mathcal{V}_{i}$, such two matrices have the following
expressions (see e.g. \cite{M-Horizon} and Refs. therein):
\begin{align}
\mathcal{M}(\mathcal{N})& =\Omega \left( \mathcal{V\bar{V}}^{T}+\mathcal{\bar{V}V}^{T}+\mathcal{V}_{i}\,g^{i\bar{\jmath}}\mathcal{\bar{V}}_{\bar{\jmath}}^{T}+\mathcal{\bar{V}}_{\bar{\jmath}}g^{\bar{\jmath}i}\mathcal{V}_{i}^{T}\right) \Omega \, \label{M(N)-2}\\
\mathcal{M}(\mathcal{F})& =\Omega \left( \mathcal{V\bar{V}}^{T}+\mathcal{\bar{V}V}^{T}-\mathcal{V}_{i}\,g^{i\bar{\jmath}}\mathcal{\bar{V}}_{\bar{\jmath}}^{T}-\mathcal{\bar{V}}_{\bar{\jmath}}g^{\bar{\jmath}i}\mathcal{V}_{i}^{T}\right) \Omega \,\ \label{M(F)}
\end{align}where\begin{equation}
\Omega :=\left(
\begin{array}{cc}
0 & -\mathbb{I}_{n+1} \\
\mathbb{I}_{n+1} & 0\end{array}\right)
\end{equation}is the symplectic metric. Thus:
\begin{eqnarray}
\frac{1}{2}\mathcal{Q}^{T}\mathcal{M}\left( \mathcal{F}\right) \mathcal{G}+%
\frac{\mathbf{i}}{2}\left\langle \mathcal{Q},\mathcal{G}\right\rangle &=&-%
\mathcal{Z}\overline{\mathcal{L}}+\mathcal{Z}_{i}\overline{\mathcal{L}}^{i};
\\
-\frac{1}{2}\mathcal{Q}^{T}\mathcal{M}\left( \mathcal{N}\right) \mathcal{G}-%
\frac{\mathbf{i}}{2}\left\langle \mathcal{Q},\mathcal{G}\right\rangle &=&%
\mathcal{Z}\overline{\mathcal{L}}+\mathcal{L}_{i}\overline{\mathcal{Z}}^{i}.
\end{eqnarray}%
By denoting with
\begin{eqnarray}
\mathfrak{H} &:&=e^{-K/2}\mathcal{V};  \label{hol-sec} \\
\mathfrak{H}_{i} &:&=e^{-K/2}\mathcal{V}_{i},  \label{hol-seci}
\end{eqnarray}%
the holomorphic symplectic sections (such that $\partial _{\bar{\imath}}%
\mathfrak{H}=0$ and $\partial _{\bar{\imath}}\mathfrak{H}_{j}=0$), using the
properties (cf. e.g. \cite{K-rev, AoB})
\begin{eqnarray}
\left\langle \mathfrak{H},\overline{\mathfrak{H}}\right\rangle &=&-\mathbf{i}%
e^{-K};  \label{g1} \\
\left\langle \mathfrak{H}_{i},\overline{\mathfrak{H}}_{\bar{\jmath}%
}\right\rangle &=&\mathbf{i}e^{-K}g_{i\overline{\bar{\jmath}}},
\end{eqnarray}%
and defining the superpotential $\mathcal{W}$ and \textquotedblleft
gauging-superpotential" $\mathcal{Y}$ respectively as
\begin{eqnarray}
\mathcal{W} &:&=e^{-K/2}\mathcal{Z},~\mathcal{W}_{i}:=e^{-K/2}\mathcal{Z}%
_{i}, \\
\mathcal{Y} &:&=e^{-K/2}\mathcal{L},~\mathcal{Y}_{i}:=e^{-K/2}\mathcal{L}%
_{i},
\end{eqnarray}%
(\ref{111})-(\ref{113}) can be rewritten as follows:
\begin{eqnarray}
\frac{1}{2}\left\langle \mathcal{Q},\mathcal{G}\right\rangle &=&-\mathbf{i}%
e^{2K}\text{Im}\left( \mathcal{W}\overline{\mathcal{Y}}\right) \left\langle
\mathfrak{H},\overline{\mathfrak{H}}\right\rangle -\mathbf{i}e^{2K}\text{Im}%
\left( \mathcal{W}_{i}\overline{\mathcal{Y}}_{\bar{\jmath}}\right)
\left\langle \mathfrak{H}^{\bar{\jmath}},\overline{\mathfrak{H}}%
^{i}\right\rangle ; \\
-\frac{1}{2}\mathcal{Q}^{T}\mathcal{M}\left( \mathcal{N}\right) \mathcal{G}
&=&\mathbf{i}e^{2K}\text{Re}\left( \mathcal{W}\overline{\mathcal{Y}}\right)
\left\langle \mathfrak{H},\overline{\mathfrak{H}}\right\rangle -\mathbf{i}%
e^{2K}\text{Re}\left( \mathcal{W}_{i}\overline{\mathcal{Y}}_{\bar{\jmath}%
}\right) \left\langle \mathfrak{H}^{\bar{\jmath}},\overline{\mathfrak{H}}%
^{i}\right\rangle ; \\
\frac{1}{2}\mathcal{Q}^{T}\mathcal{M}\left( \mathcal{F}\right) \mathcal{G}
&=&-\mathbf{i}e^{2K}\text{Re}\left( \mathcal{W}\overline{\mathcal{Y}}\right)
\left\langle \mathfrak{H},\overline{\mathfrak{H}}\right\rangle -\mathbf{i}%
e^{2K}\text{Re}\left( \mathcal{W}_{i}\overline{\mathcal{Y}}_{\bar{\jmath}%
}\right) \left\langle \mathfrak{H}^{\bar{\jmath}},\overline{\mathfrak{H}}%
^{i}\right\rangle .
\end{eqnarray}

\section{\label{symm SKG}Symmetric very special geometry and quartic
2-polarizations}

The above identities hold in the projective special K\"{a}hler geometry of
the vector multiplets' scalar manifold $M_{v}$, regardless of the data
specifying such a manifold.

Instead, we will now specialize the treatment by assuming $M_{v}$ to be a
\textit{symmetric} (homogeneous) coset space, whose (local) geometry is
given by a cubic holomorphic prepotential
\begin{equation}
F\left( X\right) :=\frac{1}{3!}d_{ijk}\frac{X^{i}X^{j}X^{k}}{X^{0}},
\label{F}
\end{equation}
with the cubic symmetric constant tensor $d_{ijk}$ satisfying the so-called
\textit{adjoint identity},
\begin{equation}
d_{i(kl|}d_{j|mn)}d^{ijp}=\frac{4}{3}\delta_{(k}^{p}d^{\phantom p}_{lmn)},
\label{adj-id}
\end{equation}
or equivalently
\begin{equation}
C_{i(kl|}C_{j|mn)}\overline{C}^{ijp}=\frac{4}{3}\delta _{(k}^{p}C^{\phantom %
p}_{lmn)},  \label{add}
\end{equation}
where $C_{ijk}$ is the K\"{a}hler covariantly holomorphic rank-3 symmetric
tensor occurring in the identities
\begin{eqnarray}
D_{i}\mathcal{V}_{j} &=&\mathbf{i}C_{ijk}\overline{\mathcal{V}}^{k}, \\
R_{i\bar{\jmath}k\overline{l}} &=&-g_{i\overline{\bar{\jmath}}}g_{k\overline{%
l}}-g_{i\overline{l}}g_{k\bar{\jmath}}+C_{ikm}\overline{C}_{\bar{\jmath}
\overline{l}\overline{p}}g^{m\overline{p}},
\end{eqnarray}
with $R_{i\bar{\jmath}k\overline{l}}$ denoting the Riemann tensor of $M_{v}$%
. The $d_{ijk}$'s and duality structures of the corresponding $M_{v}$'s have
been classified in \cite{dWVP} and \cite{dWVVP}.

In this framework, the ring of invariant homogeneous polynomials under the
non-transitive action of the electric-magnetic duality group on its
representation space $\mathbf{R}$ in which both the aforementioned
symplectic vectors $\mathcal{Q}$ (\ref{Q}) and $\mathcal{G}$ (\ref{G}) sit,
is granted to be one-dimensional, and finitely generated by a \textit{%
primitive}\footnote{%
Primitivity of $I_{4}$, i.e. the fact that the corresponding invariant
tensor $K_{MNPQ}$ cannot be reduced in terms of other lower-rank tensors,
generally holds for all symmetric $M_{v}$'s characterized by the cubic
holomorphic prepotential (\ref{F}). However, in a peculiar sub-class of BPS
black holes, treated in Sec. \ref{note-BPS}, $I_{4}$ becomes a \textit{%
perfect square} (and so are all its non-vanishing 2-polarizations; cf. (\ref%
{ja1})-(\ref{ja5})).} quartic homogeneous polynomial, denoted by $I_{4}$ and
associated to the rank-4 completely symmetric invariant tensor $K_{MNPQ}$
\cite{Marrani} (see also \cite{Alek} and \cite{Kac-80}); for instance,
considering the symplectic vector $\mathcal{Q}$ (\ref{Q})$\in \mathbf{R}$,
one can define
\begin{equation}
I_{4}\left( \mathcal{Q},\mathcal{Q},\mathcal{Q},\mathcal{Q}\right) :=\frac{1%
}{2}K_{MNPQ}\mathcal{Q}^{M}\mathcal{Q}^{N}\mathcal{Q}^{P}\mathcal{Q}^{Q}.
\end{equation}
The explicit expression of the rank-4 invariant symmetric tensor $%
K_{MNPQ}=K_{(MNPQ)}$ is given by (D.1) of \cite{Special Road} in the
so-called 4D/5D special coordinate symplectic frame \cite{dWVVP,CFM}, as
well as by Eq. (5.36) of \cite{CFMZ1} and by (4.4)-(4.14) of Sec. 4.3 of
\cite{FMY-Inv} in a way independent from the symplectic frame (and
manifestly invariant under diffeomorphisms in $M_{v}$).

For the treatment given in the present paper, we will need to explicitly
compute the \textit{2-polarizations} of $I_{4}$ \cite{FMOSY, ADFMT, FMY-Inv,
VG-Plethysm} :
\begin{equation}
I_{4}\left( \mathcal{Q}+\mathcal{G},\mathcal{Q}+\mathcal{G},\mathcal{Q}+%
\mathcal{G},\mathcal{Q}+\mathcal{G}\right) =:\mathbf{I}_{2}+4\mathbf{I}%
_{1}+6 \mathbf{I}_{0}+4\mathbf{I}_{-1}+\mathbf{I}_{-2},
\end{equation}
where\footnote{%
Note that throughout our treatment $\left\vert \mathcal{Z}%
_{i}\right\vert^{2} $ and $\left\vert \mathcal{L}_{i}\right\vert ^{2}$ are
shorthand for $\sum_{i,j=1}^{n}\mathcal{Z}_{i}\overline{\mathcal{Z}}_{\bar{%
\jmath}}g^{i\bar{\jmath}}$ and $\sum_{i,j=1}^{n}\mathcal{L}_{i}\overline{%
\mathcal{L}}_{\bar{\jmath}}g^{i\bar{\jmath}}$, respectively (unless
otherwise specified).}
\begin{eqnarray}
\mathbf{I}_{2} &:&=I_{4}\left( \mathcal{Q},\mathcal{Q},\mathcal{Q},\mathcal{Q%
}\right) =\frac{1}{2}K_{MNPQ}\mathcal{Q}^{M}\mathcal{Q}^{N}\mathcal{Q}^{P}
\mathcal{Q}^{Q}  \label{I2} \\
&=&-\left( p^{0}q_{0}+p^{i}q_{i}\right) ^{2}+\frac{2}{3}
q_{0}d_{ijk}p^{i}p^{j}p^{k}-\frac{2}{3}
p^{0}d^{ijk}q_{i}q_{j}q_{k}+d_{ijk}d^{ilm}p^{j}p^{k}q_{l}q_{m}  \label{I2-1}
\\
&=&\left( \left\vert \mathcal{Z}\right\vert ^{2}-\left\vert \mathcal{Z}%
_{i}\right\vert ^{2}\right) ^{2}-\frac{4}{3}\text{Im}\left( \mathcal{Z}
\overline{C}_{\bar{\imath}\bar{\jmath}\overline{k}}\mathcal{Z}^{\bar{\imath}%
} \mathcal{Z}^{\overline{\bar{\jmath}}}\mathcal{Z}^{\overline{k}}\right)
-g^{i\overline{\bar{\jmath}}}C_{ikl}\overline{C}_{\overline{\bar{\jmath}}%
\overline{m}\overline{n}}\overline{\mathcal{Z}}^{k}\overline{\mathcal{Z}}%
^{l} \mathcal{Z}^{\overline{m}}\mathcal{Z}^{\overline{n}};  \label{I2-2}
\end{eqnarray}
\begin{eqnarray}
\mathbf{I}_{1} &:&=I_{4}\left( \mathcal{Q},\mathcal{Q},\mathcal{Q},\mathcal{G%
}\right) =\frac{1}{2}K_{MNPQ}\mathcal{Q}^{M}\mathcal{Q}^{N}\mathcal{Q}^{P}%
\mathcal{G}^{Q}  \label{I1} \\
&=&-\frac{1}{2}\left[ \left( p^{0}\right) ^{2}q_{0}g_{0}+p^{0}g^{0}q_{0}^{2}%
\right] -\frac{1}{2}\left( p^{i}p^{j}q_{i}g_{j}+p^{i}g^{j}q_{i}q_{j}\right)
\notag \\
&&-\frac{1}{2}\left(
p^{0}q_{0}p^{i}g_{i}+p^{0}q_{0}g^{i}q_{i}+p^{0}g_{0}p^{i}q_{i}+g^{0}q_{0}p^{i}q_{i}\right)
\notag \\
&&+\frac{1}{6}\left(
g_{0}d_{ijk}p^{i}p^{j}p^{k}+3q_{0}d_{ijk}p^{i}p^{j}g^{k}\right) -\frac{1}{6}%
\left( g^{0}d^{ijk}q_{i}q_{j}q_{k}+3p^{0}d^{ijk}q_{i}q_{j}g_{k}\right)
\notag \\
&&+\frac{1}{2}d_{ijk}d^{ilm}\left(p^{j}p^{k}q_{l}g_{m}+p^{j}g^{k}q_{l}q_{m}%
\right)  \label{I1-1} \\
&=&\frac{1}{2}\left( \left\vert \mathcal{Z}\right\vert ^{2}-\left\vert
\mathcal{Z}_{i}\right\vert ^{2}\right) \left( \mathcal{Z}\overline{\mathcal{L%
}}+\overline{\mathcal{Z}}\mathcal{L}-\mathcal{Z}^{\overline{\bar{\jmath}}}
\overline{\mathcal{L}}_{\overline{\bar{\jmath}}}-\overline{\mathcal{Z}}^{j}
\mathcal{L}_{j}\right)  \notag \\
&&-\frac{1}{3}\text{Im}\left[ \left( 3\mathcal{ZL}^{\overline{k}}+\mathcal{LZ%
}^{\overline{k}}\right) \overline{C}_{\bar{\imath}\overline{\bar{\jmath}}
\overline{k}}\mathcal{Z}^{\bar{\imath}}\mathcal{Z}^{\overline{\bar{\jmath}}} %
\right] -\frac{1}{2}g^{i\overline{\bar{\jmath}}}C_{ikl}\overline{C}_{%
\overline{\bar{\jmath}}\overline{m}\overline{n}}\overline{\mathcal{Z}}^{k}
\mathcal{Z}^{\overline{n}}\left( \overline{\mathcal{Z}}^{l}\mathcal{L}^{%
\overline{m}}+\overline{\mathcal{L}}^{l}\mathcal{Z}^{\overline{m}}\right) ;
\end{eqnarray}
\begin{eqnarray}
\mathbf{I}_{0} &:&=I_{4}\left( \mathcal{Q},\mathcal{Q},\mathcal{G},\mathcal{G%
}\right) =\frac{1}{2}K_{MNPQ}\mathcal{Q}^{M}\mathcal{Q}^{N}\mathcal{G}^{P}%
\mathcal{G}^{Q}  \label{I0} \\
&=&-\frac{1}{6}\left[ \left( p^{0}\right) ^{2}g_{0}^{2}+\left(
g^{0}\right)^{2}q_{0}^{2}+4p^{0}g^{0}q_{0}g_{0}\right]  \notag \\
&&-\frac{1}{6}\left[ \left( p^{i}g_{i}\right) ^{2}+\left(
g^{i}q_{i}\right)^{2}+2p^{i}g^{j}q_{i}g_{j}+2p^{i}g^{j}q_{j}g_{i}\right]
\notag \\
&&-\frac{1}{3}%
\left(p^{0}q_{0}g^{i}g_{i}+p^{0}g_{0}p^{i}g_{i}+p^{0}g_{0}g^{i}q_{i}+g^{0}q_{0}p^{i}g_{i}+g^{0}q_{0}g^{i}q_{i}+g^{0}g_{0}p^{i}q_{i}\right)
\notag \\
&&+\frac{1}{3}\left(g_{0}d_{ijk}p^{i}p^{j}g^{k}+q_{0}d_{ijk}p^{i}g^{j}g^{k}%
\right) -\frac{1}{3}\left(
g^{0}d^{ijk}q_{i}q_{j}g_{k}+p^{0}d^{ijk}q_{i}g_{j}g_{k}\right)  \notag \\
&&+\frac{1}{6}d_{ijk}d^{ilm}\left(
p^{j}p^{k}g_{l}g_{m}+4p^{j}g^{k}q_{l}g_{m}+g^{j}g^{k}q_{l}q_{m}\right) \\
&=&\frac{1}{3}\left( \left\vert \mathcal{Z}\right\vert ^{2}-\left\vert
\mathcal{Z}_{i}\right\vert ^{2}\right) \left( \left\vert \mathcal{L}%
\right\vert ^{2}-\left\vert \mathcal{L}_{i}\right\vert ^{2}\right) +\frac{1}{%
6}\left( \mathcal{Z}\overline{\mathcal{L}}+\overline{\mathcal{Z}}\mathcal{L}%
- \mathcal{Z}^{\bar{\imath}}\overline{\mathcal{L}}_{\bar{\imath}}-\overline{%
\mathcal{Z}}^{i}\mathcal{L}_{i}\right) ^{2}  \notag \\
&&-\frac{2}{3}\text{Im}\left[ \left( \mathcal{ZL}^{\overline{k}}+\mathcal{LZ}%
^{\overline{k}}\right) \overline{C}_{\bar{\imath}\bar{\jmath}\overline{k}}
\mathcal{Z}^{\bar{\imath}}\mathcal{L}^{\bar{\jmath}}\right]  \notag \\
&&-\frac{1}{6}g^{i\bar{\jmath}}C_{ikl}\overline{C}_{\bar{\jmath}\overline{m}
\overline{n}}\left( 4\overline{\mathcal{Z}}^{k}\mathcal{Z}^{\overline{m}}
\overline{\mathcal{L}}^{l}\mathcal{L}^{\overline{n}}+\overline{\mathcal{Z}}%
^{k} \overline{\mathcal{Z}}^{l}\mathcal{L}^{\overline{m}}\mathcal{L}^{%
\overline{n}}+\mathcal{Z}^{\overline{m}}\mathcal{Z}^{\overline{n}}\overline{%
\mathcal{L}}^{k}\overline{\mathcal{L}}^{l}\right) ;
\end{eqnarray}
\begin{eqnarray}
\mathbf{I}_{-1} &:&=I_{4}\left( \mathcal{Q},\mathcal{G},\mathcal{G},\mathcal{%
G}\right) =\frac{1}{2}K_{MNPQ}\mathcal{Q}^{M}\mathcal{G}^{N}\mathcal{G}^{P}
\mathcal{G}^{Q}  \label{I-1} \\
&=&-\frac{1}{2}\left[ \left( g^{0}\right) ^{2}q_{0}g_{0}+p^{0}g^{0}g_{0}^{2}%
\right] -\frac{1}{2}\left( p^{i}g^{j}g_{i}g_{j}+g^{i}g^{j}q_{i}g_{j}\right)
\notag \\
&&-\frac{1}{2}%
\left(p^{0}g_{0}g^{i}g_{i}+g^{0}q_{0}g^{i}g_{i}+g^{0}g_{0}p^{i}g_{i}+g^{0}g_{0}g^{i}q_{i}\right)
\notag \\
&&+\frac{1}{6}\left(3g_{0}d_{ijk}p^{i}g^{j}g^{k}+q_{0}d_{ijk}g^{i}g^{j}g^{k}%
\right) -\frac{1}{6}\left(
3g^{0}d^{ijk}q_{i}g_{j}g_{k}+p^{0}d^{ijk}g_{i}g_{j}g_{k}\right)  \notag \\
&&+\frac{1}{2}d_{ijk}d^{ilm}\left(p^{j}g^{k}g_{l}g_{m}+g^{j}g^{k}q_{l}g_{m}%
\right) \\
&=&\frac{1}{2}\left( \left\vert \mathcal{L}\right\vert ^{2}-\left\vert
\mathcal{L}_{i}\right\vert ^{2}\right) \left( \mathcal{Z}\overline{\mathcal{L%
}}+\overline{\mathcal{Z}}\mathcal{L}-\mathcal{Z}^{\bar{\jmath}} \overline{%
\mathcal{L}}_{\bar{\jmath}}-\overline{\mathcal{Z}}^{j}\mathcal{L}_{j}\right)
\notag \\
&&-\frac{1}{3}\text{Im}\left[ \left( \mathcal{ZL}^{\overline{k}}+3\mathcal{LZ%
}^{\overline{k}}\right) \overline{C}_{\bar{\imath}\bar{\jmath}\overline{k}}
\mathcal{L}^{\bar{\imath}}\mathcal{L}^{\bar{\jmath}}\right] -\frac{1}{2}g^{i%
\bar{\jmath}}C_{ikl} \overline{C}_{\bar{\jmath}\overline{m}\overline{n}}%
\overline{\mathcal{L}}^{k}\mathcal{L}^{\overline{n}}\left( \mathcal{Z}^{%
\overline{m}}\overline{\mathcal{L}}^{l}+\overline{\mathcal{Z}}^{l}\mathcal{L}%
^{\overline{m}}\right) ;
\end{eqnarray}
\begin{eqnarray}
\mathbf{I}_{-2} &:&=I_{4}\left( \mathcal{G},\mathcal{G},\mathcal{G},\mathcal{%
G}\right) =\frac{1}{2}K_{MNPQ}\mathcal{G}^{M}\mathcal{G}^{N}\mathcal{G}^{P}%
\mathcal{G}^{Q}  \label{I-2} \\
&=&-\left( g^{0}g_{0}+g^{i}g_{i}\right) ^{2}+\frac{2}{3}%
g_{0}d_{ijk}g^{i}g^{j}g^{k}-\frac{2}{3}
g^{0}d^{ijk}g_{i}g_{j}g_{k}+d_{ijk}d^{ilm}g^{j}g^{k}g_{l}g_{m}  \label{I-2-1}
\\
&=&\left( \left\vert \mathcal{L}\right\vert ^{2}-\left\vert \mathcal{L}%
_{i}\right\vert ^{2}\right) ^{2}-\frac{4}{3}\text{Im}\left( \mathcal{L}
\overline{C}_{\bar{\imath}\bar{\jmath}\overline{k}}\mathcal{L}^{\bar{\imath}}%
\mathcal{L}^{\bar{\jmath}}\mathcal{L}^{\overline{k}}\right) -g^{i\bar{\jmath}%
}C_{ikl}\overline{C}_{\bar{\jmath}\overline{m}\overline{n}}\overline{%
\mathcal{L}}^{k} \overline{\mathcal{L}}^{l}\mathcal{L}^{\overline{m}}%
\mathcal{L}^{\overline{n}}.  \label{I-2-2}
\end{eqnarray}
Notice that
\begin{eqnarray}
\mathbf{I}_{-1} &=&\left. \mathbf{I}_{1}\right\vert _{\mathcal{Q}%
\leftrightarrow \mathcal{G}}; \\
\mathbf{I}_{-2} &=&\left. \mathbf{I}_{2}\right\vert _{\mathcal{Q}\rightarrow
\mathcal{G}}; \\
\mathbf{I}_{0} &=&\left. \mathbf{I}_{0}\right\vert _{\mathcal{Q}
\leftrightarrow \mathcal{G}}.
\end{eqnarray}
Furthermore, $\mathbf{I}_{2}$, $\mathbf{I}_{1}$, $\mathbf{I}_{0}$, $\mathbf{I%
}_{-1}$ and $\mathbf{I}_{-2}$ form a spin-2 (quintet) representation of the
would-be horizontal symmetry $SL_{h}(2,\mathbb{R})$ acting on the doublet $%
\left( \mathcal{Q},\mathcal{G}\right) ^{T}$ \cite{FMOSY}.


\section{\label{BPSBPS}BPS black hole entropy...}

In \cite{DG}, within the assumption of mutual non-locality\footnote{%
Actually, \cite{DG} only dealt with spherical horizon ($\kappa =1$). The
derivation of BPS conditions for hyperbolic horizon ($\kappa =-1$) was done
in \cite{KPR}.}
\begin{equation}
\left\langle \mathcal{G},\mathcal{Q}\right\rangle =-\kappa ,  \label{nonloc}
\end{equation}
the general form of the ($\frac{1}{4}$-)BPS conditions were obtained to read
\begin{eqnarray}
\mathcal{Z} &=&\mathbf{i}\kappa S\mathcal{L};  \label{BPS1} \\
\mathcal{Z}_{i} &=&\mathbf{i}\kappa S\mathcal{L}_{i},  \label{BPS2}
\end{eqnarray}
where $S$ denotes the Bekenstein-Hawking entropy\footnote{%
In units of $\pi $, as understood throughout all the treatment.} of the
extremal BPS black hole solution. Note that, by virtue of the identity (\ref%
{111}), the mutual non-locality condition (\ref{nonloc}) can be rewritten as
\begin{equation}
2\text{Im}\left( \mathcal{Z}\overline{\mathcal{L}}-\mathcal{Z}_{i}\overline{%
\mathcal{L}}^{i}\right) =-\kappa .  \label{nonloc2}
\end{equation}

From (\ref{BPS1})-(\ref{BPS2}), one obtains the following expressions of the
Bekenstein-Hawking entropy $S$ of BPS extremal black holes (no sum on
repeated indices)
\begin{equation}
S=-\mathbf{i}\kappa \frac{\mathcal{Z}}{\mathcal{L}}=-\mathbf{i}\kappa \frac{%
\mathcal{Z}_{j}}{\mathcal{L}_{j}},~\forall j,  \label{S-BPS}
\end{equation}%
implying
\begin{equation}
S=-\mathbf{i}\kappa \frac{\mathcal{Z}}{\mathcal{L}}\frac{\overline{\mathcal{L%
}}}{\overline{\mathcal{L}}}=-\mathbf{i}\kappa \frac{\mathcal{Z}\overline{%
\mathcal{L}}}{\left\vert \mathcal{L}\right\vert ^{2}}=-\frac{\mathbf{i}%
\kappa }{2\left\vert \mathcal{L}\right\vert ^{2}}\left( \mathcal{Z}\overline{%
\mathcal{L}}-\overline{\mathcal{Z}}\mathcal{L}\right) =\frac{\kappa \text{Im}%
\left( \mathcal{Z}\overline{\mathcal{L}}\right) }{\left\vert \mathcal{L}%
\right\vert ^{2}},  \label{SSS1}
\end{equation}%
and
\begin{equation}
S=-\mathbf{i}\kappa \frac{\mathcal{Z}_{j}\overline{\mathcal{L}}^{j}}{\mathcal{L}_{j}\overline{\mathcal{L}}^{j}}=\frac{-\mathbf{i}\kappa }{2\mathcal{L}_{j}\overline{\mathcal{L}}^{j}}\left( \mathcal{Z}_{j}\overline{\mathcal{L}}^{j}-\overline{\mathcal{Z}}_{\bar{\jmath}}\mathcal{L}^{\bar{\jmath}}\right) =\frac{\kappa \text{Im}\left( \mathcal{Z}_{j}\overline{\mathcal{L}}^{j}\right) }{\mathcal{L}_{j}\overline{\mathcal{L}}^{j}},
\label{SSS2}
\end{equation}where no summation on repeated indices is performed. From (%
\ref{SSS1}) and (\ref{SSS2}), one obtains
\begin{gather}
\left\vert \mathcal{L}\right\vert ^{2}S-\left\vert \mathcal{L}%
_{i}\right\vert ^{2}S=\kappa \text{Im}\left( \mathcal{Z}\overline{\mathcal{L}%
}-\mathcal{Z}_{i}\overline{\mathcal{L}}^{i}\right) =\frac{\kappa }{2}%
\left\langle \mathcal{G},\mathcal{Q}\right\rangle ; \\
\Updownarrow  \notag \\
S\left( \left\vert \mathcal{L}\right\vert ^{2}-\left\vert \mathcal{L}%
_{i}\right\vert ^{2}\right) =\frac{\kappa }{2}\left\langle \mathcal{G},%
\mathcal{Q}\right\rangle ,  \label{P}
\end{gather}%
where we recall that $n$ is the number of vector multiplets (or,
equivalently, the complex dimension of $M_{v}$), and the identity (\ref{111}%
) has been used in the last step.

We should here also recall another expression for the BPS entropy in the
case $\kappa =1$, obtained in \cite{DG} by studying the near-horizon
dynamics :
\begin{equation}
S=2\left( \left\vert \mathcal{Z}_{i}\right\vert ^{2}-\left\vert \mathcal{Z}
\right\vert ^{2}\right) =\frac{1}{2\left( \left\vert \mathcal{L}%
_{i}\right\vert ^{2}-\left\vert \mathcal{L}\right\vert ^{2}\right) }=
\mathcal{Q}^{T}\mathcal{M}(\mathcal{F})\mathcal{Q}=\frac{1}{\mathcal{G}^{T}
\mathcal{M}(\mathcal{F})\mathcal{G}},  \label{S-BPS2}
\end{equation}
implying
\begin{equation}
S=\sqrt{\frac{\left\vert \mathcal{Z}\right\vert ^{2}-\left\vert \mathcal{Z}%
_{i}\right\vert ^{2}}{\left\vert \mathcal{L}\right\vert ^{2}-\left\vert
\mathcal{L}_{i}\right\vert ^{2}}}= \sqrt{\frac{\mathcal{Q}^{T}\mathcal{M}(%
\mathcal{F})\mathcal{Q}}{\mathcal{G}^{T}\mathcal{M}(\mathcal{F})\mathcal{G}}}%
,  \label{S-BPS3}
\end{equation}
where we have recalled the symplectic-invariant quadratic form of projective
special K\"{a}hler geometry defined by the matrix $\mathcal{M}(\mathcal{F})$ %
(see (\ref{M(N)})-(\ref{M(F)})). We also observe that (\ref{P}%
) (with $\kappa =1$) and (\ref{S-BPS2}) consistently imply (\ref{nonloc})
with $\kappa =1$, because
\begin{equation}
S\left( \left\vert \mathcal{L}\right\vert ^{2}-\left\vert \mathcal{L}%
_{i}\right\vert ^{2}\right) =\frac{1}{2}\left\langle \mathcal{G},\mathcal{Q}
\right\rangle \Leftrightarrow \left\langle \mathcal{G},\mathcal{Q}
\right\rangle =-1.
\end{equation}

By plugging (\ref{BPS1})-(\ref{BPS2}) into (\ref{id-2}), one obtains
\begin{equation}
\mathbf{i}\mathcal{Q}=\kappa \,S\mathcal{G}+2\mathcal{Z}\overline{\mathcal{V}%
}-2\mathcal{Z}^{\bar{\imath}}\overline{\mathcal{V}}_{\bar{\imath}},
\label{Q1-BPS}
\end{equation}
a relation which holds at the event horizon of the BPS extremal black hole.%
\newline

Within these conventions, we can write the following symplectic products,
that will be useful later,
\begin{gather}
\langle \mathcal{V},\overline{\mathcal{V}}\rangle =-\mathbf{i},\qquad
\left\langle \mathcal{V}_{i},\overline{\mathcal{V}}_{\overline{\bar{\jmath}}%
}\right\rangle =\mathbf{i}g_{i\overline{\bar{\jmath}}},\qquad \left\langle
\mathcal{V}_{i},\overline{\mathcal{V}}^{j}\right\rangle =\mathbf{i}%
\delta_{i}^{\,\,j},\qquad \left\langle \overline{\mathcal{V}}_{\bar{\imath}},%
\mathcal{V}^{\overline{\bar{\jmath}}}\right\rangle =-\mathbf{i}\delta _{\bar{%
\imath}}^{\,\,\bar{\jmath}},  \label{SKG-ids-1} \\
\left\langle \overline{\mathcal{V}},\mathcal{V}_{i}\right\rangle=\left%
\langle \overline{\mathcal{V}}_{\bar{\imath}},\mathcal{V}\right\rangle=0,%
\qquad \left\langle \mathcal{V}_{i},\mathcal{V}_{j}\right\rangle
=\left\langle \overline{\mathcal{V}}_{\bar{\imath}},\overline{\mathcal{V}}_{%
\bar{\jmath}}\right\rangle =0,\qquad \left\langle \mathcal{V},\mathcal{V}%
_{i}\right\rangle =\left\langle \overline{\mathcal{V}}\,, \overline{\mathcal{%
V}}_{\bar{\imath}}\right\rangle =0.  \label{SKG-ids-2}
\end{gather}
We note that, by using the BPS relation (\ref{Q1-BPS}), the contractions
between $\mathbf{i}\mathcal{Q}$ and the symplectic sections allow to
retrieve again the BPS relations (\ref{BPS1})-(\ref{BPS2}):
\begin{gather}
\mathbf{i}\left\langle \mathcal{V},\mathcal{Q}\right\rangle =-\mathbf{i}
\mathcal{Z}=\kappa S\mathcal{L}\Leftrightarrow \mathcal{Z}=\mathbf{i}\kappa
S \mathcal{L};  \label{BPS-1} \\
\mathbf{i}\left\langle \mathcal{V}_{i},\mathcal{Q}\right\rangle =-\mathbf{i}
\mathcal{Z}_{i}=\kappa S\mathcal{L}_{i}\Leftrightarrow \mathcal{Z}_{i}=%
\mathbf{i}\kappa S\mathcal{L}_{i}.  \label{BPS-2}
\end{gather}

Moreover, ($\frac{1}{4}$-)BPS conditions (\ref{BPS1})-(\ref{BPS2}) yield
\begin{eqnarray}
\left\langle \mathcal{Q},\mathcal{G}\right\rangle &=&-2\kappa
S\left(\left\vert \mathcal{L}\right\vert ^{2}-\left\vert \mathcal{L}%
_{i}\right\vert^{2}\right) ;  \label{first} \\
\mathcal{Q}^{T}\mathcal{M}\left( \mathcal{N}\right) \mathcal{G} &=&\mathcal{Q%
}^{T}\mathcal{M}\left( \mathcal{F}\right) \mathcal{G}=0.
\end{eqnarray}
Note that (\ref{nonloc}) and (\ref{first}) yield the generalization of the
last term of the r.h.s. of (\ref{S-BPS2}) to $\kappa =\pm 1$, namely
\begin{equation}
S=\frac{\kappa }{2\left( \left\vert \mathcal{L}_{i}\right\vert^{2}-\left%
\vert \mathcal{L}\right\vert ^{2}\right) }=\frac{\kappa }{\mathcal{G}^{T}%
\mathcal{M}\left( \mathcal{N}\right) \mathcal{G}}.  \label{S-gen-k}
\end{equation}
Remarkably, in Sec. \ref{BPS-gen} we will obtain a generalization, given by (%
\ref{resss}), of the first term in the r.h.s. of (\ref{S-BPS2}) holding for
both cases $\kappa =\pm 1$.


\section{\label{rela}...and its relations with quartic 2-polarizations}

The BPS conditions and their properties discussed above hold in the
projective special K\"{a}hler geometry of the vector multiplets' scalar
manifold $M_{v}$, regardless of the data specifying such a manifold.

Once again, we will now specialize the treatment by assuming $M_{v}$ to be a
\textit{symmetric} (homogeneous) coset space, associated to the cubic
holomorphic prepotential $F$ (\ref{F}). In this framework, we are going to
determine the relations among the BPS\ black hole entropy $S$ and the
various 2-polarizations of the quartic invariant introduced in Sec. \ref%
{symm SKG}.

In order to do this, we start and consider the contraction of the duality
invariant quartic structure $\frac{1}{2}K_{MNPQ}$ with the \textquotedblleft
algebraic BPS conditions\textquotedblright\ given by (\ref{Q1-BPS}). To this
aim, from (\ref{Q1-BPS}) we get
\begin{equation}
\mathcal{Q}^{M}+\mathbf{i}\kappa \,S\mathcal{G}^{M}=2\mathbf{i}\left( -%
\mathcal{Z}\overline{\mathcal{V}}^{M}+\mathcal{Z}^{\bar{\imath}}\overline{%
\mathcal{V}}_{\bar{\imath}}^{M}\right) ,
\end{equation}
whose l.h.s. and r.h.s. can then be contracted as follows:
\begin{eqnarray}
\left\langle \mathcal{Q}+\mathbf{i}\kappa \,S\mathcal{G},\mathcal{Q}+\mathbf{%
i}\kappa \,S\mathcal{G}\right\rangle &=&0; \\
\left( \mathcal{Q}^{M}+\mathbf{i}\kappa \,S\mathcal{G}^{M}\right) ^{T}%
\mathcal{M}(\mathcal{N})\left( \mathcal{Q}^{M}+\mathbf{i}\kappa \,S\mathcal{G%
}^{M}\right) &=&0; \\
\left( \mathcal{Q}^{M}+\mathbf{i}\kappa \,S\mathcal{G}^{M}\right) ^{T}%
\mathcal{M}(\mathcal{F})\left( \mathcal{Q}^{M}+\mathbf{i}\kappa \,S\mathcal{G%
}^{M}\right) &=&0,
\end{eqnarray}
and
\begin{eqnarray}
&&\frac{1}{2}K_{MNPQ}\left( \mathcal{Q}^{M}+\mathbf{i}\kappa \,S\mathcal{G}%
^{M}\right) \left( \mathcal{Q}^{N}+\mathbf{i}\kappa \,S\mathcal{G}%
^{N}\right) \left( \mathcal{Q}^{P}+\mathbf{i}\kappa \,S\mathcal{G}%
^{P}\right) \left( \mathcal{Q}^{Q}+\mathbf{i}\kappa \,S\mathcal{G}^{Q}\right)
\notag \\
&=&8\cdot \frac{1}{2}K_{MNPQ}\left( -\mathcal{Z}\overline{\mathcal{V}}^{M}+%
\mathcal{Z}^{\bar{\imath}}\overline{\mathcal{V}}_{\bar{\imath}}^{M}\right)
\left( -\mathcal{Z}\overline{\mathcal{V}}^{N}+\mathcal{Z}^{\bar{\imath}}
\overline{\mathcal{V}}_{\bar{\imath}}^{N}\right) \left( -\mathcal{Z}
\overline{\mathcal{V}}^{P}+\mathcal{Z}^{\bar{\imath}}\overline{\mathcal{V}}_{%
\bar{\imath}}^{P}\right) \left( -\mathcal{Z}\overline{\mathcal{V}}^{Q}+
\mathcal{Z}^{\bar{\imath}}\overline{\mathcal{V}}_{\bar{\imath}}^{Q}\right) .
\notag \\
&&  \label{rel}
\end{eqnarray}

Let us start from the l.h.s. of Eq. (\ref{rel}), which, by recalling the
definitions (\ref{I2})-(\ref{I-2}), reads
\begin{eqnarray}
&&\frac{1}{2}K_{MNPQ}\left( \mathcal{Q}^{M}+\mathbf{i}\kappa \,S\mathcal{G}%
^{M}\right) \left( \mathcal{Q}^{N}+\mathbf{i}\kappa \,S\mathcal{G}%
^{N}\right) \left( \mathcal{Q}^{P}+\mathbf{i}\kappa \,S\mathcal{G}%
^{P}\right) \left( \mathcal{Q}^{Q}+\mathbf{i}\kappa \,S\mathcal{G}^{Q}\right)
\notag \\
&=&\frac{1}{2}K_{MNPQ}\mathcal{Q}^{M}\mathcal{Q}^{N}\mathcal{Q}^{P}\mathcal{Q%
}^{Q}+4\mathbf{i}\kappa \,S\cdot \frac{1}{2}K_{MNPQ}\mathcal{Q}^{M}\mathcal{Q%
}^{N}\mathcal{Q}^{P}\mathcal{G}^{Q}  \notag \\
&&-6S^{2}\cdot \frac{1}{2}K_{MNPQ}\mathcal{Q}^{M}\mathcal{Q}^{N}\mathcal{G}%
^{P}\mathcal{G}^{Q}  \notag \\
&&-4\mathbf{i}\kappa S^{3}\cdot \frac{1}{2}K_{MNPQ}\mathcal{Q}^{M}\mathcal{G}%
^{N}\mathcal{G}^{P}\mathcal{G}^{Q}+S^{4}\cdot \frac{1}{2}K_{MNPQ}\mathcal{G}%
^{M}\mathcal{G}^{N}\mathcal{G}^{P}\mathcal{G}^{Q}  \notag \\
&=&\mathbf{I}_{2}-6S^{2}\mathbf{I}_{0}+S^{4}\mathbf{I}_{-2}+4\mathbf{i}
\kappa \,S\left( \mathbf{I}_{1}-S^{2}\mathbf{I}_{-1}\right) .  \label{rel1}
\end{eqnarray}
On the other hand, the r.h.s. of Eq. (\ref{rel}) reads
\begin{eqnarray}
&&8\cdot \frac{1}{2}K_{MNPQ}\left( -\mathcal{Z}\overline{\mathcal{V}}^{M}+%
\mathcal{Z}^{\bar{\imath}}\overline{\mathcal{V}}_{\bar{\imath}}^{M}\right)
\left( -\mathcal{Z}\overline{\mathcal{V}}^{N}+\mathcal{Z}^{\bar{\imath}}
\overline{\mathcal{V}}_{\bar{\imath}}^{N}\right) \left( -\mathcal{Z}
\overline{\mathcal{V}}^{P}+\mathcal{Z}^{\bar{\imath}}\overline{\mathcal{V}}_{%
\bar{\imath}}^{P}\right) \left( -\mathcal{Z}\overline{\mathcal{V}}^{Q}+%
\mathcal{Z}^{\bar{\imath}}\overline{\mathcal{V}}_{\bar{\imath}}^{Q}\right)
\notag \\
&=&8\mathcal{Z}^{4}\cdot \frac{1}{2}K_{MNPQ}\overline{\mathcal{V}}^{M}
\overline{\mathcal{V}}^{N}\overline{\mathcal{V}}^{P}\overline{\mathcal{V}}%
^{Q}-32\mathcal{Z}^{3}\mathcal{Z}^{\bar{\imath}}\cdot \frac{1}{2}K_{MNPQ}
\overline{\mathcal{V}}^{M}\overline{\mathcal{V}}^{N}\overline{\mathcal{V}}%
^{P}\overline{\mathcal{V}}_{\bar{\imath}}^{Q}  \notag \\
&&+48\mathcal{Z}^{2}\mathcal{Z}^{\bar{\imath}}\mathcal{Z}^{\bar{\jmath}%
}\cdot \frac{1}{2}K_{MNPQ}\overline{\mathcal{V}}^{M}\overline{\mathcal{V}}%
^{N}\overline{\mathcal{V}}_{\bar{\imath}}^{P}\overline{\mathcal{V}}_{\bar{%
\jmath}}^{Q} -32\mathcal{ZZ}^{\bar{\imath}}\mathcal{Z}^{\bar{\jmath}}%
\mathcal{Z}^{\bar{k}}\cdot \frac{1}{2}K_{MNPQ}\overline{\mathcal{V}}^{M}%
\overline{\mathcal{V}}_{\bar{\imath}}^{N}\overline{\mathcal{V}}_{\bar{\jmath}%
}^{P} \overline{\mathcal{V}}_{\bar{k}}^{Q}  \notag \\
&&+8\mathcal{Z}^{\bar{\imath}}\mathcal{Z}^{\bar{\jmath}}\mathcal{Z}^{\bar{k}%
} \mathcal{Z}^{\bar{l}}\cdot \frac{1}{2}\Omega _{MNPQ}\overline{\mathcal{V}}%
_{ \bar{\imath}}^{M}\overline{\mathcal{V}}_{\bar{\jmath}}^{N}\overline{%
\mathcal{V}}_{\bar{k}}^{P} \overline{\mathcal{V}}_{\bar{l}}^{Q}.
\label{rel2}
\end{eqnarray}

The vanishing of each term of the r.h.s. (\ref{rel2}) of Eq. (\ref{rel}) can
be proved without performing any computation\footnote{%
An explicit computation is presented in App. \ref{App-Explicit}.}, as
follows. Through the expressions (\ref{I2})-(\ref{I-2-2}), the five
two-centered invariants $\mathbf{I}_{2}$, $\mathbf{I}_{1}$, $\mathbf{I}_{0}$%
, $\mathbf{I}_{-1}$ and $\mathbf{I}_{-2}$ are quartic \textit{homogeneous}
polynomials in the respective variables:
\begin{eqnarray}
\mathbf{I}_{2} &:&=\frac{1}{2}K_{MNPQ}\mathcal{Q}^{M}\mathcal{Q}^{N}\mathcal{%
Q}^{P}\mathcal{Q}^{Q}=\left. \mathbf{I}_{2}\left( \mathcal{Z},\mathcal{Z}%
_{i},\overline{\mathcal{Z}},\overline{\mathcal{Z}}_{\bar{\imath}}\right)
\right\vert _{\mathcal{Z}:=\left\langle \mathcal{Q},\mathcal{V}%
\right\rangle,~\mathcal{Z}_{i}:=\left\langle \mathcal{Q},\mathcal{V}%
_{i}\right\rangle };  \label{I-two} \\
\mathbf{I}_{1} &:&=\frac{1}{2}K_{MNPQ}\mathcal{Q}^{M}\mathcal{Q}^{N}\mathcal{%
Q}^{P}\mathcal{G}^{Q}  \notag \\
&=&\left. \mathbf{I}_{1}\left( \mathcal{Z},\mathcal{Z}_{i},\overline{%
\mathcal{Z}},\overline{\mathcal{Z}}_{\bar{\imath}},\mathcal{L},\mathcal{L}%
_{i},\overline{\mathcal{L}},\overline{\mathcal{L}}_{\bar{\imath}},\right)
\right\vert _{\mathcal{Z}:=\left\langle \mathcal{Q},\mathcal{V}%
\right\rangle,~\mathcal{Z}_{i}:=\left\langle \mathcal{Q},\mathcal{V}%
_{i}\right\rangle ,~\mathcal{L}:=\left\langle \mathcal{G},\mathcal{V}%
\right\rangle ,~\mathcal{L}_{i}:= \left\langle \mathcal{G},\mathcal{V}%
_{i}\right\rangle };  \label{I-one} \\
\mathbf{I}_{0} &:&=\frac{1}{2}K_{MNPQ}\mathcal{Q}^{M}\mathcal{Q}^{N}\mathcal{%
G}^{P}\mathcal{G}^{Q}  \notag \\
&=&\left. \mathbf{I}_{0}\left( \mathcal{Z},\mathcal{Z}_{i},\overline{%
\mathcal{Z}},\overline{\mathcal{Z}}_{\bar{\imath}},\mathcal{L},\mathcal{L}%
_{i},\overline{\mathcal{L}},\overline{\mathcal{L}}_{\bar{\imath}},\right)
\right\vert _{\mathcal{Z}:=\left\langle \mathcal{Q},\mathcal{V}%
\right\rangle,~\mathcal{Z}_{i}:=\left\langle \mathcal{Q},\mathcal{V}%
_{i}\right\rangle ,~ \mathcal{L}:=\left\langle \mathcal{G},\mathcal{V}%
\right\rangle ,~\mathcal{L}_{i}:=\left\langle \mathcal{G},\mathcal{V}%
_{i}\right\rangle };  \label{I-zero} \\
\mathbf{I}_{-1} &:&=\frac{1}{2}K_{MNPQ}\mathcal{Q}^{M}\mathcal{G}^{N}
\mathcal{G}^{P}\mathcal{G}^{Q}=\left. \mathbf{I}_{1}\right\vert _{\mathcal{Z}
\leftrightarrow \mathcal{L},~\mathcal{Z}_{i}\leftrightarrow \mathcal{L}_{i}};
\label{I-onee} \\
\mathbf{I}_{-2} &:&=\frac{1}{2}K_{MNPQ}\mathcal{G}^{M}\mathcal{G}^{N}
\mathcal{G}^{P}\mathcal{G}^{Q}=\left. \mathbf{I}_{2}\right\vert _{\mathcal{Z}
\rightarrow \mathcal{L},~\mathcal{Z}_{i}\rightarrow \mathcal{L}_{i}}.
\label{I-twoo}
\end{eqnarray}
Thus, one can proceed and evaluate
\begin{eqnarray}
\overline{\frac{1}{2}K_{MNPQ}\overline{\mathcal{V}}^{M}\overline{\mathcal{V}}%
^{N}\overline{\mathcal{V}}^{P}\overline{\mathcal{V}}^{Q}} &=&\frac{1}{2}
K_{MNPQ}\mathcal{V}^{M}\mathcal{V}^{N}\mathcal{V}^{P}\mathcal{V}^{Q} \overset%
{\text{(\ref{I-two})}}{=}\mathbf{I}_{2}\left( \mathbb{Y},\mathbb{Y}_{i},
\overline{\mathbb{Y}},\overline{\mathbb{Y}}_{\bar{\imath}}\right) =0;
\label{11} \\
\overline{\frac{1}{2}K_{MNPQ}\overline{\mathcal{V}}^{M}\overline{\mathcal{V}}%
^{N}\overline{\mathcal{V}}^{P}\overline{\mathcal{V}}_{\bar{\imath}}^{Q}} &=&
\frac{1}{2}K_{MNPQ}\mathcal{V}^{M}\mathcal{V}^{N}\mathcal{V}^{P}\mathcal{V}%
_{i}^{Q}\overset{\text{(\ref{I-one})}}{=}  \notag \\
&=&\mathbf{I}_{1}\left( \mathbb{Y},\mathbb{Y}_{j},\overline{\mathbb{Y}},
\overline{\mathbb{Y}}_{\bar{\jmath}},\mathbb{X}_{i},\mathbb{X}_{ij},
\overline{\mathbb{X}}_{\bar{\imath}},\overline{\mathbb{X}}_{\bar{\imath}\bar{%
\jmath}},\right) =0;  \label{12} \\
\overline{\frac{1}{2}K_{MNPQ}\overline{\mathcal{V}}^{M}\overline{\mathcal{V}}%
^{N}\overline{\mathcal{V}}_{\bar{\imath}}^{P}\overline{\mathcal{V}}_{\bar{%
\jmath}}^{Q}} &=&\frac{1}{2}K_{MNPQ}\mathcal{V}^{M}\mathcal{V}^{N}\mathcal{V}%
_{i}^{P} \mathcal{V}_{j}^{Q}\overset{\text{(\ref{I-zero})}}{=}  \notag \\
&=&\mathbf{I}_{0}\left( \mathbb{Y},\mathbb{Y}_{k},\overline{\mathbb{Y}},
\overline{\mathbb{Y}}_{\bar{k}},\mathbb{X}_{\mathbf{a}},\mathbb{X}_{\mathbf{a%
}k},\overline{\mathbb{X}}_{\mathbf{\bar{a}}},\overline{\mathbb{X}}_{\mathbf{%
\bar{a}}\bar{k}}\right) =0;  \label{13} \\
\overline{\frac{1}{2}K_{MNPQ}\overline{\mathcal{V}}^{M}\overline{\mathcal{V}}%
_{\bar{\imath}}^{N}\overline{\mathcal{V}}_{\bar{\jmath}}^{P}\overline{%
\mathcal{V}}_{\bar{k}}^{Q}} &=&\frac{1}{2}K_{MNPQ}\overline{\mathcal{V}}^{M}
\mathcal{V}_{i}^{N}\mathcal{V}_{j}^{P}\mathcal{V}_{k}^{Q}\overset{\text{(\ref%
{I-onee})}}{=}  \notag \\
&=&\left. \mathbf{I}_{1}\left( \mathbb{Y},\mathbb{Y}_{l},\overline{\mathbb{Y}%
},\overline{\mathbb{Y}}_{\bar{l}},\mathbb{X}_{\mathbf{b}},\mathbb{X}_{%
\mathbf{b}l},\overline{\mathbb{X}}_{\overline{\mathbf{b}}}, \overline{%
\mathbb{X}}_{\overline{\mathbf{b}}\bar{l}}\right) \right\vert _{\mathbb{Y}
\leftrightarrow \mathbb{X}_{\mathbf{b}},~\mathbb{Y}_{l}\leftrightarrow
\mathbb{X}_{\mathbf{b}l}}=0;  \label{14} \\
\overline{\frac{1}{2}K_{MNPQ}\overline{\mathcal{V}}_{\bar{\imath}}^{M}
\overline{\mathcal{V}}_{\bar{\jmath}}^{N}\overline{\mathcal{V}}_{\bar{k}%
}^{P} \overline{\mathcal{V}}_{\bar{l}}^{Q}} &=&\frac{1}{2}K_{MNPQ}\mathcal{V}%
_{i}^{M}\mathcal{V}_{j}^{N}\mathcal{V}_{k}^{P}\mathcal{V}_{l}^{Q}\overset{%
\text{(\ref{I-twoo})}}{=}  \notag \\
&=&\left. \mathbf{I}_{2}\left( \mathbb{Y},\mathbb{Y}_{m},\overline{\mathbb{Y}%
},\overline{\mathbb{Y}}_{\overline{m}}\right) \right\vert _{\mathbb{Y}
\rightarrow \mathbb{X}_{\mathbf{c}},~\mathbb{Y}_{m}\rightarrow \mathbb{X}_{%
\mathbf{c}m}}=0,  \label{15}
\end{eqnarray}
where $\mathbf{a}=i,j$, $\mathbf{b}=i,j,k$, and $\mathbf{c}=i,j,k,l$.
Crucially, in the last step of Eqs. (\ref{11})-(\ref{15}) the homogeneity of
the (suitable $n$-polarizations, with\footnote{%
Rigorously speaking, only the (1- and)2-polarizations of the quartic
structure $I_{4}$ are explicitly known. Nevertheless, this is immaterial for
the reasoning made here, because only the homogeneity (of degree 4) matters.}
$n=1,2,3,4$, of the) polynomials $\mathbf{I}_{2}$, $\mathbf{I}_{1}$, $%
\mathbf{I}_{0}$, $\mathbf{I}_{-1}$ and $\mathbf{I}_{-2}$ has been used,
implying the vanishing of (\ref{11})-(\ref{15}), because
\begin{eqnarray}
\mathbb{Y} &:&=\left\langle \mathcal{V},\mathcal{V}\right\rangle =0; \\
\mathbb{Y}_{m} &:&=\left\langle \mathcal{V},U_{m}\right\rangle =0; \\
\mathbb{X}_{\mathbf{c}} &:&=\left\langle U_{\mathbf{c}},\mathcal{V}
\right\rangle =0; \\
\mathbb{X}_{\mathbf{c}m} &:&=\left\langle U_{\mathbf{c}},U_{m}\right\rangle
=0,
\end{eqnarray}
as a consequence of the identities (\ref{SKG-ids-1})-(\ref{SKG-ids-2}).

Then, one can re-consider the equation (\ref{rel}),
\begin{equation}
\text{(\ref{rel})}\Leftrightarrow \mathbf{I}_{2}-6S^{2}\mathbf{I}_{0}+S^{4}
\mathbf{I}_{-2}+4\mathbf{i}\kappa \,S\left( \mathbf{I}_{1}-S^{2}\mathbf{I}%
_{-1}\right) =0,
\end{equation}
and obtain two relations between the quartic 2-polarizations and the (square
of) BPS entropy, namely
\begin{equation}
\mathbf{I}_{2}-6S^{2}\mathbf{I}_{0}+S^{4}\mathbf{I}_{-2}=0\Leftrightarrow
S^{2}\equiv S_{\pm }^{2}=\frac{3\mathbf{I}_{0}}{\mathbf{I}_{-2}}\pm \frac{%
\sqrt{36\mathbf{I}_{0}^{2}-4\mathbf{I}_{2}\mathbf{I}_{-2}}}{2\mathbf{I}_{-2}}
\label{rell-1}
\end{equation}
and
\begin{equation}
\mathbf{I}_{1}-S^{2}\mathbf{I}_{-1}=0\Leftrightarrow S^{2}=\frac{\mathbf{I}%
_{1}}{\mathbf{I}_{-1}}.  \label{rell-2}
\end{equation}
In turn, the consistency of such two expressions yield a polynomial cubic
constraint among the quartic 2-polarizations for BPS black holes :
\begin{gather}
\pm \sqrt{36\mathbf{I}_{0}^{2}-4\mathbf{I}_{2}\mathbf{I}_{-2}}=2\mathbf{I}%
_{-2}\left( \frac{\mathbf{I}_{1}}{\mathbf{I}_{-1}}-\frac{3\mathbf{I}_{0}}{%
\mathbf{I}_{-2}}\right) ; \\
\Downarrow  \notag \\
9\mathbf{I}_{0}^{2}\mathbf{I}_{-1}^{2}-\mathbf{I}_{2}\mathbf{I}_{-1}^{2}%
\mathbf{I}_{-2}=\left( \mathbf{I}_{1}\mathbf{I}_{-2}-3\mathbf{I}_{0}\mathbf{I%
}_{-1}\right) ^{2}=\mathbf{I}_{1}^{2}\mathbf{I}_{-2}^{2}+9\mathbf{I}_{0}^{2}
\mathbf{I}_{-1}^{2}-6\mathbf{I}_{1}\mathbf{I}_{0}\mathbf{I}_{-1}\mathbf{I}%
_{-2}; \\
\Updownarrow  \notag \\
-\mathbf{I}_{2}\mathbf{I}_{-1}^{2}\mathbf{I}_{-2}=\mathbf{I}_{1}^{2}\mathbf{I%
}_{-2}^{2}-6\mathbf{I}_{1}\mathbf{I}_{0}\mathbf{I}_{-1}\mathbf{I}_{-2}; \\
\Updownarrow _{\mathbf{I}_{-2}\neq 0}  \notag \\
\mathbf{I}_{2}\mathbf{I}_{-1}^{2}-6\mathbf{I}_{1}\mathbf{I}_{0}\mathbf{I}%
_{-1}+\mathbf{I}_{1}^{2}\mathbf{I}_{-2}=0.  \label{rell-3}
\end{gather}
To make contact with literature, by setting $\mathbf{I}_{2}$, $\mathbf{I}%
_{1} $, $-6\mathbf{I}_{0}$, $-\mathbf{I}_{-1}$ and $\mathbf{I}_{-2}$
respectively equal to $a_{0}$, $a_{2}$, $a_{4}$, $a_{6}$ and $a_{8}$, formul%
\ae\ (\ref{rell-1}), (\ref{rell-2}) and (\ref{rell-3}) respectively match
Eqs. (3.26), (3.27) and (3.28) of \cite{H1}, and moreover formul\ae\ (\ref%
{I2})-(\ref{I-2-2}) accomplish the task mentioned below Eq. (3.28)
therein.\bigskip

Consequently, for BPS black holes (in models with \textit{symmetric} vector
multiplets' scalar manifolds $M_{v}$), as far as the evaluation of the
2-polarizations of the quartic duality invariant $I_{4}$ and their relation
with BPS black hole entropy are concerned, three possibilities may arise:


\subsection{General case}

\begin{equation}
\mathbf{I}_{1}\neq 0\overset{\text{(\ref{rell-2})}}{\Leftrightarrow }\mathbf{%
I}_{-1}\neq 0.  \label{pc2}
\end{equation}
Both the expressions (\ref{rell-1}) and (\ref{rell-2}) hold true, with the
constraint (\ref{rell-3}).


\subsection{Vanishing of $\mathbf{I}_{2}$, $\mathbf{I}_{-2}$ or $\mathbf{I}%
_{0}$}

\begin{equation}
\mathbf{I}_{2}=0\overset{\text{(\ref{rell-3})}}{\Leftrightarrow }\mathbf{I}%
_{-2}=0\overset{\text{(\ref{rell-3})}}{\Leftrightarrow }\mathbf{I}_{0}=0.
\end{equation}
In this case (\ref{rell-1}) is meaningless, and only the expression (\ref%
{rell-2}) holds true.


\subsection{Vanishing of $\mathbf{I}_{1}$ or $\mathbf{I}_{-1}$}

\begin{equation}
\mathbf{I}_{1}=0\overset{\text{(\ref{rell-2})}}{\Leftrightarrow }\mathbf{I}%
_{-1}=0.  \label{pc1}
\end{equation}
In this case (\ref{rell-2}) is meaningless, and only the expression (\ref%
{rell-1}) holds true.


\subsubsection{\label{note-BPS}A noteworthy BPS sub-class}

A remarkable sub-class of BPS critical points, satisfying (\ref{BPS1})-(\ref%
{BPS2}), is characterized by the further condition
\begin{equation}
C_{ijk}\overline{\mathcal{L}}^{j}\overline{\mathcal{L}}^{k}=0,~\forall i,
\label{BPSsub1}
\end{equation}
which implies also
\begin{equation}
C_{ijk}\overline{\mathcal{Z}}^{j}\overline{\mathcal{Z}}^{k}=-S^{2}C_{ijk}
\overline{\mathcal{L}}^{j}\overline{\mathcal{L}}^{k}=0,~\forall i.
\label{BPSsub2}
\end{equation}
Thus, when $M_{v}$ is symmetric, at the BPS critical points which further
satisfy (\ref{BPSsub1})-(\ref{BPSsub2}), the 2-polarizations (\ref{I2})-(\ref%
{I-2-2}) of the quartic duality invariant $I_{4}$ read as follows\footnote{%
Even if it does not belong to the class of symmetric manifolds $M_{v}$'s
with cubic holomorphic prepotential (\ref{F}), the class of the so-called
\textit{minimally coupled} models of $\mathcal{N}=2$, $D=4$ supergravity
have $\overline{\mathbb{CP}}^{n}$, and thus symmetric, scalar manifolds \cite%
{Luciani}, and the corresponding quartic structure is \textit{non-primitive}%
, because it is reducible in terms of a quadratic symmetric invariant
structure (cf. e.g. \cite{FMO-minimal}, as well the treatment of Secs. 3 and
4 of \cite{FMY-Inv-Str}). In this class of models the condition (\ref%
{BPSsub1}) holds globally (and not only at BPS\ attractors) because $%
C_{ijk}=0$ globally. The BPS sub-class under consideration indeed
encompasses \textit{all} BPS critical points in such models. For the case $%
n=1$, see Sec. \ref{CP1}.}:
\begin{eqnarray}
\mathbf{I}_{2} &=&\left( \left\vert \mathcal{Z}\right\vert ^{2}-\left\vert
\mathcal{Z}_{i}\right\vert ^{2}\right) ^{2}=S^{4}\left( \left\vert \mathcal{L%
}\right\vert ^{2}-\left\vert \mathcal{L}_{i}\right\vert ^{2}\right) ^{2};
\label{ja1} \\
\mathbf{I}_{1} &=&0; \\
\mathbf{I}_{0} &=&\frac{1}{3}S^{2}\left( \left\vert \mathcal{L}%
\right\vert^{2}-\left\vert \mathcal{L}_{i}\right\vert ^{2}\right) ^{2}; \\
\mathbf{I}_{-1} &=&0; \\
\mathbf{I}_{-2} &=&\left( \left\vert \mathcal{L}\right\vert ^{2}-\left\vert
\mathcal{L}_{i}\right\vert ^{2}\right) ^{2},  \label{ja5}
\end{eqnarray}
yielding the relation
\begin{equation}
\mathbf{I}_{0}^{2}=\frac{\mathbf{I}_{2}\mathbf{I}_{-2}}{9}\Rightarrow 3
\mathbf{I}_{0}=\sqrt{\mathbf{I}_{2}\mathbf{I}_{-2}},  \label{rrel}
\end{equation}
as well as the simple expression of BPS entropy,
\begin{equation}
S^{4}=\frac{\mathbf{I}_{2}}{\mathbf{I}_{-2}}.  \label{rrel2}
\end{equation}

Interestingly, (\ref{rrel}) is the very condition of vanishing of the
radicand in the square root in Eq. (\ref{rell-1}), which indeed simplifies
(removing the inherent \textquotedblleft $\pm $\textquotedblright\
branching) down to
\begin{equation}
S^{2}=\frac{3\mathbf{I}_{0}}{\mathbf{I}_{-2}}\pm \frac{\sqrt{36\mathbf{I}%
_{0}^{2}-4\mathbf{I}_{2}\mathbf{I}_{-2}}}{2\mathbf{I}_{-2}}=\sqrt{\frac{%
\mathbf{I}_{2}}{\mathbf{I}_{-2}}},  \label{Ent}
\end{equation}
which is nothing but (\ref{rrel2}). In fact, the BPS sub-class under
consideration, defined by (\ref{rrel}), satisfies (\ref{pc1}).


\section{\label{Sec-Eff}Effective black hole potential formalism}

So far, we have been considering only BPS attractors; a generalization of
the whole treatment to encompass all classes of extremal BH attractors,
including the non-BPS ones\footnote{Non-BPS extremal BH solutions in supergravity with $U(1)$ FI gaugings have been discussed in literature, for instance in \cite{Gnecchi:2012kb}, in which the attractor mechanism and the scalar flow have been described by a first order formalism exploiting a suitably defined fake superpotential.}, can be achieved by exploiting the so-called
\textit{effective black hole potential} formalism. Indeed, regardless of
the specific data of the projective special K\"{a}hler geometry of the
vector multiplets's scalar manifolds as well as from the quaternionic K\"{a}%
hler geometry of the hypermultiplets' scalar manifolds, from the treatment
of \cite{BFMY-FI}, then extended to Abelian gaugings of hypermultiplets in
\cite{Chimento} and made manifestly symplectic-invariant in \cite{KPR}, the
near-horizon attractor dynamics of the equations of motion is known to be
governed by an \textit{effective black hole potential} function%
\footnote{In the specific example of the magnetic $STU$ model treated in Sec. \ref{STU}
and App. \ref{STU-Details}, the introduction of $V_{\text{eff}}$ is recalled
in App. \ref{App-Eff}.} $V_{\text{eff}}$, whose critical points can be
related to $\mathcal{Q}$'s (\ref{Q}) supporting extremal black hole
solutions in the $U(1)$ FI gauging of $\mathcal{N}=2$, $D=4$ supergravity
specified by $\mathcal{G}$ (\ref{G}). As specified at the start of this
paper, we will not be considering the coupling to hypermultiplets. As
resulting from Eqs. (\ref{Veff})-(\ref{S}), which we report here for
simplicity's sake, the Bekenstein-Hawking \cite{Bek, Haw} black hole entropy
$S$ (in units of $\pi $, as always understood) is expressed by\footnote{%
The evaluation at $\partial V_{\text{eff}}=0$ (which corresponds to $%
\partial _{i}V_{\text{eff}}=\partial _{u}V_{\text{eff}}=0$ $\forall
i,\forall u$) is understood throughout.}
\begin{eqnarray}
V_{\text{eff}} &:&\mathbf{=}\frac{1-\kappa \sqrt{\kappa ^{2}-4VV_{BH}}}{2V}
\label{VV} \\
S &=&\kappa V_{\text{eff}}\mathbf{=}\frac{\kappa -\sqrt{\kappa ^{2}-4VV_{BH}}%
}{2V},  \label{SS}
\end{eqnarray}%
where the (manifestly symplectic) effective black hole potential $V_{BH}$ in
the ungauged case \cite{AM2} and the manifestly symplectic-invariant gauge
potential $V$ \cite{DG} are defined by
\begin{eqnarray}
V_{BH} &:&=\left\vert \mathcal{Z}\right\vert ^{2}+\left\vert \mathcal{Z}%
_{i}\right\vert ^{2}=-\frac{1}{2}\mathcal{Q}^{T}\mathcal{M}\left( \mathcal{N}%
\right) \mathcal{Q};  \label{VVBH} \\
V &:&=-3\left\vert \mathcal{L}\right\vert ^{2}+\left\vert \mathcal{L}%
_{i}\right\vert ^{2}=\frac{1}{2}\mathcal{G}^{T}\mathcal{M}\left( \mathcal{F}%
\right) \mathcal{G}-2\left\vert \mathcal{L}\right\vert ^{2}.  \label{V}
\end{eqnarray}%
Note how $V_{BH}$ is non-negative by definition, whereas $V$ can have any
sign; in particular, the critical points of $V$ (evaluated at spatial
infinity) define the cosmological constant $\Lambda $ of the asymptotical
geometry of the black hole solution. For $\kappa =1$ (spherical horizon),
the (at least) local minima of $V_{\text{eff}}$ support extremal black
holes, whereas for $\kappa =-1$ (hyperbolic horizon) the (at least) local
maxima of $V_{\text{eff}}$ support extremal black holes. Moreover, we will
see below that for $\kappa =-1$ the effective potential $V_{\text{eff}}$
does not pertain to the entropy itself, but rather to the entropy density.
It is here worth pointing out the consistency condition for $V_{\text{eff}}$
(and thus for $S$),
\begin{equation}
1-4V_{BH}V\geqslant 0.  \label{pre-conds}
\end{equation}


\subsection{$\protect\kappa =1$}

This case has spherical near-horizon geometry $S^{2}$. The angular integral
is finite,
\begin{equation}
\int_{0}^{2\pi }d\varphi \int_{0}^{\pi }\sin \theta d\theta =4\pi ,
\end{equation}
and the Bekenstein-Hawking entropy-area formula holds,
\begin{equation}
\frac{S}{\pi }=\frac{A_{S^{2}}}{4\pi }=r_{H}^{2}=\left. V_{\text{eff}%
}\right\vert _{\partial V_{\text{eff}}=0},
\end{equation}
where $A_{S^{2}}=4\pi r_{H}^{2}$ is the area of the event horizon surface $%
S^{2}$ of radius $r_{H}$, and $\left. V_{\text{eff}}\right\vert _{\partial
V_{\text{eff}}=0}>0$ necessarily. Explicitly, it holds that
\begin{equation}
\frac{S}{\pi }=V_{\text{eff}}\mathbf{=}\frac{1-\sqrt{1-4VV_{BH}}}{2V}>0~%
\text{for~}\left\{
\begin{array}{l}
V<0; \\
\text{or} \\
V>0:1-4VV_{BH}\geqslant 0.%
\end{array}
\right.
\end{equation}
Note that in this case the symplectic vector of charges has magnetic and
electric components defined as (cf. (\ref{Q}))
\begin{eqnarray}
\mathcal{Q} &:&=\left( p^{\Lambda },q_{\Lambda }\right) ^{T}, \\
\text{where~}p^{\Lambda } &:&=\frac{1}{4\pi }\int_{S^{2}}F^{\Lambda},~q_{%
\Lambda }:=\frac{1}{4\pi }\int_{S^{2}}G_{\Lambda }.
\end{eqnarray}


\subsection{$\protect\kappa =-1$}

This case has hyperbolic near-horizon geometry $H^{2}$. The angular integral
\cite{Chimento}
\begin{equation}
\mathbf{V}:=\int_{H^{2}}\sinh \theta d\theta \wedge d\varphi
\end{equation}
diverges, and thus, strictly speaking, the black hole entropy $\mathbb{S}$
is infinite. However, for $\kappa =-1$ one can define the \textit{entropy
density}
\begin{equation}
S:=\frac{\mathbb{S}}{\mathbf{V}}=-\left. V_{\text{eff}}\right\vert_{\partial
V_{\text{eff}}=0},  \label{S-k=-1}
\end{equation}
which is finite and positive, and being given by the opposite of the
critical value of $V_{\text{eff}}$; thus, it must necessarily hold that $%
\left. V_{\text{eff}}\right\vert _{\partial V_{\text{eff}}=0}<0$.
Explicitly,
\begin{equation}
S=-V_{\text{eff}}\mathbf{=}\frac{-1-\sqrt{1-4VV_{BH}}}{2V},  \label{SS-1}
\end{equation}
which is not consistent for $V>0$, because in this case it would entail a
negative entropy density:
\begin{equation}
S=-\frac{\left( 1+\sqrt{1-4\left\vert V\right\vert V_{BH}}\right) }{%
2\left\vert V\right\vert }\overset{!}{<}0.
\end{equation}
Therefore, for $\kappa =-1$ the relation (\ref{SS}), or equivalently (\ref%
{SS-1}), is consistent only for $V<0$, for which
\begin{equation}
S=\frac{\left( 1+\sqrt{1+4\left\vert V\right\vert V_{BH}}\right) }{%
2\left\vert V\right\vert }>0.
\end{equation}
This is in line with the observation below (5.38) of \cite{Chimento}. Note
that in this case the symplectic vector of charges has magnetic and electric
components defined as (cf. (\ref{Q}))
\begin{equation}
\mathcal{Q}:=\left( p^{\Lambda },q_{\Lambda }\right) ^{T},
\end{equation}
where $p^{\Lambda }$ and $q_{\Lambda }$ are actually \textit{charge densities%
}, defined by the following expressions (cf. (3.10)-(3.11) of \label{KPR})
\begin{eqnarray}
p^{\Lambda } &:&=\frac{1}{\mathbf{V}}\int_{H^{2}}F^{\Lambda }\mathbf{;} \\
q_{\Lambda } &:&=\frac{1}{\mathbf{V}}\int_{H^{2}}G_{\Lambda }\mathbf{.}
\end{eqnarray}


\section{General properties of $V_{\text{eff}}$}

In $\mathcal{N}=2$, $D=4$ supergravity coupled to vector multiplets and with
$U(1)$ FI gauging, regardless of the specific data of the projective special
K\"{a}hler geometry of the vector multiplets' scalar manifolds, the
attractor flow in the near-horizon limit is governed by the critical points
(respectively minima for $\kappa =1$ and maxima for $\kappa =-1$) of the
\textit{effective black hole potential}\footnote{%
Note that, apart from the redefinitions $V_{BH~\text{them}}\rightarrow \frac{%
V_{BH~\text{us}}}{\left( 8\pi \right) ^{2}}$ and $V_{\text{them}}\rightarrow
\frac{V_{\text{us}}}{2}$ (cf. the last footnote of App. \ref{App-Eff}), the $%
V_{\text{eff}}$ defined in \cite{Chimento} is $\kappa $ times the $V_{\text{%
eff}}$ defined by (\ref{Veff}) or (\ref{VV}).} $V_{\text{eff}}$ (\ref{VV})
\cite{BFMY-FI, Chimento}, such that the attractors are critical points of $%
V_{\text{eff}}$, satisfying \cite{BFMY-FI}
\begin{equation}
\partial _{i}V_{\text{eff}}=\frac{2V^{2}\partial
_{i}V_{BH}-\left(2V_{BH}V+\kappa \sqrt{1-4V_{BH}V}-1\right) \partial _{i}V}{%
2V^{2}\sqrt{1-4V_{BH}V}}=0,~\forall i,  \label{crit}
\end{equation}
where
\begin{eqnarray}
\partial _{i}V_{BH} &=&2\overline{\mathcal{Z}}\mathcal{Z}_{i}+\mathbf{i}
C_{ijk}\overline{\mathcal{Z}}^{j}\overline{\mathcal{Z}}^{k};  \label{dVBH} \\
\partial _{i}V &=&-2\overline{\mathcal{L}}\mathcal{L}_{i}+\mathbf{i}C_{ijk}%
\overline{\mathcal{L}}^{j}\overline{\mathcal{L}}^{k}.  \label{dV}
\end{eqnarray}
Note how $V_{\text{eff}}$ can have any sign. Indeed, as recalled below (\ref%
{VVBH})-(\ref{V}), it holds that
\begin{eqnarray}
\left. V_{BH}\right\vert _{\partial V_{BH}=0} &=&S_{\text{ungauged}}>0; \\
\left. V\right\vert _{\partial V=0} &=&:\Lambda \gtreqless 0,
\end{eqnarray}
where $S_{\text{ungauged}}$ denotes the Bekenstein-Hawking entropy (in units
of $\pi $, as understood throughout) of the extremal black hole in ungauged $%
\mathcal{N}=2$, $D=4$ Maxwell-Einstein supergravity. The Bekenstein-Hawking
entropy $S$ of static extremal black hole solutions (with spherical or
hyperbolic near-horizon geometry, respectively corresponding to $\kappa =1$
and $\kappa =-1$) in $U(1)$ FI gauged $\mathcal{N}=2$, $D=4$ supergravity is
given by (\ref{SS}), in which we recall once more that the evaluation at the
horizon of the extremal black hole solution under consideration
(corresponding to the evaluation at $\partial _{i}V_{\text{eff}}=0$ (\ref%
{crit})), will be understood.

Note that the ungauged limit \cite{BFMY-FI}
\begin{equation}
\lim_{V\rightarrow 0}\left. V_{\text{eff}}\right\vert _{\partial _{i}V_{%
\text{eff}}=0}=\left. V_{BH}\right\vert _{\partial _{i}V_{BH}=0}=S_{\text{%
ungauged}}  \label{ungauged}
\end{equation}
exists only for $\kappa =1$ \cite{AM2}. Moreover, from the treatment given
at the start of Sec. \ref{Sec-Eff} at the critical points of $V_{\text{eff}}$
the following consistency conditions must hold:
\begin{equation}
\left\{
\begin{array}{l}
1-4V_{BH}V\geqslant 0; \\
V\neq 0.%
\end{array}
\right.  \label{conds}
\end{equation}
If such two conditions hold, then
\begin{equation}
\partial _{i}V_{\text{eff}}=0\Leftrightarrow 2V^{2}\partial_{i}V_{BH}-\left(
2V_{BH}V+\kappa \sqrt{1-4V_{BH}V}-1\right) \partial_{i}V=0,~\forall i.
\label{condd}
\end{equation}

Let us also notice that the saturation of the consistency bound (\ref%
{pre-conds}) corresponds to
\begin{equation}
1-4V_{BH}V=0\Leftrightarrow V=\frac{1}{4V_{BH}}.
\end{equation}
If such a saturation holds, the Bekenstein-Hawking black hole entropy $S$ of
the extremal black hole reads (manifestly specifying the evaluation at
critical points of $V_{\text{eff}}$) reads
\begin{equation}
\left. S\right\vert _{1-4V_{BH}V=0}=\left. \frac{\kappa }{2V}%
\right\vert_{\partial _{i}V_{\text{eff}}=0}=2\kappa \left. V_{BH}\right\vert
_{\partial_{i}V_{\text{eff}}=0},
\end{equation}
which however (for $\kappa =1$) is generally not the double of $S_{\text{%
ungauged}}=\left. V_{BH}\right\vert _{\partial _{i}V_{BH}=0}$ (cf. (\ref%
{ungauged})), because for $\kappa =1$ in general it holds that
\begin{equation}
\left. V_{BH}\right\vert _{\partial _{i}V_{\text{eff}}=0}\neq \left.
V_{BH}\right\vert _{\partial _{i}V_{BH}=0}.
\end{equation}


\section{\label{BPS-gen}Generalization of BPS entropy formula (\protect\ref%
{S-BPS2}) to $\protect\kappa =\pm 1$}

The BPS critical points generally satisfy the criticality conditions (\ref%
{crit}) as well as the BPS conditions (\ref{BPS1})-(\ref{BPS2}). By virtue
of the latter, at BPS critical points it holds that
\begin{equation}
V\overset{\text{(\ref{V})}}{=}-3\left\vert \mathcal{L}\right\vert^{2}+\left%
\vert \mathcal{L}_{i}\right\vert ^{2}=\frac{1}{S^{2}}\left(-3\left\vert
\mathcal{Z}\right\vert ^{2}+\left\vert \mathcal{Z}_{i}\right\vert
^{2}\right) .
\end{equation}
Therefore, for $\kappa =\pm 1$, by virtue of (\ref{SS}) and (\ref{VV}), the
BPS Bekenstein-Hawking entropy $S$ satisfies the following equation\footnote{%
The eveluation at the BPS conditions (\ref{BPS1})-(\ref{BPS2}) will be
henceforth understood.}:
\begin{gather}
S=\kappa V_{\text{eff}}=\frac{\kappa -\sqrt{1-\frac{4}{S^{2}}%
\left(\left\vert \mathcal{Z}\right\vert ^{2}+\left\vert \mathcal{Z}%
_{i}\right\vert^{2}\right) \left( -3\left\vert \mathcal{Z}\right\vert
^{2}+\left\vert \mathcal{Z}_{i}\right\vert ^{2}\right) }}{\frac{2}{S^{2}}%
\left( -3\left\vert \mathcal{Z}\right\vert ^{2}+\left\vert \mathcal{Z}%
_{i}\right\vert^{2}\right) };  \notag \\
\Updownarrow  \notag \\
\frac{2}{S}\left( -3\left\vert \mathcal{Z}\right\vert ^{2}+\left\vert
\mathcal{Z}_{i}\right\vert ^{2}\right) =\kappa -\sqrt{1-\frac{4}{S^{2}}
\left( \left\vert \mathcal{Z}\right\vert ^{2}+\left\vert \mathcal{Z}%
_{i}\right\vert ^{2}\right) \left( -3\left\vert \mathcal{Z}%
\right\vert^{2}+\left\vert \mathcal{Z}_{i}\right\vert ^{2}\right) };  \notag
\\
\Downarrow  \notag \\
1+\frac{4}{S^{2}}\left( -3\left\vert \mathcal{Z}\right\vert ^{2}+\left\vert
\mathcal{Z}_{i}\right\vert ^{2}\right) ^{2}-\frac{4\kappa }{S}\left(
-3\left\vert \mathcal{Z}\right\vert ^{2}+\left\vert \mathcal{Z}%
_{i}\right\vert ^{2}\right) =1-\frac{4}{S^{2}}\left( \left\vert \mathcal{Z}
\right\vert ^{2}+\left\vert \mathcal{Z}_{i}\right\vert ^{2}\right)
\left(-3\left\vert \mathcal{Z}\right\vert ^{2}+\left\vert \mathcal{Z}%
_{i}\right\vert ^{2}\right) ;  \notag \\
\Updownarrow  \notag \\
S=2\kappa \left( \left\vert \mathcal{Z}_{i}\right\vert ^{2}-\left\vert
\mathcal{Z}\right\vert ^{2}\right) =\kappa \mathcal{Q}^{T}\mathcal{M}(%
\mathcal{F})\mathcal{Q},  \label{resss}
\end{gather}
with consistency conditions given by
\begin{gather}
\left\vert \mathcal{Z}\right\vert ^{2}-\left\vert \mathcal{Z}%
_{i}\right\vert^{2}\lessgtr 0\Leftrightarrow \left\vert \mathcal{L}%
\right\vert^{2}-\left\vert \mathcal{L}_{i}\right\vert ^{2}\lessgtr 0,~\text{%
for~}\kappa=\pm 1; \\
\Updownarrow  \notag \\
\mathcal{Q}^{T}\mathcal{M}(\mathcal{F})\mathcal{Q}\gtrless 0\Leftrightarrow
\mathcal{G}^{T}\mathcal{M}(\mathcal{F})\mathcal{G}\gtrless 0,~\text{for~}%
\kappa =\pm 1.
\end{gather}
As announced below Eq. (\ref{S-gen-k}), Eq. (\ref{resss}), which holds true
regardless of the specific data of the projective special K\"{a}hler
geometry of the vector multiplets' scalar manifold $M_{v}$, provides the
generalization to $\kappa =\pm 1$ of the first term in the r.h.s. of (\ref%
{S-BPS2}) (which hold only for $\kappa =1$). By collecting Eqs. (\ref%
{S-gen-k}) and (\ref{resss}), one can thus write that for $\kappa =\pm 1$
the BPS entropy reads
\begin{equation}
S=2\kappa \left( \left\vert \mathcal{Z}_{i}\right\vert ^{2}-\left\vert
\mathcal{Z}\right\vert ^{2}\right) =\frac{\kappa }{2\left( \left\vert
\mathcal{L}_{i}\right\vert ^{2}-\left\vert \mathcal{L}\right\vert^{2}\right)
},  \label{S-gen-k-2}
\end{equation}
which thus generalizes (\ref{S-BPS2}) (obtained in \cite{DG} for $\kappa =-1$%
) to $\kappa =\pm 1$, and still implies (\ref{S-BPS3}).

Finally, let us remark that, by virtue of the BPS conditions (\ref{BPS1})-(%
\ref{BPS2}), at the BPS critical points of $V_{\text{eff}}$ the gradients of
$V_{BH}$ and $V$ become proportional (of a factor $-S^{2}$), namely :
\begin{gather}
\partial _{i}V_{BH}=S^{2}\left( 2\overline{\mathcal{L}}\mathcal{L}_{i}-%
\mathbf{i}C_{ijk}\overline{\mathcal{L}}^{j}\overline{\mathcal{L}}%
^{k}\right)=-S^{2}\partial _{i}V;  \label{BPS-crit} \\
\Updownarrow  \notag \\
S^{2}=-\frac{\partial _{i}V_{BH}}{\partial _{i}V},~\forall i\text{%
~(no~summation~on~}i\text{)}.  \label{BPS-crit-2}
\end{gather}
Again, (\ref{BPS-crit}) and (\ref{BPS-crit-2}) hold for any BPS extremal
black hole (with $\kappa =\pm 1$) in $\mathcal{N}=2$, $D=4$ $U(1)$ FI gauged
supergravity coupled to vector multiplets, regardless of their scalar
manifold $M_{v}$.


\section{\label{Class-VBH}Classification of critical points of $V_{BH}$}

The critical points of $V_{BH}$ pertain to extremal black hole attractors in
the ungauged limit, and by construction of $V_{BH}$ they are placed at the
(unique) event horizon of the extremal black hole; from (\ref{dVBH}), they
satisfy \cite{on-Some-Props}
\begin{equation}
\partial _{i}V_{BH}=2\overline{\mathcal{Z}}\mathcal{Z}_{i}+\mathbf{i}C_{ijk}
\overline{\mathcal{Z}}^{j}\overline{\mathcal{Z}}^{k}=0,~\forall i.
\label{dVBH=0}
\end{equation}
After \cite{FMY-FD}, the relation between $\mathbf{I}_{2}$ (\ref{I2})-(\ref%
{I2-2}) and $V_{BH}$ (\ref{VVBH}) at the critical points of $V_{BH}$ itself
reads
\begin{equation}
\mathbf{I}_{2}=V_{BH}^{2}-\frac{32}{3}\left\vert \mathcal{Z}%
\right\vert^{2}\left\vert \mathcal{Z}_{i}\right\vert ^{2}.
\end{equation}
From the treatment of \cite{on-Some-Props} (see also \cite{AoB} or \cite%
{K-rev} for a complete list of references), three classes of critical points
of $V_{BH}$ exist, namely:

\begin{description}
\item[$V_{BH}$.\textbf{I}] $\mathcal{Z}_{i}=0~\forall i$, and $\mathcal{Z}
\neq 0$ (for $\kappa =1$, corresponding to the would-be $\frac{1}{2}$-BPS
critical points in the ungauged limit), yielding
\begin{eqnarray}
V_{BH} &=&\left\vert \mathcal{Z}\right\vert ^{2}; \\
\mathbf{I}_{2} &=&V_{BH}^{2}=\left\vert \mathcal{Z}\right\vert ^{4}.
\end{eqnarray}

\item[$V_{BH}$.\textbf{II}] $\mathcal{Z}=0$, and $C_{ijk}\overline{\mathcal{Z%
}}^{j}\overline{\mathcal{Z}}^{k}=0~\forall i$ (for $\kappa =1$,
corresponding to the would-be non-BPS $Z=0$ critical points in the ungauged
limit), yielding
\begin{eqnarray}
V_{BH} &=&\left\vert \mathcal{Z}_{i}\right\vert ^{2}; \\
\mathbf{I}_{2} &=&V_{BH}^{2}=\left\vert \mathcal{Z}_{i}\right\vert ^{4}.
\end{eqnarray}

\item[$V_{BH}$.\textbf{III}] $\mathcal{Z}\neq 0$ and $\mathcal{Z}_{i}\neq 0$%
, such that (\ref{dVBH=0}) holds true (for $\kappa =1$, corresponding to the
would-be non-BPS $Z\neq 0$ critical points in the ungauged limit), yielding
(cfr. Secs. 4-6 of \cite{BMR1} and Refs. therein, and \cite{CFMZ1})
\begin{equation}
\left\vert \mathcal{Z}_{i}\right\vert ^{2}=3\left\vert \mathcal{Z}%
\right\vert ^{2}+\Delta _{\mathcal{Z}},  \label{Rule-3}
\end{equation}
and thus
\begin{eqnarray}
V_{BH} &=&4\left\vert \mathcal{Z}\right\vert ^{2}+\Delta _{\mathcal{Z}}; \\
\mathbf{I}_{2} &=&-16\left\vert \mathcal{Z}\right\vert ^{4}+\Delta _{%
\mathcal{Z}}^{2}-\frac{8}{3}\Delta _{\mathcal{Z}}\left\vert \mathcal{Z}%
\right\vert ^{2},
\end{eqnarray}
where $\Delta _{\mathcal{Z}}$ is defined as\footnote{%
Note that $\Delta _{\mathcal{Z}}$ (\ref{DeltaZ}) is generally complex, but
at critical points of $V_{BH}$ it is real, and it is such that $\left.
V_{BH}\right\vert _{\partial V_{BH}=0}\geqslant 0$.}
\begin{equation}
\Delta _{\mathcal{Z}}:=-\frac{1}{4}\frac{\left( D_{m}\overline{D}_{(\bar{%
\imath}}\overline{C}_{\bar{\jmath}\bar{k}\bar{l})}\right) \overline{\mathcal{%
Z}}^{m}\mathcal{Z}^{\bar{\imath}}\mathcal{Z}^{\bar{\jmath}}\mathcal{Z}^{\bar{%
k}} \mathcal{Z}^{\bar{l}}}{\overline{N}_{3}\left( \mathcal{Z}\right) },
\label{DeltaZ}
\end{equation}
where $N_{3}$ is the cubic form related to the tensor $C_{ijk}$ of special
geometry,
\begin{equation}
N_{3}(\overline{\mathcal{Z}})\equiv N_{3}(\overline{\mathcal{Z}},\overline{%
\mathcal{Z}},\overline{\mathcal{Z}}):=C_{ijk}\overline{\mathcal{Z}}^{i}%
\overline{\mathcal{Z}}^{j}\overline{\mathcal{Z}}^{k}.
\end{equation}
Note that $\Delta _{\mathcal{Z}}=0$ (\textit{at least} in) in symmetric
scalar manifolds (because in those cases $D_{m}\overline{D}_{(\bar{\imath}}
\overline{C}_{\bar{\jmath}\bar{k}\bar{l})}=0$ identically).
\end{description}


\section{\label{Class-V}Classification of critical points of $V$}

In an analogous way, one can classify the critical points of the (manifestly
symplectic invariant) potential $V$ of the Abelian $U(1)$ FI gauging in $%
\mathcal{N}=2$, $D=4$ supergravity. When placing the critical points of $V$
at the spatial asymptotical background of the extremal black hole solution,
they determine the type of flux vacua; in other words, the critical value of
$V$ at the asymptotical background determines the cosmological constant $%
\Lambda $. On the other hand, we will see in Sec. \ref{Class} that the
critical points of $V$ placed at the event horizon of the extremal black
hole will be relevant for the classification of the \textbf{class I} of
critical points of $V_{\text{eff}}$.

From (\ref{dV}), the critical points of $V$ satisfy
\begin{equation}
\partial _{i}V=-2\overline{\mathcal{L}}\mathcal{L}_{i}+\mathbf{i}C_{ijk}
\overline{\mathcal{L}}^{j}\overline{\mathcal{L}}^{k}=0,~\forall i.
\label{dV=0}
\end{equation}
From the definition of $\mathbf{I}_{-2}$ (\ref{I-2})-(\ref{I-2-2}) and $V$ (%
\ref{V}), at the critical points of $V$ it holds that%
\begin{equation}
\mathbf{I}_{-2}=V^{2}+\frac{8}{3}\left\vert \mathcal{L}\right\vert^{2}V=V%
\left( V+\frac{8}{3}\left\vert \mathcal{L}\right\vert ^{2}\right)=\left(
-3\left\vert \mathcal{L}\right\vert ^{2}+\left\vert \mathcal{L}%
_{i}\right\vert ^{2}\right) \left( -\frac{1}{3}\left\vert \mathcal{L}
\right\vert ^{2}+\left\vert \mathcal{L}_{i}\right\vert ^{2}\right) .
\end{equation}

Three classes of critical points of $V$ exist, namely

\begin{description}
\item[$V$.\textbf{I}] $\mathcal{L}_{i}=0~\forall i$, and $\mathcal{L}\neq 0$%
, yielding
\begin{eqnarray}
V &=&-3\left\vert \mathcal{L}\right\vert ^{2}; \\
\mathbf{I}_{-2} &=&\left\vert \mathcal{L}\right\vert ^{4}.
\end{eqnarray}
\end{description}

If placed at spatial infinity, this class would correspond to supersymmetric
anti-de Sitter (AdS$_{4}$) vacua ($\Lambda <0$).

\begin{description}
\item[$V$.\textbf{II}] $\mathcal{L}=0$, and $C_{ijk}\overline{\mathcal{L}}%
^{j}\overline{\mathcal{L}}^{k}=0~\forall i$, yielding%
\begin{eqnarray}
V &=&\left\vert \mathcal{L}_{i}\right\vert ^{2}; \\
\mathbf{I}_{-2} &=&\left\vert \mathcal{L}_{i}\right\vert ^{4}.
\end{eqnarray}
If placed at spatial infinity, this class would correspond to de Sitter (dS$%
_{4}$) vacua ($\Lambda >0$).

\item[$V$.\textbf{III}] $\mathcal{L}\neq 0$ and $\mathcal{L}_{i}\neq 0$,
such that (\ref{dV=0}) holds true. It can be proven that (see App. \ref%
{App-comp})
\begin{equation}
\left\vert \mathcal{L}_{i}\right\vert ^{2}=3\left\vert \mathcal{L}
\right\vert ^{2}+\Delta _{\mathcal{L}},  \label{Rule-3-2}
\end{equation}
and thus
\begin{eqnarray}
V &=&\Delta _{\mathcal{L}}; \\
\mathbf{I}_{-2} &=&\Delta _{\mathcal{L}}^{2}+\frac{8}{3}\left\vert \mathcal{L%
}\right\vert ^{2}\Delta _{\mathcal{L}},
\end{eqnarray}
where, analogously to (\ref{DeltaZ}), $\Delta _{\mathcal{L}}$ is defined as%
\footnote{%
Note that $\Delta _{\mathcal{L}}$ (\ref{DeltaL}) is generally complex, but
at critical points of $V$ it is real.}
\begin{equation}
\Delta _{\mathcal{L}}:=-\frac{1}{4}\frac{\left( D_{m}\overline{D}_{(\bar{%
\imath}}\overline{C}_{\bar{\jmath}\bar{k}\bar{l})}\right) \overline{\mathcal{%
L}}^{m}\mathcal{L}^{\bar{\imath}}\mathcal{L}^{\bar{\jmath}}\mathcal{L}^{\bar{%
k}}\mathcal{L}^{\bar{l}}}{\overline{N}_{3}(\mathcal{L})},  \label{DeltaL}
\end{equation}
where
\begin{equation}
N_{3}(\overline{\mathcal{L}})\equiv N_{3}(\overline{\mathcal{L}},\overline{%
\mathcal{L}},\overline{\mathcal{L}}):=C_{ijk}\overline{\mathcal{L}}^{i}%
\overline{\mathcal{L}}^{j}\overline{\mathcal{L}}^{k}.  \label{NN3}
\end{equation}
Note that $\Delta _{\mathcal{L}}=0$ (\textit{at least} in) in symmetric
scalar manifolds (because in those cases $D_{m}\overline{D}_{(\bar{\imath}}
\overline{C}_{\bar{\jmath}\bar{k}\bar{l})}=0$ identically). If placed at
spatial infinity, this class would correspond to dS$_{4}$ vacua, Minkowski$%
_{4}$ vacua or AdS$_{4}$ vacua depending on whether $\Lambda \gtreqqless
0\Leftrightarrow \Delta _{\mathcal{L}}\gtreqqless 0$.
\end{description}

Thus, it should be remarked that, (\textit{at least}) for symmetric vector
multiplets' scalar manifolds, \textit{each} class of flux vacua is
associated to \textit{only one} class of critical points of $V$ (placed at
the asymptotical background):
\begin{equation}
\begin{array}{cccc}
\text{supersymmetric~AdS}_{4}\text{ vacua} & \left( \Lambda <0\right) &
\Leftrightarrow & \text{class~}V\text{.\textbf{I}}; \\
\text{dS}_{4}\text{~vacua} & \left( \Lambda >0\right) & \Leftrightarrow &
\text{class~}V\text{.\textbf{II}}; \\
\text{Minkowski}_{4}\text{~vacua} & \left( \Lambda =0\right) &
\Leftrightarrow & \text{class~}V\text{.\textbf{III}}.%
\end{array}
\text{ }
\end{equation}


\section{\label{Class}Classification of critical points of $V_{\text{eff}}$}

For $\kappa =\pm 1$, from (\ref{condd}), regardless of $M_{v}$, only two
classes of critical points of $V_{\text{eff}}$ exist, placed at the (unique)
event horizon of the extremal black hole; namely\footnote{\textit{A priori},
(\ref{condd}) would imply that a third class of critical points,
characterized by $\partial _{i}V_{BH}=0$ and $\partial _{i}V\neq 0$, but
with $2V_{BH}V=1-\kappa \sqrt{1-4V_{BH}V}$, might exist. However, this class
is not consistent with the equations of motion; see \label{BFMY-FI} and Eq.
(5.35) of \cite{Chimento}.} :

\begin{description}
\item[\textbf{Class I}] corresponds to critical points of $V_{\text{eff}}$
which are critical points of both $V_{BH}$ and $V$, as well:
\begin{equation}
\left.
\begin{array}{r}
\partial _{i}V_{BH}=0; \\
\partial _{i}V=0;%
\end{array}
\right\} \Rightarrow \partial _{i}V_{\text{eff}}=0,~\forall i.  \label{uno}
\end{equation}
Again, the placement of the critical points of $V_{BH}$ and $V$ is at the
(unique) event horizon of the extremal black hole. The nine sub-classes of
\textbf{class I} will be listed and discussed below. Note that (\ref{uno})
is trivially consistent with (\ref{BPS-crit}); thus, we anticipate that the
\textbf{class I} of critical points of $V_{\text{eff}}$ includes various
sub-classes admitting BPS critical points (namely, sub-classes \textbf{I.1},
\textbf{I.5} and \textbf{I.9}, at which (\ref{BPS-crit-2}) is meaningless,
of course; see below).

\item[\textbf{Class II}] corresponds to critical points of $V_{\text{eff}}$
which are \textit{not} critical points of $V_{BH}$ \textit{nor} of $V$, with
the gradients of $V_{BH}$ and of $V$ being proportional:
\begin{equation}
\partial _{i}V_{BH}=\frac{\left( 2V_{BH}V+\kappa \sqrt{1-4V_{BH}V}-1\right)}{%
2V^{2}}\partial _{i}V\overset{\text{(\ref{VV})}}{=}\frac{\left( V_{BH}-V_{%
\text{eff}}\right) }{V}\partial _{i}V\Rightarrow \partial _{i}V_{\text{eff}%
}=0,~\forall i.  \label{due}
\end{equation}
For what concerns the \textbf{BPS sector}, by exploiting the BPS\ conditions
(\ref{BPS1})-(\ref{BPS2}) within this class, and using (\ref{SS}), (\ref{VV}%
) and (\ref{BPS-crit-2}), one trivially obtains that the entropy of the BPS
extremal black holes of \textbf{class II} satisfies the square of (\ref{SS}%
):
\begin{equation}
S^{2}\overset{\text{(\ref{BPS-crit-2}),~}\forall i}{=}-\frac{%
\partial_{i}V_{BH}}{\partial _{i}V}=\frac{\left( 1-2V_{BH}V-\kappa \sqrt{%
1-4V_{BH}V}\right) }{2V^{2}}=V_{\text{eff}}^{2}.  \label{due-bis}
\end{equation}
\end{description}


\subsection{\label{Class-I}\textbf{Class I}}

The \textbf{class I} of critical points splits into $9$ sub-classes, given
by the combinatorial product (denoted by \textquotedblleft $\otimes $") of
classes of critical points of $V_{BH}$ and $V$:
\begin{equation}
\underset{\text{crit.~pts~of~}V_{BH}}{\left\{
\begin{array}{l}
V_{BH}\text{.1} \\
V_{BH}\text{.2} \\
V_{BH}\text{.3}%
\end{array}
\right. }\otimes \underset{\text{crit.~pts~of~}V}{\left\{
\begin{array}{l}
V\text{.1} \\
V\text{.2} \\
V\text{.3}%
\end{array}
\right. }=9~\text{sub-classes~of~\textbf{class~I}~of~crit.~pts~of~}V_{\text{%
eff}}\text{.}
\end{equation}


\subsubsection*{\textbf{I.1}}

This sub-class is given by \textquotedblleft $V_{BH}$.1$\otimes V$%
.1\textquotedblright , and thus it is characterized by all covariant
derivatives of $\mathcal{Z}$ and $\mathcal{L}$ vanishing,
\begin{equation}
\forall i,\left\{
\begin{array}{l}
\mathcal{Z}_{i}=0; \\
\mathcal{L}_{i}=0,%
\end{array}
\right.  \label{def-1}
\end{equation}
yielding
\begin{equation}
\begin{array}{l}
V_{BH}=\left\vert \mathcal{Z}\right\vert ^{2}>0,~~~\underset{\text{%
[if~placed~at spatial~infinity :~AdS}_{4}\text{]}}{V=-3\left\vert \mathcal{L}
\right\vert ^{2}<0;} \\
~ \\
S=\kappa V_{\text{eff}}=\frac{-\kappa +\sqrt{1+12\left\vert \mathcal{Z}
\right\vert ^{2}\left\vert \mathcal{L}\right\vert ^{2}}}{6\left\vert
\mathcal{L}\right\vert ^{2}}>0.%
\end{array}%
\end{equation}

\textbf{Ungauged limit:}
\begin{equation}
\lim_{\left\vert \mathcal{L}\right\vert \rightarrow 0}S=\frac{%
-\kappa+1+6\left\vert \mathcal{Z}\right\vert ^{2}\left\vert \mathcal{L}%
\right\vert^{2}}{6\left\vert \mathcal{L}\right\vert ^{2}}=\left\{%
\begin{array}{l}
\left\vert \mathcal{Z}\right\vert ^{2}~\left( \kappa =1\right); \\
\\
\frac{-1+3\left\vert \mathcal{Z}\right\vert ^{2}\left\vert \mathcal{L}%
\right\vert ^{2}}{3\left\vert \mathcal{L}\right\vert ^{2}}~\left(
\kappa=-1\right) ;%
\end{array}
\right.
\end{equation}
since the limit $\left\vert \mathcal{L}\right\vert \rightarrow 0$
corresponds to the limit $V\rightarrow 0^{-}$, from (\ref{ungauged}) one can
conclude that only the $\kappa =1$ case is allowed (in other words, the $%
\kappa =-1$ consistency condition $-1+3\left\vert \mathcal{Z}%
\right\vert^{2}\left\vert \mathcal{L}\right\vert ^{2}\geqslant 0$ never
holds). When $M_{v}$ is symmetric, the 2-polarizations of the quartic
invariant (\ref{I2})-(\ref{I-2-2}) respectively read
\begin{eqnarray}
\mathbf{I}_{2} &=&\left\vert \mathcal{Z}\right\vert ^{4};  \label{I_2} \\
\mathbf{I}_{1} &=&\left\vert \mathcal{Z}\right\vert ^{2}\text{Re}\left(%
\mathcal{Z}\overline{\mathcal{L}}\right) ; \\
\mathbf{I}_{0} &=&\frac{1}{3}\left\vert \mathcal{Z}\right\vert^{2}\left\vert
\mathcal{L}\right\vert ^{2}+\frac{2}{3}\text{Re}^{2}\left(\mathcal{Z}%
\overline{\mathcal{L}}\right) ; \\
\mathbf{I}_{-1} &=&\left\vert \mathcal{L}\right\vert ^{2}\text{Re}\left(%
\mathcal{Z}\overline{\mathcal{L}}\right) ; \\
\mathbf{I}_{-2} &=&\left\vert \mathcal{L}\right\vert ^{4}.  \label{I_(-2)}
\end{eqnarray}

\textbf{BPS sector} : the BPS critical points of this sub-class further
enjoy the following relations:
\begin{equation}
V_{BH}=S^{2}\left\vert \mathcal{L}\right\vert ^{2}>0,~~~V=-3\left\vert
\mathcal{L}\right\vert ^{2}<0
\end{equation}
and
\begin{gather}
S=\kappa V_{\text{eff}}=\frac{-\kappa +\sqrt{1+12S^{2}\left\vert \mathcal{L}%
\right\vert ^{4}}}{6\left\vert \mathcal{L}\right\vert ^{2}};  \notag \\
\Downarrow  \notag \\
12\left\vert \mathcal{L}\right\vert ^{2}S\left( 2S\left\vert \mathcal{L}%
\right\vert ^{2}+\kappa \right) =0\overset{S\neq 0\text{,~}\mathcal{L}\neq 0}%
{\Leftrightarrow }S=-\frac{\kappa }{2\left\vert \mathcal{L}\right\vert ^{2}}
=-2\kappa \left\vert \mathcal{Z}\right\vert ^{2},  \label{1.1S}
\end{gather}
which is also obtained from (\ref{resss}) by using (\ref{def-1}). This
expression is possible only for $\kappa =-1$: only extremal black holes with
hyperbolic horizon topology can be BPS, within this sub-class. When $M_{v}$
is symmetric, the 2-polarizations of the quartic invariant (\ref{I_2})-(\ref%
{I_(-2)}) respectively read
\begin{eqnarray}
\mathbf{I}_{2} &=&S^{4}\left\vert \mathcal{L}\right\vert ^{4}; \\
\mathbf{I}_{1} &=&0; \\
\mathbf{I}_{0} &=&\frac{S^{2}}{3}\left\vert \mathcal{L}\right\vert ^{4}; \\
\mathbf{I}_{-1} &=&0; \\
\mathbf{I}_{-2} &=&\left\vert \mathcal{L}\right\vert ^{4}.
\end{eqnarray}

\textbf{Summary} : the sub-class \textbf{I.1} describes extremal black holes
with AdS$_{4}$ asymptotics (\textit{at least} in the doubly-extremal case).
Both spherical and hyperbolic horizon geometries are allowed; however, the
BPS subsector has only hyperbolic ($\kappa =-1$) near-horizon geometry.
Thus, BPS doubly-extremal\footnote{%
In the extremal but not doubly-extremal case, namely when the scalars are
running, the asymptotics depends on whether the horizon attractor values of
scalars and their values at spatial infinity (i.e., in the asymptotic
background) belong to the same class of critical points of $V$, or not. In
the former case, the asymptotics is still AdS$_{4}$ and the comment below
(3.17) of \cite{DG} gets generalized to any extremal (BPS) black hole; in
the latter case; the asymptotics will be Minkowski$_{4}$ or dS$_{4}$.} black
holes with spherical symmetry and AdS$_{4}$ asymptotics cannot exist, in
this sub-class : the comment below (3.17) of \cite{DG}, explaining the
results of \cite{DG-1, DG-2, DG-3}, is retrieved.


\subsubsection*{\textbf{I.2}}

This sub-class is given by \textquotedblleft $V_{BH}$.1$\otimes V$%
.2\textquotedblright , and thus it is characterized by
\begin{equation}
\forall i,\left\{%
\begin{array}{l}
\mathcal{Z}_{i}=0; \\
\mathcal{L}=0,\qquad C_{ijk}\overline{\mathcal{L}}^{j}\overline{\mathcal{L}}%
^{k}=0.%
\end{array}
\right.  \label{def-2}
\end{equation}
This forbids the existence of a BPS subsector, and moreover yields
\begin{equation}
\begin{array}{l}
V_{BH}=\left\vert \mathcal{Z}\right\vert ^{2}>0,~~~\underset{\text{%
[if~placed~at spatial~infinity :~dS}_{4}\text{]}}{V=\left\vert \mathcal{L}%
_{i}\right\vert ^{2}>0;} \\
~ \\
S=\kappa V_{\text{eff}}=\frac{\kappa -\sqrt{1-4\left\vert \mathcal{Z}
\right\vert ^{2}\left\vert \mathcal{L}_{i}\right\vert ^{2}}}{2\left\vert
\mathcal{L}_{i}\right\vert ^{2}}>0,%
\end{array}%
\end{equation}
which forbids $\kappa =-1$ (i.e., hyperbolic near-horizon geometry). The
consistency bound for this sub-class is
\begin{equation}
1-4\left\vert \mathcal{Z}\right\vert ^{2}\left\vert \mathcal{L}%
_{i}\right\vert ^{2}\geqslant 0.  \label{cond}
\end{equation}

\textbf{Saturation of consistency bound (\ref{cond}):} when
\begin{equation}
2\left\vert \mathcal{L}_{i}\right\vert ^{2}=\frac{1}{2\left\vert \mathcal{Z}%
\right\vert ^{2}},
\end{equation}
the bound (\ref{cond}) is saturated, and the entropy boils down to
\begin{equation}
\left. S\right\vert _{1-4\left\vert \mathcal{Z}\right\vert ^{2}\left\vert
\mathcal{L}_{i}\right\vert ^{2}=0}=\frac{\kappa }{2\left\vert \mathcal{L}%
_{i}\right\vert ^{2}}=2\kappa \left\vert \mathcal{Z}\right\vert ^{2},
\end{equation}
which necessarily implies $\kappa =1$.

\textbf{Ungauged limit:}
\begin{equation}
\lim_{\left\vert \mathcal{L}_{i}\right\vert \rightarrow 0}S=\frac{%
-\kappa+1+2\left\vert \mathcal{Z}\right\vert ^{2}\left\vert \mathcal{L}%
_{i}\right\vert ^{2}}{2\left\vert \mathcal{L}_{i}\right\vert ^{2}}=\left\{
\begin{array}{l}
\left\vert \mathcal{Z}\right\vert ^{2}~\left( \kappa =1\right) ; \\
\\
\frac{-1+\left\vert \mathcal{Z}\right\vert ^{2}\left\vert \mathcal{L}%
_{i}\right\vert ^{2}}{\left\vert \mathcal{L}_{i}\right\vert ^{2}}%
~\left(\kappa =-1\right) .%
\end{array}
\right.
\end{equation}
Again, $\kappa =-1$ cannot hold in the ungauged limit, because the entropy
positivity condition ($-1+\left\vert \mathcal{Z}\right\vert ^{2}\left\vert
\mathcal{L}_{i}\right\vert ^{2}\geqslant 0$) is not consistent with (\ref%
{cond}) : in the ungauged limit only a spherical horizon is allowed. When $%
M_{v}$ is symmetric, the 2-polarizations of the quartic invariant (\ref{I2}%
)-(\ref{I-2-2}) respectively read
\begin{eqnarray}
\mathbf{I}_{2} &=&\left\vert \mathcal{Z}\right\vert ^{4}; \\
\mathbf{I}_{1} &=&0; \\
\mathbf{I}_{0} &=&-\frac{1}{3}\left\vert \mathcal{Z}\right\vert^{2}\left%
\vert \mathcal{L}_{i}\right\vert ^{2}; \\
\mathbf{I}_{-1} &=&0; \\
\mathbf{I}_{-2} &=&\left\vert \mathcal{L}_{i}\right\vert ^{4}.
\end{eqnarray}

\textbf{Summary} : the sub-class \textbf{I.2} describes non-supersymmetric
extremal black holes with spherical near-horizon geometry and dS$_{4}$
asymptotics (in the doubly-extremal case, or when the classes of critical
points of $V$ - to which horizon scalars resp. asymptotic scalars belong -
coincide; cf. footnote 17, which will be understood throughout), and
characterized by the bound (\ref{cond}).


\subsubsection*{\textbf{I.3}}

This sub-class is given by \textquotedblleft $V_{BH}$.1$\otimes V$%
.3\textquotedblright , and thus it is characterized by
\begin{equation}
\forall i,\left\{
\begin{array}{l}
\mathcal{Z}_{i}=0; \\
\mathcal{L}_{i}=\frac{\mathbf{i}}{2\mathcal{\bar{L}}}C_{ijk}\overline{%
\mathcal{L}}^{j}\overline{\mathcal{L}}^{k}.%
\end{array}
\right.
\end{equation}
This forbids the existence of a BPS subsector, and moreover yields
\begin{equation}
\begin{array}{l}
V_{BH}=\left\vert \mathcal{Z}\right\vert ^{2}>0,~~~V=\Delta _{\mathcal{L}};
\\
~ \\
S=\kappa V_{\text{eff}}=\frac{\kappa -\sqrt{1-4\left\vert \mathcal{Z}%
\right\vert ^{2}\Delta _{\mathcal{L}}}}{2\Delta _{\mathcal{L}}},%
\end{array}%
\end{equation}
within the following consistency conditions:
\begin{eqnarray}
\Delta _{\mathcal{L}} &\leqslant &\frac{1}{4|\mathcal{Z}|^{2}};
\label{cond2} \\
\Delta _{\mathcal{L}} &\neq &0.  \label{cond3}
\end{eqnarray}
(\ref{cond3}) implies that this sub-class does not exist when $M_{v}$ is
symmetric (and whenever $D_{m}\overline{D}_{(\bar{\imath}}\overline{C}_{\bar{%
\jmath}\bar{k}\bar{l})}=0$; cf. discussion in Sec. \ref{Class-V}). Moreover,
there is no an asymptotic Minkowski solution in this sub-class\footnote{%
This holds in the doubly-extremal case, or when the classes of critical
points of $V$ - to which horizon scalars resp. asymptotic scalars belong -
coincide.}.

\textbf{Saturation of consistency bound (\ref{cond2}) :} when the bound (\ref%
{cond2}) is saturated, the entropy boils down to a very simple expression,
valid only for $\kappa =1$,
\begin{equation}
S=\frac{1}{2V}=2|\mathcal{Z}|^{2},
\end{equation}
which represents a dS$_{4}$ extremal black hole with spherical symmetry.

\textbf{Ungauged limit:}
\begin{equation}
\lim_{\Delta _{\mathcal{L}}\rightarrow 0}S=\frac{\kappa -1+2\left\vert
\mathcal{Z}\right\vert ^{2}\Delta _{\mathcal{L}}}{2\Delta _{\mathcal{L}}}
=\left\{
\begin{array}{l}
\left\vert \mathcal{Z}\right\vert ^{2}~\left( \kappa =1\right) ; \\
\\
\frac{-1+\left\vert \mathcal{Z}\right\vert ^{2}\Delta _{\mathcal{L}}}{%
\Delta_{\mathcal{L}}}~\left( \kappa =-1\right);%
\end{array}
\right.
\end{equation}
in such a limit, the hyperbolic geometry would further constrain the
attractor such that
\begin{equation}
\left\{
\begin{array}{l}
-1+\left\vert \mathcal{Z}\right\vert ^{2}\Delta _{\mathcal{L}}\geqslant 0;
\\
\Delta _{\mathcal{L}}>0;%
\end{array}
\right. ~\text{or~}\left\{
\begin{array}{l}
-1+\left\vert \mathcal{Z}\right\vert ^{2}\Delta _{\mathcal{L}}\leqslant 0;
\\
\Delta _{\mathcal{L}}<0;%
\end{array}
\right.  \label{conddd}
\end{equation}
however, again, the limit $\Delta _{\mathcal{L}}\rightarrow 0$ corresponds
to the limit $V\rightarrow 0$, and thus, from (\ref{ungauged}), only the $%
\kappa =1$ case is allowed, and therefore conditions (\ref{conddd}) never
hold. When $M_{v}$ is symmetric, the 2-polarizations of the quartic
invariant (\ref{I2})-(\ref{I-2-2}) respectively read
\begin{eqnarray}
\mathbf{I}_{2} &=&\left\vert \mathcal{Z}\right\vert ^{4}; \\
\mathbf{I}_{1} &=&\left\vert \mathcal{Z}\right\vert ^{2}\text{Re}\left(%
\mathcal{Z}\overline{\mathcal{L}}\right) ; \\
\mathbf{I}_{0} &=&-\frac{1}{3}\left\vert \mathcal{Z}\right\vert
^{2}\left(2\left\vert \mathcal{L}\right\vert ^{2}+\Delta _{\mathcal{L}%
}\right) +\frac{2}{3}\text{Re}^{2}\left( \mathcal{Z}\overline{\mathcal{L}}%
\right) ; \\
\mathbf{I}_{-1} &=&-\left( 2\left\vert \mathcal{L}\right\vert ^{2}+\Delta _{%
\mathcal{L}}\right) \text{Re}\left( \mathcal{Z}\overline{\mathcal{L}}\right);
\\
\mathbf{I}_{-2} &=&\Delta _{\mathcal{L}}^{2}+\frac{8}{3}\left\vert \mathcal{L%
}\right\vert ^{2}\Delta _{\mathcal{L}}.
\end{eqnarray}

\textbf{Summary} : the sub-class \textbf{I.3} describes asymptotically
non-flat and non-supersymmetric extremal black holes characterized by the
bound (\ref{cond2}), as well as by (\ref{cond3}). The spatial asymptotics of
the extremal black hole is controlled by the asymptotic, critical value of $%
V $ : by assuming that such a value belongs to the class $V$.3, it
corresponds to the asymptotical value of $\Delta _{\mathcal{L}}$. In turn,
the evaluation of $\Delta _{\mathcal{L}}$ at the event horizon determines
the near-horizon geometry : when $\Delta _{\mathcal{L}}>0$, (\ref{cond2})
must hold and only $\kappa =1$ is allowed; on the other hand, when $\Delta _{%
\mathcal{L}}<0$ (\ref{cond2}) is automatically satisfied, and no restriction
on the horizon geometry holds.


\subsubsection*{\textbf{I.4}}

This sub-class is given by \textquotedblleft $V_{BH}$.2$\otimes V$%
.1\textquotedblright, and thus it is characterized by
\begin{equation}
\forall i,\left\{
\begin{array}{l}
\mathcal{Z}=0,\qquad C_{ijk}\overline{\mathcal{Z}}^{i}\overline{\mathcal{Z}}%
^{j}=0; \\
\mathcal{L}_{i}=0.%
\end{array}
\right.
\end{equation}
This forbids the existence of a BPS subsector, and moreover yields
\begin{equation}
\begin{array}{l}
V_{BH}=\left\vert \mathcal{Z}_{i}\right\vert ^{2}>0,~~~\underset{\text{%
[if~placed~at spatial~infinity :~AdS}_{4}\text{]}}{V=-3\left\vert \mathcal{L}%
\right\vert ^{2}<0;} \\
~ \\
S=\kappa V_{\text{eff}}=\frac{-\kappa +\sqrt{1+12\left\vert \mathcal{Z}%
_{i}\right\vert ^{2}\left\vert \mathcal{L}\right\vert ^{2}}}{6\left\vert
\mathcal{L}\right\vert ^{2}}>0.%
\end{array}%
\end{equation}

\textbf{Ungauged limit:}
\begin{equation}
\lim_{\left\vert \mathcal{L}\right\vert \rightarrow 0}S=\frac{%
-\kappa+1+6\left\vert \mathcal{Z}_{i}\right\vert ^{2}\left\vert \mathcal{L}%
\right\vert ^{2}}{6\left\vert \mathcal{L}\right\vert ^{2}}=\left\{
\begin{array}{l}
\left\vert \mathcal{Z}_{i}\right\vert ^{2}~\left( \kappa =1\right) ; \\
\\
\frac{-1+3\left\vert \mathcal{Z}_{i}\right\vert ^{2}\left\vert \mathcal{L}%
\right\vert ^{2}}{3\left\vert \mathcal{L}\right\vert ^{2}}~\left(
\kappa=-1\right) ,%
\end{array}
\right.
\end{equation}
in this limit, the hyperbolic geometry ($\kappa =-1$) would further
constrain the attractor such that $-1+3\left\vert \mathcal{Z}%
_{i}\right\vert^{2}\left\vert \mathcal{L}\right\vert ^{2}\geqslant 0$;
however, the limit $\left\vert \mathcal{L}\right\vert \rightarrow 0$
corresponds to the limit $V\rightarrow 0^{-}$, and thus, from (\ref{ungauged}%
), one can conclude that only the $\kappa =1$ case is allowed (in other
words, the $\kappa =-1$ consistency condition $-1+3\left\vert \mathcal{Z}%
_{i}\right\vert^{2}\left\vert \mathcal{L}\right\vert ^{2}\geqslant 0$ never
holds). When $M_{v}$ is symmetric, the 2-polarizations of the quartic
invariant (\ref{I2})-(\ref{I-2-2}) respectively read
\begin{eqnarray}
\mathbf{I}_{2} &=&\left\vert \mathcal{Z}_{i}\right\vert ^{4}; \\
\mathbf{I}_{1} &=&0; \\
\mathbf{I}_{0} &=&-\frac{1}{3}\left\vert \mathcal{Z}_{i}\right\vert^{2}\left%
\vert \mathcal{L}\right\vert ^{2}; \\
\mathbf{I}_{-1} &=&0; \\
\mathbf{I}_{-2} &=&\left\vert \mathcal{L}\right\vert ^{4}.
\end{eqnarray}

\textbf{Summary}: the sub-class \textbf{I.4} describes only asymptotically
AdS$_{4}$ and non-supersymmetric extremal black holes, with no restriction
on the near-horizon geometry (the observation done in Footnote 18 holds true
here, as well).


\subsubsection*{\textbf{I.5}}

This sub-class is given by \textquotedblleft $V_{BH}$.2$\otimes V$%
.2\textquotedblright , and thus it is characterized by
\begin{equation}
\forall i,\left\{
\begin{array}{l}
\mathcal{Z}=0,\qquad C_{ijk}\bar{\mathcal{Z}}^{j}\bar{\mathcal{Z}}^{k}=0; \\
\mathcal{L}=0,\qquad C_{ijk}\bar{\mathcal{L}}^{j}\bar{\mathcal{L}}^{k}=0,%
\end{array}
\right.
\end{equation}
yielding
\begin{equation}
\begin{array}{l}
V_{BH}=\left\vert \mathcal{Z}_{i}\right\vert ^{2}>0,~~~\underset{\text{%
[if~placed~at spatial~infinity :~dS}_{4}\text{]}}{V=\left\vert \mathcal{L}%
_{i}\right\vert ^{2}>0;} \\
~ \\
S=\kappa V_{\text{eff}}=\frac{\kappa -\sqrt{1-4\left\vert \mathcal{Z}%
_{i}\right\vert ^{2}\left\vert \mathcal{L}_{i}\right\vert ^{2}}}{2\left\vert%
\mathcal{L}_{i}\right\vert ^{2}}>0,%
\end{array}%
\end{equation}
which does not allow for flat ($\kappa =0$) or hyperbolic ($\kappa =-1$)
near-horizon geometry, within the conditions :
\begin{eqnarray}
1-4\left\vert \mathcal{Z}_{i}\right\vert ^{2}\left\vert \mathcal{L}%
_{i}\right\vert ^{2} &\geqslant &0;  \label{cond4} \\
\left\vert \mathcal{L}_{i}\right\vert ^{2} &\neq &0.  \label{cond5}
\end{eqnarray}

\textbf{Saturation of consistency bound (\ref{cond4}) : }when (\ref{cond4})
is saturated, the entropy boils down to a very simple expression, valid only
for $\kappa =1$,
\begin{equation}
S=\frac{\kappa }{2\left\vert \mathcal{L}_{i}\right\vert ^{2}}=2\kappa
\left\vert \mathcal{Z}_{i}\right\vert ^{2},  \label{cond6}
\end{equation}
which represents an extremal black hole with spherical horizon and dS$_{4}$
asymptotics (the observation done in Footnote 18 holds true here, as well).

\textbf{Ungauged limit }:
\begin{equation}
\lim_{\left\vert \mathcal{L}_{i}\right\vert \rightarrow 0}S=\frac{%
\kappa-1+2\left\vert \mathcal{Z}_{i}\right\vert ^{2}\left\vert \mathcal{L}%
_{i}\right\vert ^{2}}{2\left\vert \mathcal{L}_{i}\right\vert ^{2}}=\left\{
\begin{array}{l}
\left\vert \mathcal{Z}_{i}\right\vert ^{2}~\left( \kappa =1\right) ; \\
\\
\frac{-1+\left\vert \mathcal{Z}_{i}\right\vert ^{2}\left\vert \mathcal{L}%
_{i}\right\vert ^{2}}{\left\vert \mathcal{L}_{i}\right\vert ^{2}}%
~\left(\kappa =-1\right) .%
\end{array}
\right.
\end{equation}
Again, the case $\kappa =-1$ cannot hold, because the entropy positivity
condition ($-1+\left\vert \mathcal{Z}\right\vert ^{2}\left\vert \mathcal{L}%
_{i}\right\vert ^{2}\geqslant 0$) is not consistent with (\ref{cond4}) : in
the ungauged limit only spherical horizon is allowed. When $M_{v}$ is
symmetric, the 2-polarizations of the quartic invariant (\ref{I2})-(\ref%
{I-2-2}) respectively read
\begin{eqnarray}
\mathbf{I}_{2} &=&\left\vert \mathcal{Z}_{i}\right\vert ^{4};  \label{unoo}
\\
\mathbf{I}_{1} &=&\left\vert \mathcal{Z}_{i}\right\vert ^{2}\text{Re}\left(%
\mathcal{Z}^{\bar{\jmath}}\overline{\mathcal{L}}_{\bar{\jmath}}\right) ; \\
\mathbf{I}_{0} &=&\frac{1}{3}\left\vert \mathcal{Z}_{i}\right\vert^{2}\left%
\vert \mathcal{L}_{i}\right\vert ^{2}+\frac{2}{3}\text{Re}^{2}\left(
\mathcal{Z}^{\bar{\jmath}}\overline{\mathcal{L}}_{\bar{\jmath}}\right) -%
\frac{2}{3}g^{i\bar{\jmath}}C_{ikl}\overline{C}_{\bar{\jmath} \overline{m}%
\overline{n}}\overline{\mathcal{Z}}^{k}\overline{\mathcal{L}}^{l}\mathcal{Z}%
^{\overline{m}}\mathcal{L}^{\overline{n}}; \\
\mathbf{I}_{-1} &=&\left\vert \mathcal{L}_{i}\right\vert ^{2}\text{Re}\left(%
\mathcal{Z}^{\bar{\jmath}}\overline{\mathcal{L}}_{\bar{\jmath}}\right) ; \\
\mathbf{I}_{-2} &=&\left\vert \mathcal{L}_{i}\right\vert ^{4}.
\label{cinquee}
\end{eqnarray}

\textbf{BPS sector}: the BPS critical points of this sub-class saturate the
consistency condition (\ref{cond4}); indeed, they enjoy the following
relations:
\begin{eqnarray}
V_{BH} &=&S^{2}\left\vert \mathcal{L}_{i}\right\vert ^{2}>0; \\
V &=&\left\vert \mathcal{L}_{i}\right\vert ^{2}>0,
\end{eqnarray}
which imply (\ref{cond6}), clearly valid only for a spherical horizon
topology. When $M_{v}$ is symmetric, the 2-polarizations of the quartic
invariant (\ref{unoo})-(\ref{cinquee}) can be further simplified as follows:
\begin{eqnarray}
\mathbf{I}_{2} &=&S^{4}\left\vert \mathcal{L}_{i}\right\vert ^{4}; \\
\mathbf{I}_{1} &=&0; \\
\mathbf{I}_{0} &=&-S^{2}\left( \left\vert \mathcal{L}_{i}\right\vert ^{4}+%
\frac{2}{3}\mathcal{R}\left( \mathcal{L}\right) \right) ; \\
\mathbf{I}_{-1} &=&0; \\
\mathbf{I}_{-2} &=&\left\vert \mathcal{L}_{i}\right\vert ^{4},
\end{eqnarray}
where $\mathcal{R}\left( \mathcal{L}\right) $ denotes the sectional
curvature evaluated on $\mathcal{L}$'s,
\begin{equation}
\mathcal{R}\left( \mathcal{L}\right) \equiv \mathcal{R}\left( \overline{%
\mathcal{L}},\mathcal{L},\overline{\mathcal{L}},\mathcal{L}\right) :=R_{i%
\bar{\jmath}k\overline{l}}\overline{\mathcal{L}}^{i}\mathcal{L}^{\overline{%
\bar{\jmath}}} \overline{\mathcal{L}}^{k}\mathcal{L}^{\overline{l}},
\label{sect-curv}
\end{equation}
with $R_{i\overline{\bar{\jmath}}k\overline{l}}$ denoting the Riemann tensor
of the vector multiplets' scalar manifold.

\textbf{Summary} : the sub-class describes only \textquotedblleft large",
asymptotically dS$_{4}$ extremal black holes characterized by the bound (\ref%
{cond4}), as well as by (\ref{cond5}), and having only spherical ($\kappa =1$%
) near horizon geometry.
At the horizon, such black holes can be supersymmetric ($\frac{1}{4}
$-BPS).


\subsubsection*{\textbf{I.6}}

This sub-class is given by \textquotedblleft $V_{BH}$.2$\otimes V$%
.3\textquotedblright , and thus it is characterized by
\begin{equation}
\forall i,\left\{
\begin{array}{l}
\mathcal{Z}=0,\qquad C_{ijk}\overline{\mathcal{Z}}^{j}\overline{\mathcal{Z}}%
^{k}=0; \\
\mathcal{L}_{i}=\frac{\mathbf{i}}{2\overline{\mathcal{L}}}C_{ijk}\overline{%
\mathcal{L}}^{j}\overline{\mathcal{L}}^{k}.%
\end{array}
\right.
\end{equation}
This forbids the existence of a BPS subsector, and moreover yields
\begin{equation}
\begin{array}{l}
V_{BH}=\left\vert \mathcal{Z}_{i}\right\vert ^{2}>0,~~~V=\Delta _{\mathcal{L}%
}; \\
~ \\
S=\kappa V_{\text{eff}}=\frac{\kappa -\sqrt{1-4\left\vert \mathcal{Z}%
_{i}\right\vert ^{2}\Delta _{\mathcal{L}}}}{2\Delta _{\mathcal{L}}},%
\end{array}%
\end{equation}
within the following conditions:
\begin{eqnarray}
\Delta _{\mathcal{L}} &\leqslant &\frac{1}{4|\mathcal{Z}_{i}|^{2}};
\label{condd2} \\
\Delta _{\mathcal{L}} &\neq &0.  \label{condd3}
\end{eqnarray}
Thus, since $\Delta _{\mathcal{L}}\neq 0$, this class does not exist when $%
M_{v}$ is symmetric (and whenever $D_{m}\overline{D}_{(\bar{\imath}}%
\overline{C}_{\bar{\jmath}\bar{k}\bar{l})}=0$; cf. discussion in Sec. \ref%
{Class-V}). Moreover, there is no an asymptotic Minkowski solution in this
sub-class (the observation done in Footnote 18 holds true here, as well).

\textbf{Saturation of consistency bound (\ref{condd2}) : }when the bound (%
\ref{cond2}) is saturated, the entropy boils down to a very simple
expression, valid only for $\kappa =1$,
\begin{equation}
S=\frac{\kappa }{2\Delta _{\mathcal{L}}}=2|\mathcal{Z}_{i}|^{2},
\end{equation}
which represents an extremal black hole with spherical horizon and dS$_{4}$
asymptotics (again, the observation done in Footnote 18 holds true here, as
well).

\textbf{Ungauged limit}:
\begin{equation}
\lim_{\Delta _{\mathcal{L}}\rightarrow 0}S=\frac{\kappa -1+2\left\vert
\mathcal{Z}_{i}\right\vert ^{2}\Delta _{\mathcal{L}}}{2\Delta _{\mathcal{L}}}%
=\left\{
\begin{array}{l}
\left\vert \mathcal{Z}_{i}\right\vert ^{2}~\left( \kappa =1\right) ; \\
\\
\frac{-1+\left\vert \mathcal{Z}_{i}\right\vert ^{2}\Delta _{\mathcal{L}}}{%
\Delta _{\mathcal{L}}}~\left( \kappa =-1\right) .%
\end{array}
\right.
\end{equation}
\textit{A priori}, the hyperbolic geometry further constrains the attractor
such that
\begin{equation}
\left\{
\begin{array}{l}
-1+\left\vert \mathcal{Z}\right\vert ^{2}\Delta _{\mathcal{L}}\geqslant 0;
\\
\Delta _{\mathcal{L}}>0;%
\end{array}
\right. ~\text{or~}\left\{
\begin{array}{l}
-1+\left\vert \mathcal{Z}\right\vert ^{2}\Delta _{\mathcal{L}}\leqslant 0;
\\
\Delta _{\mathcal{L}}<0;%
\end{array}
\right.
\end{equation}
however, the limit $\Delta _{\mathcal{L}}\rightarrow 0$ corresponds to the
limit $V\rightarrow 0$, and thus, from (\ref{ungauged}), one can conclude
that the $\kappa =1$ case is allowed (in other words, the $\kappa =-1$
consistency condition $\frac{-1+\left\vert \mathcal{Z}_{i}\right\vert^{2}%
\Delta _{\mathcal{L}}}{\Delta _{\mathcal{L}}}\geqslant 0$ never holds). When
$M_{v}$ is symmetric, the 2-polarizations of the quartic invariant (\ref{I2}%
)-(\ref{I-2-2}) respectively read
\begin{eqnarray}
\mathbf{I}_{2} &=&\left\vert \mathcal{Z}_{i}\right\vert ^{4}; \\
\mathbf{I}_{1} &=&\left\vert \mathcal{Z}_{i}\right\vert ^{2}\text{Re}\left(%
\mathcal{Z}^{\bar{\jmath}}\overline{\mathcal{L}}_{\bar{\jmath}}\right) ; \\
\mathbf{I}_{0} &=&-\frac{2}{3}\left\vert \mathcal{Z}_{i}\right\vert^{2}\left%
\vert \mathcal{L}\right\vert ^{2}-\frac{2}{3}\mathcal{R}\left(\overline{%
\mathcal{Z}},\mathcal{Z},\overline{\mathcal{L}},\mathcal{L}\right) +\frac{5}{%
3}\text{Re}^{2}\left( \mathcal{Z}^{\bar{\jmath}}\overline{\mathcal{L}}_{\bar{%
\jmath}}\right) +\text{Im}^{2}\left( \mathcal{Z}^{\bar{\jmath}}\overline{%
\mathcal{L}}_{\bar{\jmath}}\right) ; \\
\mathbf{I}_{-1} &=&-\left( 4\left\vert \mathcal{L}\right\vert ^{2}+\Delta _{%
\mathcal{L}}\right) \text{Re}\left( \mathcal{Z}^{\bar{\jmath}}\overline{%
\mathcal{L}}_{\bar{\jmath}}\right) -\text{Im}\left( \mathcal{L}\overline{N}%
_{3} \left( \mathcal{L},\mathcal{L},\mathcal{Z}\right) \right) -\text{Re}%
\left( \mathcal{R}\left( \overline{\mathcal{L}},\mathcal{Z},\overline{%
\mathcal{L}},\mathcal{L}\right) \right) ; \\
\mathbf{I}_{-2} &=&\Delta _{\mathcal{L}}^{2}+\frac{8}{3}\left\vert \mathcal{L%
}\right\vert ^{2}\Delta _{\mathcal{L}},
\end{eqnarray}
where
\begin{eqnarray}
\mathcal{R}\left( \overline{\mathcal{Z}},\mathcal{Z},\overline{\mathcal{L}},
\mathcal{L}\right) &:&=R_{i\bar{\jmath}k\overline{l}}\overline{\mathcal{Z}}%
^{i}\mathcal{Z}^{\bar{\jmath}}\overline{\mathcal{L}}^{k}\mathcal{L}^{%
\overline{l}}; \\
\mathcal{R}\left( \overline{\mathcal{L}},\mathcal{Z},\overline{\mathcal{L}},
\mathcal{L}\right) &:&=R_{i\bar{\jmath}k\overline{l}}\overline{\mathcal{L}}%
^{i}\mathcal{Z}^{\bar{\jmath}}\overline{\mathcal{L}}^{k}\mathcal{L}^{%
\overline{l}}
\end{eqnarray}
are suitable polarizations of the sectional curvature (\ref{sect-curv}), and
\begin{equation}
\overline{N}_{3}\left( \mathcal{L},\mathcal{L},\mathcal{Z}\right) :=%
\overline{C}_{\bar{\imath}\bar{\jmath}\overline{k}}\mathcal{L}^{\bar{\imath}}%
\mathcal{L}^{\bar{\jmath}}\mathcal{Z}^{\overline{k}}.
\end{equation}

\textbf{Summary}: the sub-class \textbf{I.6} describes only
\textquotedblleft large", asymptotically non-flat and non-supersymmetric
extremal black holes characterized by the bound (\ref{condd2}), as well as
by (\ref{condd3}). The spatial asymptotics of the extremal black hole is
controlled by the asymptotic, critical value of $V$: by assuming that such a
value belongs to the class $V$.3, it corresponds to the asymptotical value
of $\Delta _{\mathcal{L}}$. In turn, the evaluation of $\Delta _{\mathcal{L}%
} $ at the event horizon determines the near-horizon geometry : when $%
\Delta_{\mathcal{L}}>0$, (\ref{condd2}) must hold and only $\kappa =1$ is
allowed; on the other hand, when $\Delta _{\mathcal{L}}<0$ (\ref{condd2}) is
automatically satisfied, and no restriction on the horizon geometry holds.


\subsubsection*{\textbf{I.7}}

This sub-class is given by \textquotedblleft $V_{BH}$.3$\otimes V$%
.1\textquotedblright , and thus it is characterized by
\begin{equation}
\forall i,\left\{
\begin{array}{c}
\mathcal{Z}_{i}=-\frac{\mathbf{i}}{2\overline{\mathcal{Z}}}C_{ijk}\overline{%
\mathcal{Z}}^{j}\overline{\mathcal{Z}}^{k}; \\
\mathcal{L}_{i}=0.%
\end{array}
\right.
\end{equation}
This forbids the existence of a BPS subsector, and moreover yields
\begin{equation}
\begin{array}{l}
V_{BH}=4\left\vert \mathcal{Z}\right\vert ^{2}+\Delta _{\mathcal{Z}}>0,~~~%
\underset{\text{[if~placed~at spatial~infinity :~AdS}_{4}\text{]}}{%
V=-3\left\vert \mathcal{L}\right\vert ^{2}<0;} \\
~ \\
S=\kappa V_{\text{eff}}=\frac{-\kappa +\sqrt{1+12\left( 4\left\vert \mathcal{%
Z}\right\vert ^{2}+\Delta _{\mathcal{Z}}\right) \left\vert \mathcal{L}%
\right\vert ^{2}}}{6\left\vert \mathcal{L}\right\vert ^{2}}>0.%
\end{array}%
\end{equation}

\textbf{Ungauged limit}:
\begin{equation}
\lim_{\left\vert \mathcal{L}\right\vert \rightarrow 0}S=\frac{%
-\kappa+1+6\left( 4\left\vert \mathcal{Z}\right\vert ^{2}+\Delta _{\mathcal{Z%
}}\right) \left\vert \mathcal{L}\right\vert ^{2}}{6\left\vert \mathcal{L}
\right\vert ^{2}}=\left\{
\begin{array}{l}
4\left\vert \mathcal{Z}\right\vert ^{2}+\Delta _{\mathcal{Z}}~\left(
\kappa=1\right) ; \\
\\
\frac{-1+3\left( 4\left\vert \mathcal{Z}\right\vert ^{2}+\Delta _{\mathcal{Z}%
}\right) \left\vert \mathcal{L}\right\vert ^{2}}{3\left\vert \mathcal{L}%
\right\vert ^{2}}~\left( \kappa =-1\right) .%
\end{array}
\right.
\end{equation}
The hyperbolic geometry would further constrain the attractor such that $%
-1+3\left( 4\left\vert \mathcal{Z}\right\vert ^{2}+\Delta _{\mathcal{Z}%
}\right) \left\vert \mathcal{L}\right\vert ^{2}\geqslant 0$; however, the
limit $\left\vert \mathcal{L}\right\vert \rightarrow 0$ corresponds to the
limit $V\rightarrow 0^{-}$, and thus, from (\ref{ungauged}), one can
conclude that only the $\kappa =1$ case is allowed (in other words, the $%
\kappa =-1$ consistency condition $-1+3\left( 4\left\vert \mathcal{Z}%
\right\vert ^{2}+\Delta _{\mathcal{Z}}\right) \left\vert \mathcal{L}%
\right\vert ^{2}\geqslant 0$ never holds). When $M_{v}$ is symmetric, the
2-polarizations of the quartic invariant (\ref{I2})-(\ref{I-2-2})
respectively read
\begin{eqnarray}
\mathbf{I}_{2} &=&-16\left\vert \mathcal{Z}\right\vert ^{4}+\Delta _{%
\mathcal{Z}}^{2}-\frac{8}{3}\Delta _{\mathcal{Z}}\left\vert \mathcal{Z}%
\right\vert ^{2}; \\
\mathbf{I}_{1} &=&-2\left( \left\vert \mathcal{Z}\right\vert ^{2}+\frac{%
\Delta _{\mathcal{Z}}}{2}\right) \text{Re}\left( \mathcal{Z}\overline{%
\mathcal{L}}\right) +2\left( \left\vert \mathcal{Z}\right\vert ^{2}+\frac{%
\Delta _{\mathcal{Z}}}{3}\right) \text{Re}\left( \mathcal{ZL}\right); \\
\mathbf{I}_{0} &=&-\frac{1}{3}\left( 2\left\vert \mathcal{Z}%
\right\vert^{2}+\Delta _{\mathcal{Z}}\right) \left\vert \mathcal{L}%
\right\vert ^{2}+\frac{2}{3}\text{Re}^{2}\left( \mathcal{Z}\overline{%
\mathcal{L}}\right); \\
\mathbf{I}_{-1} &=&\left\vert \mathcal{L}\right\vert ^{2}\text{Re}\left(%
\mathcal{Z}\overline{\mathcal{L}}\right) ; \\
\mathbf{I}_{-2} &=&\left\vert \mathcal{L}\right\vert ^{4}.
\end{eqnarray}

\textbf{Summary} : the sub-class \textbf{I.7} describes asymptotically AdS$%
_{4}$ and non-supersymmetric extremal black holes, with no restrictions on
the near-horizon geometry (the observation done in Footnote 18 holds true
here, as well).


\subsubsection*{\textbf{I.8}}

This sub-class is given by \textquotedblleft $V_{BH}$.3$\otimes V$%
.2\textquotedblright , and thus it is characterized by%
\begin{equation}
\forall i,\left\{
\begin{array}{c}
\mathcal{Z}_{i}=-\frac{\mathbf{i}}{2\overline{\mathcal{Z}}}C_{ijk}\overline{%
\mathcal{Z}}^{j}\overline{\mathcal{Z}}^{k}; \\
\mathcal{L}=0,\qquad C_{ijk}\overline{\mathcal{L}}^{j}\overline{\mathcal{L}}%
^{k}=0.%
\end{array}
\right.
\end{equation}
This forbids the existence of a BPS subsector, and moreover yields
\begin{equation}
\begin{array}{l}
V_{BH}=4\left\vert \mathcal{Z}\right\vert ^{2}+\Delta _{\mathcal{Z}}>0,~~~%
\underset{\text{[if~placed~at spatial~infinity :~dS}_{4}\text{]}}{%
V=\left\vert \mathcal{L}_{i}\right\vert ^{2}>0;} \\
~ \\
S=\kappa V_{\text{eff}}=\frac{\kappa -\sqrt{1-4\left( 4\left\vert \mathcal{Z}%
\right\vert ^{2}+\Delta _{\mathcal{Z}}\right) \left\vert \mathcal{L}%
_{i}\right\vert ^{2}}}{2\left\vert \mathcal{L}_{i}\right\vert ^{2}}>0,%
\end{array}%
\end{equation}
which does not allow for flat or hyperbolic near-horizon geometry, within
the condition
\begin{equation}
1-4\left( 4\left\vert \mathcal{Z}\right\vert ^{2}+\Delta _{\mathcal{Z}%
}\right) \left\vert \mathcal{L}_{i}\right\vert ^{2}\geqslant 0.
\label{condddd}
\end{equation}

\textbf{Saturation of consistency bound (\ref{condddd})} : when (\ref%
{condddd}) is saturated, the entropy boils down to a very simple expression,
valid only for $\kappa =1$,
\begin{equation}
S=\frac{1}{2\left\vert \mathcal{L}_{i}\right\vert ^{2}}=2\left( 4\left\vert
\mathcal{Z}\right\vert ^{2}+\Delta _{\mathcal{Z}}\right) ,
\end{equation}
which represents an extremal black hole with spherical horizon and dS$_{4}$
asymptotics (the observation done in Footnote 18 holds true here, as well).

\textbf{Ungauged limit}:
\begin{equation}
\lim_{\left\vert \mathcal{L}_{i}\right\vert \rightarrow 0}S=\frac{%
\kappa-1+2\left( 4\left\vert \mathcal{Z}\right\vert ^{2}+\Delta _{\mathcal{Z}%
}\right) \left\vert \mathcal{L}_{i}\right\vert ^{2}}{2\left\vert \mathcal{L}%
_{i}\right\vert ^{2}}=\left\{
\begin{array}{l}
4\left\vert \mathcal{Z}\right\vert ^{2}+\Delta _{\mathcal{Z}}~\left(
\kappa=1\right) ; \\
\\
\frac{-1+\left( 4\left\vert \mathcal{Z}\right\vert ^{2}+\Delta _{\mathcal{Z}%
}\right) \left\vert \mathcal{L}_{i}\right\vert ^{2}}{\left\vert \mathcal{L}%
_{i}\right\vert ^{2}}~\left( \kappa =-1\right) .%
\end{array}
\right.
\end{equation}
Again, the hyperbolic geometry would further constrain the attractor such
that $-1+\left( 4\left\vert \mathcal{Z}\right\vert ^{2}+\Delta _{\mathcal{Z}%
}\right) \left\vert \mathcal{L}_{i}\right\vert ^{2}\geqslant 0$; however,
the limit $\left\vert \mathcal{L}_{i}\right\vert \rightarrow 0$ corresponds
to the limit $V\rightarrow 0^{+}$, and thus, from (\ref{ungauged}), one can
conclude that only the $\kappa =1$ case is allowed (in other words, the $%
\kappa =-1$ consistency condition $-1+\left( 4\left\vert \mathcal{Z}%
\right\vert ^{2}+\Delta _{\mathcal{Z}}\right) \left\vert \mathcal{L}%
_{i}\right\vert ^{2}\geqslant 0$ never holds). When $M_{v}$ is symmetric,
the 2-polarizations of the quartic invariant (\ref{I2})-(\ref{I-2-2})
respectively read
\begin{eqnarray}
\mathbf{I}_{2} &=&-16\left\vert \mathcal{Z}\right\vert ^{4}+\Delta _{%
\mathcal{Z}}^{2}-\frac{8}{3}\Delta _{\mathcal{Z}}\left\vert \mathcal{Z}%
\right\vert ^{2}; \\
\mathbf{I}_{1} &=&-\left( 4\left\vert \mathcal{Z}\right\vert ^{2}+\Delta _{%
\mathcal{Z}}\right) \text{Re}\left( \mathcal{Z}^{\bar{\imath}}\overline{%
\mathcal{L}}_{\bar{\imath}}\right) -\text{Re}\left( \mathcal{R}\left(
\overline{\mathcal{Z}},\mathcal{L},\overline{\mathcal{Z}},\mathcal{Z}%
\right)\right) +2\text{Re}\left( \mathcal{Z}^{2}\mathcal{Z}^{\bar{\imath}}%
\overline{\mathcal{L}}_{\bar{\imath}}\right) ; \\
\mathbf{I}_{0} &=&-\frac{1}{3}\left\vert \mathcal{L}_{i}\right\vert^{2}%
\left( 4\left\vert \mathcal{Z}\right\vert ^{2}+\Delta _{\mathcal{Z}}\right) -%
\frac{2}{3}\mathcal{R}\left( \overline{\mathcal{Z}},\mathcal{Z}, \overline{%
\mathcal{L}},\mathcal{L}\right) -\frac{2}{3}\text{Im}^{2}\left(\mathcal{Z}^{%
\bar{\imath}}\overline{\mathcal{L}}_{\bar{\imath}}\right) ; \\
\mathbf{I}_{-1} &=&\left\vert \mathcal{L}_{i}\right\vert ^{2}\text{Re}\left(%
\mathcal{Z}^{\bar{\imath}}\overline{\mathcal{L}}_{\bar{\imath}}\right) ; \\
\mathbf{I}_{-2} &=&\left\vert \mathcal{L}_{i}\right\vert ^{4}.
\end{eqnarray}

\textbf{Summary} : the sub-class \textbf{I.8} describes asymptotically dS$%
_{4}$, non-supersymmetric extremal black holes characterized by the bound (%
\ref{condddd}), and having only spherical ($\kappa =1$) near-horizon
geometry (the observation done in Footnote 18 holds true here, as well).


\subsubsection*{\textbf{I.9}}

This sub-class is given by \textquotedblleft $V_{BH}$.3$\otimes V$%
.3\textquotedblright , and thus it is characterized by
\begin{equation}
\forall i,\left\{
\begin{array}{c}
\mathcal{Z}_{i}=-\frac{\mathbf{i}}{2\overline{\mathcal{Z}}}C_{ijk}\overline{%
\mathcal{Z}}^{j}\overline{\mathcal{Z}}^{k}; \\
\mathcal{L}_{i}=\frac{\mathbf{i}}{2\overline{\mathcal{L}}}C_{ijk}\overline{%
\mathcal{L}}^{j}\overline{\mathcal{L}}^{k},%
\end{array}
\right.  \label{jjj}
\end{equation}
yielding
\begin{equation}
\begin{array}{l}
V_{BH}=4\left\vert \mathcal{Z}\right\vert ^{2}+\Delta _{\mathcal{Z}%
}>0,~~~V=\Delta _{\mathcal{L}}; \\
~ \\
S=\kappa V_{\text{eff}}=\frac{\kappa -\sqrt{1-4\left( 4\left\vert \mathcal{Z}%
\right\vert ^{2}+\Delta _{\mathcal{Z}}\right) \Delta _{\mathcal{L}}}}{%
2\Delta _{\mathcal{L}}}>0,%
\end{array}%
\end{equation}
with the following conditions :
\begin{eqnarray}
1-4\left( 4\left\vert \mathcal{Z}\right\vert ^{2}+\Delta _{\mathcal{Z}%
}\right) \Delta _{\mathcal{L}} &\geqslant &0;  \label{jj} \\
\Delta _{\mathcal{L}} &\neq &0.  \label{jj-2}
\end{eqnarray}
Thus, since $\Delta _{\mathcal{L}}\neq 0$, this class does not exist when $%
M_{v}$ is symmetric (and whenever $D_{m}\overline{D}_{(\bar{\imath}}%
\overline{C}_{\bar{\jmath}\bar{k}\bar{l})}=0$; cf. discussion in Sec. \ref%
{Class-V}). Moreover, there is no an asymptotic Minkowski solution in this
sub-class (the observation done in Footnote 18 holds true here, as well).

\textbf{Saturation of consistency bound (\ref{jj})} : when (\ref{jj}) is
saturated, the entropy boils down to a very simple expression, valid only
for $\kappa =1$,
\begin{equation}
S=\frac{1}{2\Delta _{\mathcal{L}}}=2\left( 4\left\vert \mathcal{Z}
\right\vert ^{2}+\Delta _{\mathcal{Z}}\right) ,
\end{equation}
which represents an extremal black hole with spherical horizon and dS$_{4}$
asymptotics (again, the observation done in Footnote 18 holds true here, as
well).

\textbf{Ungauged limit}:
\begin{equation}
\lim_{\Delta _{\mathcal{L}}\rightarrow 0}S=\frac{\kappa
-1+2\left(4\left\vert \mathcal{Z}\right\vert ^{2}+\Delta _{\mathcal{Z}%
}\right) \Delta_{\mathcal{L}}}{2\Delta _{\mathcal{L}}}=\left\{
\begin{array}{l}
4\left\vert \mathcal{Z}\right\vert ^{2}+\Delta _{\mathcal{Z}}~\left(
\kappa=1\right) ; \\
\\
\frac{-1+\left( 4\left\vert \mathcal{Z}\right\vert ^{2}+\Delta _{\mathcal{Z}%
}\right) \Delta _{\mathcal{L}}}{\Delta _{\mathcal{L}}}~\left(
\kappa=-1\right) .%
\end{array}
\right.
\end{equation}
As in previous cases, the hyperbolic geometry would further constrains the
attractor such that $\frac{-1+\left( 4\left\vert \mathcal{Z}%
\right\vert^{2}+\Delta _{\mathcal{Z}}\right) \Delta _{\mathcal{L}}}{\Delta _{%
\mathcal{L}}}\geqslant 0$; however, the limit $\Delta _{\mathcal{L}%
}\rightarrow 0$ corresponds to the limit $V\rightarrow 0$, and thus, from (%
\ref{ungauged}), one can conclude that only the $\kappa =1$ case is allowed
(in other words, the $\kappa =-1$ consistency condition $\frac{-1+\left(
4\left\vert \mathcal{Z}\right\vert ^{2}+\Delta _{\mathcal{Z}}\right) \Delta
_{\mathcal{L}}}{\Delta _{\mathcal{L}}}\geqslant 0$ never holds). When $M_{v}$
is symmetric, the 2-polarizations of the quartic invariant (\ref{I2})-(\ref%
{I-2-2}) respectively read
\begin{eqnarray}
\mathbf{I}_{2} &=&-16\left\vert \mathcal{Z}\right\vert ^{4}+\Delta _{%
\mathcal{Z}}^{2}-\frac{8}{3}\Delta _{\mathcal{Z}}\left\vert \mathcal{Z}%
\right\vert ^{2};  \label{x} \\
\mathbf{I}_{1} &=&-\left( |\mathcal{Z}|^{2}+\frac{\Delta _{\mathcal{Z}}}{2}%
\right) \left( \mathcal{Z}\overline{\mathcal{L}}+\overline{\mathcal{Z}}%
\mathcal{L}-\mathcal{Z}^{\overline{\bar{\jmath}}}\overline{\mathcal{L}}_{%
\bar{\jmath}}- \overline{\mathcal{Z}}^{j}\mathcal{L}_{j}\right) +  \notag \\
&+&\frac{2}{3}\text{Re}\left[ \left( 3\mathcal{Z}^{2}\mathcal{L}^{\bar{\imath%
}}\overline{\mathcal{Z}}_{\bar{\imath}}+\mathcal{Z}\mathcal{L}\left( 3|%
\mathcal{Z}|^{2}+\Delta _{\mathcal{Z}}\right) \right) \right] -4|\mathcal{Z}
|^{2}\text{Re}\left( \mathcal{Z}_{i} \overline{\mathcal{L}}^{i}\right) ; \\
\mathbf{I}_{0} &=&\frac{1}{3}\left( 2|\mathcal{Z}|^{2}+\frac{\Delta _{%
\mathcal{Z}}}{2}\right) \left( 2|\mathcal{L}|^{2}+\frac{\Delta _{\mathcal{L}}%
}{2}\right) +  \notag \\
&-&\frac{4}{3}\text{Re}\left[ \mathcal{Z}\mathcal{L}\left( \mathcal{Z}^{\bar{%
\imath}}\overline{\mathcal{L}}_{\bar{\imath}}-\overline{\mathcal{Z}}_{\bar{%
\imath}}\mathcal{L}^{\bar{\imath}}\right) \right] +  \notag \\
&-&\frac{1}{6}\left[ g^{i\bar{\jmath}}C_{ikl}\overline{C}_{\overline{\bar{%
\jmath}}\overline{m}\overline{n}}4\overline{\mathcal{Z}}^{k}\mathcal{Z}^{%
\overline{m}}\overline{\mathcal{L}}^{l}\mathcal{L}^{\overline{n}}\right] +
\frac{4}{3}\text{Re}\left( \overline{\mathcal{Z}}\mathcal{L}\mathcal{Z}_{i}%
\overline{\mathcal{L}}^{i}\right); \\
\mathbf{I}_{-1} &=&-\left( |\mathcal{L}|^{2}+\frac{\Delta _{\mathcal{L}}}{2}
\right) \left( \mathcal{Z}\overline{\mathcal{L}}+\overline{\mathcal{Z}}%
\mathcal{L}-\mathcal{Z}^{\bar{\jmath}}\overline{\mathcal{L}}_{\bar{\jmath}}-
\overline{\mathcal{Z}}^{j}\mathcal{L}_{j}\right) +  \notag \\
&-&\frac{2}{3}\text{Re}\left[ \left( 3\mathcal{L}^{2}\mathcal{L}^{\bar{\imath%
}}\mathcal{Z}_{\bar{\imath}}+\mathcal{Z}\mathcal{L}\left( 3|\mathcal{L}%
|^{2}+\Delta _{\mathcal{L}}\right) \right) \right] -4|\mathcal{L}|^{2}\text{%
Re} \left( \mathcal{Z}_{i}\overline{\mathcal{L}}^{i}\right) ; \\
\mathbf{I}_{-2} &=&\Delta _{\mathcal{L}}^{2}+\frac{8}{3}\left\vert \mathcal{L%
}\right\vert ^{2}\Delta _{\mathcal{L}}.  \label{xx}
\end{eqnarray}

\textbf{BPS sector} : at the BPS critical points of this sub-class, (\ref%
{BPS2}) and the definitions (\ref{DeltaZ}) and (\ref{DeltaL}) yield%
\begin{equation}
\Delta _{\mathcal{Z}}=S^{2}\Delta _{\mathcal{L}},  \label{rell}
\end{equation}
and thus
\begin{equation}
V_{BH}=S^{2}\left( 4\left\vert \mathcal{L}\right\vert ^{2}+\Delta _{\mathcal{%
L}}\right) >0,
\end{equation}
which in turn implies
\begin{equation}
S=\frac{\kappa }{2\left( 2\left\vert \mathcal{L}\right\vert ^{2}+\Delta _{%
\mathcal{L}}\right) }=2\kappa \left( 2\left\vert \mathcal{Z}%
\right\vert^{2}+\Delta _{\mathcal{Z}}\right) ,  \label{xxx}
\end{equation}
which can be obtained by plugging (\ref{Rule-3}) and (\ref{Rule-3-2}) into (%
\ref{resss}). The expression (\ref{xxx}) constrains the near-horizon
geometry, depending on sgn$\left( \Delta _{\mathcal{Z}}\right) \overset{%
\text{(\ref{rell})}}{=}$sgn$\left( \Delta _{\mathcal{L}}\right) $. For $%
\Delta _{\mathcal{Z}}>0$, only $\kappa =1$ is allowed; for $\Delta _{%
\mathcal{Z}}<0$, the bound (\ref{jj}) yields
\begin{equation}
4\left\vert \mathcal{Z}\right\vert ^{2}>4\left\vert \mathcal{Z}%
\right\vert^{2}+\Delta _{\mathcal{Z}}\geqslant \frac{1}{4\Delta _{\mathcal{L}%
}},
\end{equation}
which in principle admits both signs of $2\left\vert \mathcal{Z}%
\right\vert^{2}+\Delta _{\mathcal{Z}}$: when $2\left\vert \mathcal{Z}%
\right\vert^{2}+\Delta _{\mathcal{Z}}>0$, once again only spherical ($\kappa
=1$) near-horizon geometry is allowed; on the other hand, when $2\left\vert
\mathcal{Z}\right\vert ^{2}+\Delta _{\mathcal{Z}}<0$, only hyperbolic ($%
\kappa =-1$) near-horizon geometry is allowed. When $M_{v}$ is symmetric,
the 2-polarizations of the quartic invariant (\ref{x})-(\ref{xx}) can be
further simplified as follows:
\begin{eqnarray}
\mathbf{I}_{2} &=&-16\left\vert \mathcal{Z}\right\vert ^{4}+\Delta _{%
\mathcal{Z}}^{2}-\frac{8}{3}\Delta _{\mathcal{Z}}\left\vert \mathcal{Z}%
\right\vert ^{2}; \\
\mathbf{I}_{1} &=&-\frac{8}{3}\frac{\kappa }{S}|\mathcal{Z}_{i}|^{2}\text{Im}%
\mathcal{Z}^{2}; \\
\mathbf{I}_{0} &=&\frac{1}{3S^{2}}\left( 2|\mathcal{Z}|^{2}+\frac{\Delta _{%
\mathcal{Z}}}{2}\right) ^{2}-\frac{8}{3}\frac{|\mathcal{Z}_{i}|^{2}}{S^{2}}%
\text{Re}\mathcal{Z}^{2}+\frac{2}{3}\frac{|\mathcal{Z}|^{2}|\mathcal{Z}%
_{i}|^{2}}{S^{2}}; \\
\mathbf{I}_{-1} &=&\frac{8}{3}\kappa S^{3}|\mathcal{Z}_{i}|^{2}\text{Im}
\mathcal{Z}^{2} ; \\
\mathbf{I}_{-2} &=&\frac{1}{S^{2}}\Delta _{\mathcal{Z}}^{2}+\frac{8}{3}%
\left\vert \mathcal{Z}\right\vert ^{2}\Delta _{\mathcal{Z}}.
\end{eqnarray}

\textbf{Summary} : the sub-class \textbf{I.9} describes asymptotically
non-flat extremal black holes characterized by the bound (\ref{jj}), as well
as by (\ref{jj-2}). The spatial asymptotics of the extremal black hole is
controlled by the asymptotic, critical value of $V$ : by assuming that such
a value belongs to the class $V$.3, it corresponds to the asymptotical value
of $\Delta _{\mathcal{L}}$. In turn, the evaluation of $\Delta _{\mathcal{L}%
} $ at the event horizon determines the near-horizon geometry : when $%
\Delta_{\mathcal{L}}>0$, (\ref{jj}) must hold and only $\kappa =1$ is
allowed; on the other hand, when $\Delta _{\mathcal{L}}<0$ (\ref{jj}) is
automatically satisfied, and no restriction on the horizon geometry holds.
Such black holes can be supersymmetric ($\frac{1}{4}$-BPS), \textit{a priori}
admitting both spherical and hyperbolic horizons, once again depending on sgn%
$\left(\Delta _{\mathcal{Z}}\right) =$sgn$\left( \Delta _{\mathcal{L}%
}\right) $: when $\Delta _{\mathcal{L}}>0$, only $\kappa =1$ is allowed,
whereas when $\Delta _{\mathcal{L}}<0$, only $\kappa =-1$ is allowed.


\subsection{\label{Class-II}\textbf{Class II}}

The \textbf{class II} of critical points of $V_{\text{eff}}$ is such that
both $\partial _{i}V_{BH}\neq 0$ and $\partial _{i}V\neq 0$, but
nevertheless $\partial _{i}V_{\text{eff}}=0$, because $\partial _{i}V_{BH}$
and $\partial _{i}V$ are suitably proportional, as given by (\ref{due}).
Note that the ungauged limit is ill-defined in this class, since it would
imply $V_{\text{eff}}\rightarrow V_{BH}$ (cfr. Eq. (\ref{ungauged})), but $\partial _{i}V_{\text{eff}}=0\rightarrow \partial _{i}V_{BH}\neq 0$.
As we will see below, the \textbf{class II} of critical points of $V_{\text{%
eff}}$ splits into 15 sub-classes.


\subsubsection{\label{Q-sector}$\mathcal{Q}$-sector}

Since (cfr. (\ref{dVBH=0}))
\begin{equation}
\partial _{i}V_{BH}=2\overline{\mathcal{Z}}\mathcal{Z}_{i}+\mathbf{i}C_{ijk}%
\overline{\mathcal{Z}}^{j}\overline{\mathcal{Z}}^{k},  \label{Cam1}
\end{equation}
one can compute\footnote{%
Note that, since $M_{v}$ has not been specified to be symmetric, the $%
\mathbf{I}_{2}$ in the second line of (\ref{sd}) may also depend on scalar
fields coordinatizing $M_{v}$ (which in (\ref{sd}) as well as in conditions $%
\mathcal{Q}$.1-$\mathcal{Q}$.3 are understood to be stabilized at the
critical points of $V_{\text{eff}}$).}
\begin{eqnarray}
\left\vert \partial _{i}V_{BH}\right\vert ^{2} &\equiv &g^{i\bar{\jmath}%
}\partial _{i}V_{BH}\partial _{\bar{\jmath}}V_{BH}=2\left\vert \mathcal{Z}%
_{i}\right\vert ^{4}+4\left\vert \mathcal{Z}\right\vert ^{2}\left\vert
\mathcal{Z}_{i}\right\vert ^{2}+\mathcal{R}\left( \mathcal{Z}\right) -4\text{%
Im}\left( \mathcal{Z}N_{3}(\overline{\mathcal{Z}},\overline{\mathcal{Z}},
\overline{\mathcal{Z}})\right)  \notag \\
&=&-\mathbf{I}_{2}+V_{BH}^{2}-\frac{16}{3}\text{Im}\left( \mathcal{Z}N_{3}(%
\overline{\mathcal{Z}},\overline{\mathcal{Z}},\overline{\mathcal{Z}}%
)\right)>0.  \label{sd}
\end{eqnarray}
From Sec. \ref{Class-VBH}, we recall that $\mathcal{Z}_{i}=0$ is a
sufficient condition for $\partial _{i}V_{BH}=0$; thus, the critical points
of $V_{\text{eff}}$ of \textbf{class II} will be characterized by the
condition
\begin{equation}
\mathcal{Z}_{i}\neq 0~\text{for \textit{at least} some }i\text{'s.}
\end{equation}

Thus, for a non-vanishing $\mathcal{Q}$, in the $\mathcal{Q}$-sector (flux
sector) we can then recognize three sub-classes of critical points of $V_{%
\text{eff}}$ of \textbf{class II}:

\begin{enumerate}
\item[$\mathcal{Q}$.\textbf{1}] Im$\left( \mathcal{Z}N_{3}(\overline{%
\mathcal{Z}},\overline{\mathcal{Z}},\overline{\mathcal{Z}})\right) =0$;

\item[$\mathcal{Q}$.\textbf{2}] $\mathbf{I}_{2}=0$;

\item[$\mathcal{Q}$.\textbf{3}] generic, with non-vanishing $\mathbf{I}_{2}$
and Im$\left( \mathcal{Z}N_{3}(\overline{\mathcal{Z}},\overline{\mathcal{Z}}%
, \overline{\mathcal{Z}})\right) $.
\end{enumerate}

\subsubsection{\label{L-sector}$\mathcal{L}$-sector}

Since (cfr. (\ref{dV=0}))
\begin{equation}
\partial _{i}V=-2\overline{\mathcal{L}}\mathcal{L}_{i}+\mathbf{i}C_{ijk}
\overline{\mathcal{L}}^{j}\overline{\mathcal{L}}^{k},  \label{Cam2}
\end{equation}
one can compute\footnote{%
Note that, since $M_{v}$ has not been specified to be symmetric, the $%
\mathbf{I}_{-2}$ in the second line of (\ref{sd2}) may also depend on scalar
fields coordinatizing $M_{v}$ (which in (\ref{sd2}) as well as in conditions
$\mathcal{L}$.1-$\mathcal{L}$.3 are understood to be stabilized at the
critical points of $V_{\text{eff}}$).}
\begin{eqnarray}
\left\vert \partial _{i}V\right\vert ^{2} &\equiv &g^{i\bar{\jmath}%
}\partial_{i}V\partial _{\bar{\jmath}}V=2\left\vert \mathcal{L}%
_{i}\right\vert^{4}+4\left\vert \mathcal{L}\right\vert ^{2}\left\vert
\mathcal{L}_{i}\right\vert ^{2}+\mathcal{R}\left( \mathcal{L}\right) +4\text{%
Im}\left( \mathcal{L}N_{3}(\overline{\mathcal{L}},\overline{\mathcal{L}},%
\overline{\mathcal{L}})\right)  \notag \\
&=&-\mathbf{I}_{-2}+V^{2}+16\left\vert \mathcal{L}\right\vert^{4}-8V\left%
\vert \mathcal{L}\right\vert ^{2}+\frac{8}{3}\text{Im}\left(\mathcal{L}N_{3}(%
\overline{\mathcal{L}},\overline{\mathcal{L}},\overline{\mathcal{L}})\right)
>0,  \label{sd2}
\end{eqnarray}
From Sec. \ref{Class-V}, we recall that $\mathcal{L}_{i}=0$ is a sufficient
condition for $\partial _{i}V=0$; thus, the critical points of $V_{\text{eff}%
}$ of \textbf{class II} will be characterized by the condition
\begin{equation}
\mathcal{L}_{i}\neq 0~\text{for \textit{at least} some }i\text{'s.}
\end{equation}
Thus, for a non-vanishing $\mathcal{L}$, in the $\mathcal{L}$-sector
(gauging sector) we can then recognize five sub-classes of critical points
of $V_{\text{eff}}$ of \textbf{class II}:

\begin{enumerate}
\item[$\mathcal{L}$.\textbf{1}] Im$\left( \mathcal{L}N_{3}(\overline{%
\mathcal{L}},\overline{\mathcal{L}},\overline{\mathcal{L}})\right) =0$;

\item[$\mathcal{L}$.\textbf{2}] $\mathbf{I}_{-2}=0$;

\item[$\mathcal{L}$.\textbf{3}] $V=0$;

\item[$\mathcal{L}$.\textbf{4}] $\mathcal{L}=0$;

\item[$\mathcal{L}$.\textbf{5}] generic, with non-vanishing $\mathbf{I}_{-2}$%
, $V$, $\mathcal{L}$ and Im$\left( \mathcal{L}N_{3}(\overline{\mathcal{L}},
\overline{\mathcal{L}},\overline{\mathcal{L}})\right) $.
\end{enumerate}


\subsubsection{General properties}

On the other hand, from (\ref{due}), at the \textbf{class II} of critical
points of $V_{\text{eff}}$ it holds that
\begin{equation}
\partial_{i}V_{BH}=\frac{\left( V_{BH}-V_{\text{eff}}\right) }{V}%
\partial_{i}V,~\forall i,  \label{sd2-2}
\end{equation}
and thus\footnote{%
Eq. (\ref{sd2-bis}) holds $\forall i$, and thus \textit{a fortiori} when
summed over $i$ (namely, when $\left\vert \partial_{i}V_{BH}\right\vert^{2} $
and $\left\vert \partial_{i}V\right\vert^{2}$ are given by (\ref{sd}) resp. (%
\ref{sd2})).}
\begin{equation}
\left\vert \partial _{i}V_{BH}\right\vert ^{2}=\frac{\left( V_{BH}-V_{\text{%
eff}}\right) ^{2}}{V^{2}}\left\vert \partial _{i}V\right\vert ^{2}.
\label{sd2-bis}
\end{equation}
Further equivalent expressions, involving $\mathbf{I}_{2}$, $\mathbf{I}_{-2}$%
, $V_{BH}$ and $V$ can be obtained by plugging (\ref{sd}) and (\ref{sd2})
into (\ref{sd2-bis}).

Moreover, one can compute:
\begin{eqnarray}
g^{i\bar{\jmath}}\partial _{i}V_{BH}\partial _{\bar{\jmath}}V &=&2\left(%
\overline{\mathcal{Z}}_{\overline{i}}\mathcal{L}^{\overline{i}}\right) ^{2}-4%
\mathcal{L}\overline{\mathcal{Z}}\mathcal{Z}_{i}\overline{\mathcal{L}}^{i}-2
\mathbf{i}\mathcal{L}N_{3}\left( \overline{\mathcal{L}},\overline{\mathcal{Z}%
},\overline{\mathcal{Z}}\right)  \notag \\
&&-2\mathbf{i}\overline{\mathcal{Z}}\overline{N}_{3}\left( \mathcal{Z},
\mathcal{L},\mathcal{L}\right) +\mathcal{R}\left( \overline{\mathcal{Z}},
\mathcal{L},\overline{\mathcal{Z}},\mathcal{L}\right) ,  \label{sd3}
\end{eqnarray}
where $N_{3}\left( \overline{\mathcal{L}},\overline{\mathcal{Z}},\overline{%
\mathcal{Z}}\right) $ and $\mathcal{R}\left( \overline{\mathcal{Z}},\mathcal{%
L},\overline{\mathcal{Z}},\mathcal{L}\right) $ denote suitable polarizations
of the cubic form associated to $C_{ijk}$ and of the sectional curvature (%
\ref{sect-curv}), respectively,
\begin{eqnarray}
N_{3}\left( \overline{\mathcal{L}},\overline{\mathcal{Z}},\overline{\mathcal{%
Z}}\right) &:&=C_{ijk}\overline{\mathcal{L}}^{i}\overline{\mathcal{Z}}^{j}
\overline{\mathcal{Z}}^{k}; \\
\mathcal{R}\left( \overline{\mathcal{Z}},\mathcal{L},\overline{\mathcal{Z}},
\mathcal{L}\right) &:&=R_{i\bar{\jmath}k\overline{l}}\overline{\mathcal{Z}}%
^{i}\mathcal{L}^{\bar{\jmath}}\overline{\mathcal{Z}}^{k}\mathcal{L}^{%
\overline{l}}.
\end{eqnarray}
While (\ref{sd}) and (\ref{sd2}) are manifestly real, (\ref{sd3}) seems a
complex quantity, but actually, it is a real one. Indeed, from (\ref{sd2-2}%
), it follows that
\begin{equation}
g^{i\bar{\jmath}}\partial _{i}V_{BH}\partial _{\bar{\jmath}}V=\frac{%
\left(V_{BH}-V_{\text{eff}}\right) }{V}\left\vert \partial _{i}V\right\vert
^{2},
\end{equation}
which is a manifestly real quantity, thus implying that%
\begin{eqnarray}
0 &=&\text{Im}\left( g^{i\bar{\jmath}}\partial _{i}V_{BH}\partial _{\bar{%
\jmath}}V\right)  \notag \\
&=&\text{Im}\left[
\begin{array}{l}
2\left( \overline{\mathcal{Z}}_{\bar{\imath}}\mathcal{L}^{\bar{\imath}%
}\right) ^{2}-4\mathcal{L}\overline{\mathcal{Z}}\mathcal{Z}_{i}\overline{%
\mathcal{L}}^{i}-2\mathbf{i}\mathcal{L}N_{3}\left( \overline{\mathcal{L}},%
\overline{\mathcal{Z}},\overline{\mathcal{Z}}\right) \\
-2\mathbf{i}\overline{\mathcal{Z}}\overline{N}_{3}\left( \mathcal{Z},%
\mathcal{L},\mathcal{L}\right) +\mathcal{R}\left( \overline{\mathcal{Z}},%
\mathcal{L},\overline{\mathcal{Z}},\mathcal{L}\right)%
\end{array}
\right] .
\end{eqnarray}

Recalling that, from (\ref{SS}), the critical values of $\kappa V_{\text{eff}%
}$ determine the Bekenstein-Hawking entropy $S$ (in units of $\pi $), the
relations (\ref{sd2-2}) allows to obtain $S$ at critical points of $\kappa
V_{\text{eff}}$ (or, equivalently, of $V_{\text{eff}}$) of the \textbf{class
II}, also in the non-supersymmetric case. Indeed, from (\ref{sd2-2}) one
obtains that ($\forall i$, no Einstein summation on dummy indices)
\begin{equation}
S=\kappa V_{BH}-\kappa V\frac{\partial _{i}V_{BH}}{\partial _{i}V}.
\label{Cam3}
\end{equation}
In order to relate $S$ to the quantities $\left\vert
\partial_{i}V_{BH}\right\vert ^{2}$ (\ref{sd}) and $\left\vert
\partial_{i}V\right\vert ^{2}$ (\ref{sd2}), one can observe that (\ref%
{sd2-bis}) entails an inhomogeneous quadratic equation\footnote{%
Eqs. (\ref{sd2-bis-bis}) and (\ref{Cam4}) hold $\forall i$, and also when $%
\left\vert \partial _{i}V_{BH}\right\vert ^{2}$ and $\left\vert
\partial_{i}V\right\vert ^{2}$ are given by (\ref{sd}) resp. (\ref{sd2}).}
in $S$,
\begin{equation}
V_{\text{eff}}^{2}-2V_{BH}V_{\text{eff}}+V_{BH}^{2}-V^{2}\frac{\left\vert
\partial _{i}V_{BH}\right\vert ^{2}}{\left\vert \partial _{i}V\right\vert^{2}%
}=0,  \label{sd2-bis-bis}
\end{equation}
whose solution reads
\begin{eqnarray}
\kappa V_{\text{eff~}\pm } &=&S_{\pm }=\kappa V_{BH}\pm \frac{1}{2}\sqrt{%
4V_{BH}^{2}-4\left( V_{BH}^{2}-V^{2}\frac{\left\vert
\partial_{i}V_{BH}\right\vert ^{2}}{\left\vert \partial _{i}V\right\vert ^{2}%
}\right) }  \notag \\
&=&\kappa V_{BH}\pm \sqrt{V^{2}\frac{\left\vert
\partial_{i}V_{BH}\right\vert ^{2}}{\left\vert \partial _{i}V\right\vert ^{2}%
}}=\kappa V_{BH}\pm \left\vert V\right\vert \sqrt{\frac{\left\vert
\partial_{i}V_{BH}\right\vert ^{2}}{\left\vert \partial _{i}V\right\vert ^{2}%
}}.  \label{Cam4}
\end{eqnarray}
The sign of the first term in the r.h.s. of (\ref{Cam4}) is $\kappa $,
whereas the sign of the second term is $\pm $. In order to maximize the
entropy, the \textquotedblleft $+$" branch should be chosen. By recalling (%
\ref{sd}) and (\ref{sd2}), one thus obtains the following expression for the
entropy $S$ at the critical points of $V_{\text{eff}}$ of \textbf{class II}
(regardless of their BPS properties and of the symmetricity\footnote{%
Again, since $M_{v}$ has not been specified to be symmetric, the $\mathbf{I}%
_{2}$ and $\mathbf{I}_{-2}$ in the second line of (\ref{S-II}) may also
depend on scalar fields coordinatizing $M_{v}$ (which in (\ref{S-II}) as
well as in Sec. \ref{Class-II} are understood to be stabilized at the
critical points of $V_{\text{eff}}$).} of $M_{v}$):
\begin{eqnarray}
S &=&\kappa V_{BH}+\left\vert V\right\vert \sqrt{\frac{2\left\vert \mathcal{Z%
}_{i}\right\vert ^{4}+4\left\vert \mathcal{Z}\right\vert ^{2}\left\vert
\mathcal{Z}_{i}\right\vert ^{2}+\mathcal{R}\left( \mathcal{Z}\right) -4\text{%
Im}\left( \mathcal{Z}N_{3}(\overline{\mathcal{Z}}, \overline{\mathcal{Z}},
\overline{\mathcal{Z}})\right) }{2\left\vert \mathcal{L}_{i}\right%
\vert^{4}+4\left\vert \mathcal{L}\right\vert ^{2}\left\vert \mathcal{L}%
_{i}\right\vert ^{2}+\mathcal{R}\left( \mathcal{L}\right) +4\text{Im}\left(
\mathcal{L}N_{3}(\overline{\mathcal{L}},\overline{\mathcal{L}},\overline{%
\mathcal{L}})\right) }} \\
&&  \notag \\
&=&\kappa V_{BH}+\left\vert V\right\vert \sqrt{\frac{-\mathbf{I}%
_{2}+V_{BH}^{2}-\frac{16}{3}\text{Im}\left( \mathcal{Z}N_{3}(\overline{%
\mathcal{Z}},\overline{\mathcal{Z}},\overline{\mathcal{Z}})\right) }{-%
\mathbf{I}_{-2}+V^{2}+16\left\vert \mathcal{L} \right\vert ^{4}-8V\left\vert
\mathcal{L}\right\vert ^{2}+\frac{8}{3}\text{Im}\left( \mathcal{L}N_{3}(%
\overline{\mathcal{L}},\overline{\mathcal{L}},\overline{\mathcal{L}})\right)}%
},  \label{S-II}
\end{eqnarray}
which can thus be evaluated in the various sub-classes of \textbf{class II}
(see below).

We should also observe that, for what concerns the \textbf{BPS sector}, one
obtains nothing new. Indeed, the BPS conditions (\ref{BPS1})-(\ref{BPS2})
plugged into (\ref{Cam1}) and (\ref{Cam2}) allow to elaborate (\ref{Cam3})
for the BPS entropy as follows:
\begin{equation}
S=\kappa V_{BH}+\kappa VS^{2}\Leftrightarrow \kappa VS^{2}-S+\kappa V_{BH}=0.
\label{quadr}
\end{equation}
Such inhomogeneous quadratic equation in $S$ is consistent with $S=\kappa V_{%
\text{eff}}$ (namely, (\ref{SS}) at the BPS critical points of $V_{\text{eff}%
}$) by suitably choosing the \textquotedblleft $\pm $" branching (in the
determination of the roots of (\ref{quadr})) and the value of $\kappa $ such
that $\pm \kappa =-1$.

From previous treatment, it follows that the \textbf{class II} of critical
points of $V_{\text{eff}}$ splits into 15 sub-classes, given by the
combinatorial product (denoted by \textquotedblleft $\otimes $") of the
possibilities in the $\mathcal{Q}$- and $\mathcal{L}$- sectors (namely,
inthe flux sector and in the gauging sector) :%
\begin{equation}
\underset{\mathcal{Q}\text{-sector}}{\left\{
\begin{array}{l}
\mathcal{Q}\text{.\textbf{1}}:\text{Im}\left( \mathcal{Z}N_{3}(\overline{%
\mathcal{Z}},\overline{\mathcal{Z}},\overline{\mathcal{Z}})\right) =0; \\
\mathcal{Q}\text{.\textbf{2}}:\mathbf{I}_{2}=0; \\
\mathcal{Q}\text{.\textbf{3}}:\text{Im}\left( \mathcal{Z}N_{3}(\overline{%
\mathcal{Z}},\overline{\mathcal{Z}},\overline{\mathcal{Z}})\right) \neq 0,~%
\mathbf{I}_{2}\neq 0;%
\end{array}
\right. }\otimes \underset{\mathcal{G}\text{-sector}}{\left\{
\begin{array}{l}
\mathcal{L}\text{.\textbf{1}}:\text{Im}\left( \mathcal{L}N_{3}(\overline{%
\mathcal{L}},\overline{\mathcal{L}},\overline{\mathcal{L}})\right) =0; \\
\mathcal{L}\text{.\textbf{2}}:\mathbf{I}_{-2}=0; \\
\mathcal{L}\text{.\textbf{3}}:V=0; \\
\mathcal{L}\text{.\textbf{4}}:\mathcal{L}=0; \\
\mathcal{L}\text{.\textbf{5}}:\text{none of Im}\left( \mathcal{L}N_{3}(%
\overline{\mathcal{L}},\overline{\mathcal{L}},\overline{\mathcal{L}})\right),%
\mathbf{I}_{-2},V\text{ and }\mathcal{L}\text{ vanishing},%
\end{array}
\right. }
\end{equation}
with the generic sub-class being given by the case\footnote{%
Throughout the present treatment, the first number denotes the sub-class in
the $\mathcal{Q}$-sector, whereas the second number denotes the sub-class in
the $\mathcal{L}$-sector.} \textquotedblleft \textbf{3}$\otimes $\textbf{5"}.


\subsubsection*{\textbf{II}.\textbf{1}}

This sub-class is given by \textquotedblleft $\mathcal{Q}$.1$\otimes
\mathcal{L}$.1\textquotedblright , and thus it is characterized by
\begin{equation}
\text{Im}\left( \mathcal{Z}N_{3}(\overline{\mathcal{Z}},\overline{\mathcal{Z}%
},\overline{\mathcal{Z}})\right) =0=\text{Im}\left( \mathcal{L}N_{3}(%
\overline{\mathcal{L}},\overline{\mathcal{L}},\overline{\mathcal{L}})\right);
\label{II.1}
\end{equation}
therefore, from (\ref{S-II}), its entropy reads
\begin{equation}
S=\kappa V_{BH}+\left\vert V\right\vert \sqrt{\frac{-\mathbf{I}%
_{2}+V_{BH}^{2}}{-\mathbf{I}_{-2}+V^{2}+16\left\vert \mathcal{L}%
\right\vert^{4}-8V\left\vert \mathcal{L}\right\vert ^{2}}}.
\end{equation}


\subsubsection*{\textbf{II}.\textbf{2}}

This sub-class is given by \textquotedblleft $\mathcal{Q}$.1$\otimes
\mathcal{L}$.2\textquotedblright , and thus it is characterized by
\begin{equation}
\left\{
\begin{array}{l}
\text{Im}\left( \mathcal{Z}N_{3}(\overline{\mathcal{Z}},\overline{\mathcal{Z}%
},\overline{\mathcal{Z}})\right) =0; \\
\mathbf{I}_{-2}=0;%
\end{array}
\right.  \label{II.2}
\end{equation}
therefore, from (\ref{S-II}), its entropy reads
\begin{equation}
S=\kappa V_{BH}+\left\vert V\right\vert \sqrt{\frac{-\mathbf{I}%
_{2}+V_{BH}^{2}}{V^{2}+16\left\vert \mathcal{L}\right\vert ^{4}-8V\left\vert
\mathcal{L}\right\vert ^{2}+\frac{8}{3}\text{Im}\left( \mathcal{L}N_{3}(%
\overline{\mathcal{L}},\overline{\mathcal{L}},\overline{\mathcal{L}})\right)}%
}.  \label{S-II.2}
\end{equation}


\subsubsection*{\textbf{II}.\textbf{3}}

This sub-class is given by \textquotedblleft $\mathcal{Q}$.1$\otimes
\mathcal{L}$.3\textquotedblright , and thus it is characterized by
\begin{equation}
\left\{
\begin{array}{l}
\text{Im}\left( \mathcal{Z}N_{3}(\overline{\mathcal{Z}},\overline{\mathcal{Z}%
},\overline{\mathcal{Z}})\right) =0; \\
V=0;%
\end{array}%
\right.   \label{II.3}
\end{equation}%
therefore, from (\ref{S-II}), its entropy reads
\begin{equation}
S=\kappa V_{BH},
\end{equation}%
which is meaningful only for $\kappa =1$, i.e. for spherical horizon.
Despite the (assumed) non-vanishingness of the gauging vector $\mathcal{G}$,
the extremal black holes of this sub-class have, formally, the same entropy
and the same asymptotical behaviour of their counterparts in the ungauged
limit; of course, such a similarity is only formal, because in general $%
\left. V_{BH}\right\vert _{\partial V_{\text{eff}}=0}\neq \left.
V_{BH}\right\vert _{\partial V_{BH}=0}$, thus their entropy will generally
be different.


\subsubsection*{\textbf{II}.\textbf{4}}

This sub-class is given by \textquotedblleft $\mathcal{Q}$.1$\otimes\mathcal{%
L}$.4\textquotedblright , and thus it is characterized by
\begin{equation}
\left\{
\begin{array}{l}
\text{Im}\left( \mathcal{Z}N_{3}(\overline{\mathcal{Z}},\overline{\mathcal{Z}%
},\overline{\mathcal{Z}})\right) =0; \\
\mathcal{L}=0;%
\end{array}
\right.  \label{II.4}
\end{equation}
therefore, from (\ref{S-II}), its entropy reads
\begin{equation}
S=\kappa V_{BH}+\left\vert \mathcal{L}_{i}\right\vert ^{2}\sqrt{\frac{-%
\mathbf{I}_{2}+V_{BH}^{2}-\frac{16}{3}\text{Im}\left( \mathcal{Z}N_{3}(%
\overline{\mathcal{Z}},\overline{\mathcal{Z}},\overline{\mathcal{Z}})\right)
}{g^{i\bar{\jmath}}C_{ikl}\overline{C}_{\bar{\jmath}\overline{m}\overline{n}}%
\overline{\mathcal{L}}^{k}\overline{\mathcal{L}}^{l}\mathcal{L}^{\overline{m}%
}\mathcal{L}^{\overline{n}}}}.
\end{equation}


\subsubsection*{\textbf{II}.\textbf{5}}

This sub-class is given by \textquotedblleft $\mathcal{Q}$.1$\otimes\mathcal{%
L}$.5\textquotedblright , and thus it is characterized by
\begin{equation}
\left\{
\begin{array}{l}
\text{Im}\left( \mathcal{Z}N_{3}(\overline{\mathcal{Z}},\overline{\mathcal{Z}%
},\overline{\mathcal{Z}})\right) =0; \\
\text{none of Im}\left( \mathcal{L}N_{3}(\overline{\mathcal{L}},\overline{%
\mathcal{L}},\overline{\mathcal{L}})\right) ,\mathbf{I}_{-2},V\text{ and }%
\mathcal{L}\text{ vanishing};%
\end{array}
\right.  \label{II.5}
\end{equation}
therefore, from (\ref{S-II}), its entropy reads
\begin{equation}
S=\kappa V_{BH}+\left\vert V\right\vert \sqrt{\frac{-\mathbf{I}%
_{2}+V_{BH}^{2}}{-\mathbf{I}_{-2}+V^{2}+16\left\vert \mathcal{L}%
\right\vert^{4}-8V\left\vert \mathcal{L}\right\vert ^{2}+\frac{8}{3}\text{Im}%
\left( \mathcal{L}N_{3}(\overline{\mathcal{L}},\overline{\mathcal{L}},%
\overline{\mathcal{L}})\right) }}.
\end{equation}
No \textbf{BPS sector} is allowed in this sub-class.


\subsubsection*{\textbf{II}.\textbf{6}}

This sub-class is given by \textquotedblleft $\mathcal{Q}$.2$\otimes
\mathcal{L}$.1\textquotedblright , and thus it is characterized by
\begin{equation}
\left\{
\begin{array}{l}
\mathbf{I}_{2}=0; \\
\text{Im}\left( \mathcal{L}N_{3}(\overline{\mathcal{L}},\overline{\mathcal{L}%
},\overline{\mathcal{L}})\right) =0;%
\end{array}
\right.  \label{II.6}
\end{equation}
therefore, from (\ref{S-II}), its entropy reads
\begin{equation}
S=\kappa V_{BH}+\left\vert V\right\vert \sqrt{\frac{V_{BH}^{2}-\frac{16}{3}%
\text{Im}\left( \mathcal{Z}N_{3}(\overline{\mathcal{Z}},\overline{\mathcal{Z}%
},\overline{\mathcal{Z}})\right) }{-\mathbf{I}_{-2}+V^{2}+16\left\vert
\mathcal{L}\right\vert ^{4}-8V\left\vert \mathcal{L}\right\vert ^{2}}}.
\end{equation}


\subsubsection*{\textbf{II}.\textbf{7}}

This sub-class is given by \textquotedblleft $\mathcal{Q}$.2$\otimes
\mathcal{L}$.2\textquotedblright, and thus it is characterized by
\begin{equation}
\left\{
\begin{array}{l}
\mathbf{I}_{2}=0; \\
\mathbf{I}_{-2}=0;%
\end{array}
\right.  \label{II.7}
\end{equation}
therefore, from (\ref{S-II}), its entropy reads
\begin{equation}
S=\kappa V_{BH}+\left\vert V\right\vert \sqrt{\frac{V_{BH}^{2}-\frac{16}{3}%
\text{Im}\left( \mathcal{Z}N_{3}(\overline{\mathcal{Z}},\overline{\mathcal{Z}%
},\overline{\mathcal{Z}})\right) }{V^{2}+16\left\vert \mathcal{L}%
\right\vert^{4}-8V\left\vert \mathcal{L}\right\vert ^{2}+\frac{8}{3}\text{Im}%
\left(\mathcal{L}N_{3}(\overline{\mathcal{L}},\overline{\mathcal{L}},%
\overline{\mathcal{L}})\right) }}.
\end{equation}


\subsubsection*{\textbf{II}.\textbf{8}}

This sub-class is given by \textquotedblleft $\mathcal{Q}$.2$\otimes
\mathcal{L}$.3\textquotedblright , and thus it is characterized by
\begin{equation}
\left\{
\begin{array}{l}
\mathbf{I}_{2}=0; \\
V=0,%
\end{array}
\right.  \label{II.8}
\end{equation}
therefore, from (\ref{S-II}), its entropy reads
\begin{equation}
S=\kappa V_{BH},
\end{equation}
which is meaningful only for $\kappa =1$, i.e. for spherical horizon.
Considerations analogous to the ones made for the sub-class \textbf{II}.\textbf{3}, hold here, as well.


\subsubsection*{\textbf{II}.\textbf{9}}

This sub-class is given by \textquotedblleft $\mathcal{Q}$.2$\otimes\mathcal{%
L}$.4\textquotedblright , and thus it is characterized by
\begin{equation}
\left\{
\begin{array}{l}
\mathbf{I}_{2}=0; \\
\mathcal{L}=0;%
\end{array}
\right.  \label{II.9}
\end{equation}
therefore, from (\ref{S-II}), its entropy reads
\begin{equation}
S=\kappa V_{BH}+\left\vert \mathcal{L}_{i}\right\vert ^{2}\sqrt{\frac{%
V_{BH}^{2}-\frac{16}{3}\text{Im}\left( \mathcal{Z}N_{3}(\overline{\mathcal{Z}%
},\overline{\mathcal{Z}},\overline{\mathcal{Z}})\right) }{g^{i\overline{\bar{%
\jmath}}} C_{ikl}\overline{C}_{\bar{\jmath}\overline{m}\overline{n}}%
\overline{\mathcal{L}}^{k}\overline{\mathcal{L}}^{l}\mathcal{L}^{\overline{m}%
}\mathcal{L}^{\overline{n}}}}.
\end{equation}


\subsubsection*{\textbf{II}.\textbf{10}}

This sub-class is given by \textquotedblleft $\mathcal{Q}$.2$\otimes\mathcal{%
L}$.5\textquotedblright , and thus it is characterized by
\begin{equation}
\left\{
\begin{array}{l}
\mathbf{I}_{2}=0; \\
\text{none of Im}\left( \mathcal{L}N_{3}(\overline{\mathcal{L}},\overline{%
\mathcal{L}},\overline{\mathcal{L}})\right) ,\mathbf{I}_{-2},V\text{ and }%
\mathcal{L}\text{ vanishing};%
\end{array}
\right.  \label{II.10}
\end{equation}
therefore, from (\ref{S-II}), its entropy reads
\begin{equation}
S=\kappa V_{BH}+\left\vert V\right\vert \sqrt{\frac{V_{BH}^{2}-\frac{16}{3}%
\text{Im}\left( \mathcal{Z}N_{3}(\overline{\mathcal{Z}},\overline{\mathcal{Z}%
},\overline{\mathcal{Z}})\right) }{-\mathbf{I}_{-2}+V^{2}+16\left\vert
\mathcal{L}\right\vert ^{4}-8V\left\vert \mathcal{L}\right\vert ^{2}+\frac{8%
}{3}\text{Im}\left( \mathcal{L}N_{3}(\overline{\mathcal{L}},\overline{%
\mathcal{L}},\overline{\mathcal{L}})\right) }}.
\end{equation}
No \textbf{BPS sector} is allowed in this sub-class.


\subsubsection*{\textbf{II}.\textbf{11}}

This sub-class is given by \textquotedblleft $\mathcal{Q}$.3$\otimes\mathcal{%
L}$.1\textquotedblright , and thus it is characterized by
\begin{equation}
\left\{
\begin{array}{l}
\text{Im}\left( \mathcal{Z}N_{3}(\overline{\mathcal{Z}},\overline{\mathcal{Z}%
},\overline{\mathcal{Z}})\right) \neq 0,~\mathbf{I}_{2}\neq 0 \\
\text{Im}\left( \mathcal{L}N_{3}(\overline{\mathcal{L}},\overline{\mathcal{L}%
},\overline{\mathcal{L}})\right) =0;%
\end{array}
\right.
\end{equation}
therefore, from (\ref{S-II}), its entropy reads
\begin{equation}
S=\kappa V_{BH}+\left\vert V\right\vert \sqrt{\frac{-\mathbf{I}%
_{2}+V_{BH}^{2}-\frac{16}{3}\text{Im}\left( \mathcal{Z}N_{3}(\overline{%
\mathcal{Z}},\overline{\mathcal{Z}},\overline{\mathcal{Z}})\right) }{-%
\mathbf{I}_{-2}+V^{2}+16\left\vert \mathcal{L} \right\vert ^{4}-8V\left\vert
\mathcal{L}\right\vert ^{2}}}.
\end{equation}
No \textbf{BPS sector} is allowed in this sub-class.


\subsubsection*{\textbf{II}.\textbf{12}}

This sub-class is given by \textquotedblleft $\mathcal{Q}$.3$\otimes\mathcal{%
L}$.2\textquotedblright , and thus it is characterized by
\begin{equation}
\left\{
\begin{array}{l}
\text{Im}\left( \mathcal{Z}N_{3}(\overline{\mathcal{Z}},\overline{\mathcal{Z}%
},\overline{\mathcal{Z}})\right) \neq 0,~\mathbf{I}_{2}\neq 0 \\
\mathbf{I}_{-2}=0;%
\end{array}
\right.
\end{equation}
therefore, from (\ref{S-II}), its entropy reads
\begin{equation}
S=\kappa V_{BH}+\left\vert V\right\vert \sqrt{\frac{-\mathbf{I}%
_{2}+V_{BH}^{2}-\frac{16}{3}\text{Im}\left( \mathcal{Z}N_{3}(\overline{%
\mathcal{Z}},\overline{\mathcal{Z}},\overline{\mathcal{Z}})\right) }{%
V^{2}+16\left\vert \mathcal{L}\right\vert ^{4}-8V\left\vert \mathcal{L}
\right\vert ^{2}+\frac{8}{3}\text{Im}\left( \mathcal{L}N_{3}(\overline{%
\mathcal{L}},\overline{\mathcal{L}},\overline{\mathcal{L}})\right) }}.
\end{equation}
No \textbf{BPS sector} is allowed in this sub-class.


\subsubsection*{\textbf{II}.\textbf{13}}

This sub-class is given by \textquotedblleft $\mathcal{Q}$.3$\otimes\mathcal{%
L}$.3\textquotedblright , and thus it is characterized by
\begin{equation}
\left\{
\begin{array}{l}
\text{Im}\left( \mathcal{Z}N_{3}(\overline{\mathcal{Z}},\overline{\mathcal{Z}%
},\overline{\mathcal{Z}})\right) \neq 0,~\mathbf{I}_{2}\neq 0 \\
V=0;%
\end{array}
\right.
\end{equation}
therefore, from (\ref{S-II}), its entropy reads
\begin{equation}
S=\kappa V_{BH},
\end{equation}
which is meaningful only for $\kappa =1$, i.e. for spherical horizon.
Similarly to sub-classes II.3 and II.8, despite the (assumed) non-vanishing
of the gauging vector $\mathcal{G}$, the extremal black holes of this
sub-class have, formally, the same entropy and the same asymptotical
behaviour of their counterparts in the ungauged limit, Again, since $\left.
V_{\text{eff}}\right\vert _{\partial V_{\text{eff}}=0}\neq \left.
V_{BH}\right\vert _{\partial V_{BH}=0}$, their entropy will generally be
different.


\subsubsection*{\textbf{II}.\textbf{14}}

This sub-class is given by \textquotedblleft $\mathcal{Q}$.3$\otimes
\mathcal{L}$.4\textquotedblright , and thus it is characterized by
\begin{equation}
\left\{
\begin{array}{l}
\text{Im}\left( \mathcal{Z}N_{3}(\overline{\mathcal{Z}},\overline{\mathcal{Z}%
},\overline{\mathcal{Z}})\right) \neq 0,~\mathbf{I}_{2}\neq 0 \\
\mathcal{L}=0;%
\end{array}
\right.
\end{equation}
therefore, from (\ref{S-II}), its entropy reads
\begin{equation}
S=\kappa V_{BH}+\left\vert \mathcal{L}_{i}\right\vert ^{2}\sqrt{\frac{-%
\mathbf{I}_{2}+V_{BH}^{2}-\frac{16}{3}\text{Im}\left( \mathcal{Z}N_{3}(%
\overline{\mathcal{Z}},\overline{\mathcal{Z}},\overline{\mathcal{Z}})\right)%
}{g^{i\overline{\bar{\jmath}}} C_{ikl}\overline{C}_{\bar{\jmath}\overline{m}%
\overline{n}}\overline{\mathcal{L}}^{k}\overline{\mathcal{L}}^{l}\mathcal{L}%
^{\overline{m}}\mathcal{L}^{\overline{n}}}}.
\end{equation}
No \textbf{BPS sector} is allowed in this sub-class.


\subsubsection*{\textbf{II}.\textbf{15}}

This sub-class is given by \textquotedblleft $\mathcal{Q}$.3$\otimes
\mathcal{L}$.5\textquotedblright , and thus it corresponds to the generic
case, in which none of Im$\left( \mathcal{Z}N_{3}(\overline{\mathcal{Z}},%
\overline{\mathcal{Z}},\overline{\mathcal{Z}})\right) $, $\mathbf{I}_{2}$, Im%
$\left( \mathcal{L}N_{3}(\overline{\mathcal{L}},\overline{\mathcal{L}},%
\overline{\mathcal{L}})\right) $, $\mathbf{I}_{-2}$, $V$ and $\mathcal{L}$
is vanishing. The entropy is thus given by the general expression (\ref{S-II}%
).


\section{\label{Ex}Taxonomy}

By way of example, in this Section, we report the main
features of some known solutions of static and extremal BHs to
$\mathcal{N}=2$, $D=4$ supergravity coupled to vector multiplets (in the STU
model, in the axion-dilaton $\overline{\mathbb{CP}}^{1}$model, and in the $%
T^{3}$ model), with $U(1)$ FI gaugings\footnote{%
In such a framework, non-extremal solutions have been discussed e.g. in \cite%
{Klemm:2012yg}, \cite{Toldo:2012ec} and \cite{Gnecchi:2012kb}.}.
It is here worth remarking that a complete taxonomy of all known solutions goes beyond the aim of the
present paper. Since the classification of each known solution requires a detailed treatment and a good deal of computations,
we will report on it in a future work.


\subsection{Electric $STU$}

We start and take under consideration the electric $STU$ model \cite{DG},
defined by the holomorphic prepotential
\begin{equation}
F=\frac{X^{1}X^{2}X^{3}}{X^{0}}.  \label{F-stu}
\end{equation}
Introducing the usual coordinates
\begin{equation}
s:=\frac{X^{1}}{X^{0}},\qquad t:=\frac{X^{2}}{X^{0}},\qquad u:=\frac{X^{1}}{%
X^{0}},
\end{equation}
the symplectic vector $\mathcal{V}$ can be written as
\begin{equation}
\mathcal{V}=e^{K/2}
\begin{pmatrix}
1, & s, & t, & u, & -stu, & tu, & su, & st%
\end{pmatrix}%
^{T},
\end{equation}
and the K{\"{a}}hler potential and the target space metric are
\begin{gather}
K=-\log \left( -8\,\text{Im}\,s\,\text{Im}\,t\,\text{Im}\,u\right) , \\
g_{s\bar{s}}=-\frac{1}{\left( s-\overline{s}\right) ^{2}},\quad g_{t\bar{t}%
}=-\frac{1}{\left( t-\overline{t}\right) ^{2}},\quad g_{u\bar{u}}=-\frac{1}{%
\left( u-\overline{u}\right) ^{2}}.  \notag
\end{gather}
We make the following choices for the charges ($i=1,2,3$)
\begin{equation}
\mathcal{G}=(\,0,\,g^{i},\,g_{0},\,0)^{T},\qquad \mathcal{Q}%
=(\,p^{0},\,0,\,0,\,q_{i})^{T},
\end{equation}
with $g^{i}=g$ and $q_{i}=q$; then, the central charges and their
derivatives are
\begin{eqnarray}
\mathcal{Z} &=&-e^{K/2}\left[ q\left( s+t+u\right) +p^{0}stu\right] ; \\
\mathcal{L} &=&e^{K/2}\left[ -g_{0}+g\left( tu+su+st\right) \right] ; \\
\mathcal{Z}_{s} &=&-\frac{\mathbf{i}e^{K/2}}{2\text{Im}s}\left[
q\left(s+t+u\right) +p^{0}stu\right] -e^{K/2}\left( q+p^{0}tu\right) ; \\
\mathcal{Z}_{t} &=&-\frac{\mathbf{i}e^{K/2}}{2\text{Im}t}\left[
q\left(s+t+u\right) +p^{0}stu\right] -e^{K/2}\left( q+p^{0}su\right) ; \\
\mathcal{Z}_{u} &=&-\frac{\mathbf{i}e^{K/2}}{2\text{Im}u}\left[
q\left(s+t+u\right) +p^{0}stu\right] -e^{K/2}\left( q+p^{0}st\right) ; \\
\mathcal{L}_{s} &=&-\frac{\mathbf{i}e^{K/2}}{2\text{Im}s}\left[%
-g_{0}+g\left( tu+su+st\right) \right] +ge^{K/2}\left( u+t\right) ; \\
\mathcal{L}_{t} &=&-\frac{\mathbf{i}e^{K/2}}{2\text{Im}t}\left[%
-g_{0}+g\left( tu+su+st\right) \right] +ge^{K/2}\left( s+u\right) ; \\
\mathcal{L}_{u} &=&-\frac{\mathbf{i}e^{K/2}}{2\text{Im}u}\left[%
-g_{0}+g\left( tu+su+st\right) \right] +ge^{K/2}\left( s+t\right) .
\end{eqnarray}
We now calculate the derivatives of the symplectic vector $\mathcal{V}$:
\begin{eqnarray}
\mathcal{V}_{s} &=&\frac{\mathbf{i}}{2\text{Im}s}\mathcal{V}+e^{K/2}
\begin{pmatrix}
0, & 1, & 0, & 0, & -tu, & 0, & u, & t%
\end{pmatrix}%
^{T}; \\
\mathcal{V}_{t} &=&\frac{\mathbf{i}}{2\text{Im}t}\mathcal{V}+e^{K/2}
\begin{pmatrix}
0, & 0, & 1, & 0, & -su, & u, & 0, & s%
\end{pmatrix}%
^{T}: \\
\mathcal{V}_{u} &=&\frac{\mathbf{i}}{2\text{Im}u}\mathcal{V}+e^{K/2}
\begin{pmatrix}
0, & 0, & 0, & 1, & -st, & t, & s, & 0%
\end{pmatrix}%
^{T}; \\
D_{s}\mathcal{V}_{s} &=&\frac{\mathbf{i}}{\text{Im}s}\mathcal{V}_{s}; \\
D_{u}\mathcal{V}_{s} &=&\frac{\mathbf{i}}{2\text{Im}u}\mathcal{V}_{s}+\frac{%
\mathbf{i}}{2\text{Im}s}\left( \mathcal{V}_{u}-\frac{\mathbf{i}}{2\text{Im}u}%
\mathcal{V}\right) +e^{K/2}
\begin{pmatrix}
0, & 0, & 0, & 0, & -t, & 0, & 1, & 0%
\end{pmatrix}%
^{T}; \\
D_{t}\mathcal{V}_{s} &=&\frac{\mathbf{i}}{2\text{Im}t}\mathcal{V}_{s}+\frac{%
\mathbf{i}}{2\text{Im}s}\left( \mathcal{V}_{t}-\frac{\mathbf{i}}{2\text{Im}t}%
\mathcal{V}\right) +e^{K/2}
\begin{pmatrix}
0, & 0, & 0, & 0, & -u, & 0, & 0, & 1%
\end{pmatrix}%
^{T}.
\end{eqnarray}
Taking the symplectic products of these derivatives we can calculate the
only non-zero element of $C_{ijk}$,
\begin{equation}
\langle \mathcal{V}_{t},D_{u}\mathcal{V}_{s}\rangle =C_{stu}=-1,
\end{equation}
then the solution belongs to Class \textbf{II.15}$\mathbf{:}=\mathcal{Q}$.
\textbf{3}$\otimes \mathcal{L}$.\textbf{5}.

Next, we take all scalar fields equal $s=t=u=-iy$ and this yields to
\begin{equation}
K=-\log (8y^{3}),\qquad e^{K/2}=\frac{1}{\sqrt{8y^{3}}},\qquad p^{0}=\frac{1%
}{g_{0}}\left( -1+3gq\right) ,
\end{equation}
while for the central charges we have
\begin{equation}
\mathcal{Z}=\mathbf{i}e^{K/2}y\left( 3q+\frac{1}{g_{0}}(1-3gq)y^{2}\right),%
\qquad \mathcal{L}=-e^{K/2}(g_{0}+3gy^{2}).
\end{equation}
Then, the non-vanishing 2-polarizations of the quartic invariant are given
by the following expressions
\begin{equation}
\mathbf{I}_{2}=-4p^{0}q^{3},\qquad \mathbf{I}_{-2}=4g_{0}g^{3},\qquad
\mathbf{I}_{0}=-\frac{1}{6}(1-12gq+24g^{2}q^{2}),
\end{equation}
and the entropy of extremal ($\frac{1}{4}$-)BPS black holes reads
\begin{equation}
S=\sqrt{\frac{3\mathbf{I}_{0}}{\mathbf{I}_{-2}}+\frac{\sqrt{36\mathbf{I}%
_{0}^{2}-4\mathbf{I}_{2}\mathbf{I}_{-2}}}{2\mathbf{I}_{-2}}}=\frac{1}{4}
\sqrt{\frac{1+2(1-4gq)\sqrt{1-16gq+48g^{2}q^{2}}-3(1-4gq)^{2}}{g_{0}g^{3}}},
\end{equation}
corresponding to (\ref{rell-1}) with the choice of the branch
\textquotedblleft $+$" (for entropy maximization). From the discussion in
Sec. \ref{rela}, one can immediately observe that the BPS extremal black
holes of this example satisfy the condition (\ref{pc1}).


\subsection{\label{STU}Magnetic $STU$}

In the previous Section, we have considered the electric $STU$ model in the
symplectic frame defined by (\ref{F-stu}), in which the quartic invariant
reads (cf. (\ref{I2})-(\ref{I2-1}))
\begin{equation}
\mathbf{I}_{2}:=I_{4}(\mathcal{Q}%
)=-(p^{0}q_{0}+p^{i}q_{i})^{2}+4q_{0}p^{1}p^{2}p^{3}-4p^{0}q_{1}q_{2}q_{3}+4(p^{1}p^{2}q_{1}q_{2}+p^{1}p^{3}q_{1}q_{3}+p^{2}p^{3}q_{2}q_{3}).
\label{I_4STU}
\end{equation}
By performing a symplectic transformation defined by the following matrix
\cite{DG}:
\begin{equation}
\mathcal{S}:=
\begin{pmatrix}
-1 & 0 & 0 & 0 & 0 & 0 & 0 & 0 \\
0 & 0 & 0 & 0 & 0 & -1 & 0 & 0 \\
0 & 0 & 0 & 0 & 0 & 0 & -1 & 0 \\
0 & 0 & 0 & 0 & 0 & 0 & 0 & -1 \\
0 & 0 & 0 & 0 & -1 & 0 & 0 & 0 \\
0 & 1 & 0 & 0 & 0 & 0 & 0 & 0 \\
0 & 0 & 1 & 0 & 0 & 0 & 0 & 0 \\
0 & 0 & 0 & 1 & 0 & 0 & 0 & 0 \\
&  &  &  &  &  &  &
\end{pmatrix}%
,
\end{equation}
one can switch to the so-called magnetic symplectic frame defined by%
\footnote{%
This symplectic frame can be obtained from the $\mathcal{N}=2$ truncation of
the $SO(8)$ gauged $\mathcal{N}=8$ supergravity \cite{CJ-1, dWN}.}
\begin{equation}
F_{STU}:=-2\mathbf{i}\sqrt{X^{0}X^{1}X^{2}X^{3}},  \label{MagneticSTU}
\end{equation}
and
\begin{align}
\mathcal{S}\left( p^{0},p^{i},q_{0},q_{i}\right) ^{T}&
=:\left(-p^{0},-q_{i},-q_{0},p^{i}\right) ^{T}; \\
\mathcal{S}
\begin{pmatrix}
1, & s, & t, & u, & -stu, & tu, & su, & st%
\end{pmatrix}%
^{T}& =:-
\begin{pmatrix}
1, & tu, & su, & st, & -stu, & -s, & -t, & -u%
\end{pmatrix}%
^{T},
\end{align}
and in which the quartic invariant reads
\begin{equation}
\mathbf{I}%
_{2}=-(p^{0}q_{0}-p^{i}q_{i})^{2}+4q_{0}q_{1}q_{2}q_{3}+4p^{0}p^{1}p^{2}p^{3}+4(p^{1}p^{2}q_{1}q_{2}+p^{1}p^{3}q_{1}q_{3}+p^{2}p^{3}q_{2}q_{3}).
\end{equation}
Now, we present a new non-BPS solution (with hyperbolic horizon) to this
prepotential ($I=1,2,3$). For details, see App. \ref{STU-Details}; note that
only the scalar $\tau _{1}$ is running, whereas $\tau _{2}$ and $\tau _{3}$
are frozen at their asymptotical values, which are critical points for $V$
itself:
\begin{gather}
ds^{2}=-A(r)dt^{2}+\frac{dr^{2}}{A(r)}+\left( r^{2}-\Delta ^{2}\right)
\left( d\theta ^{2}+\sinh ^{2}\theta d\phi ^{2}\right) ; \\
A(r):=\left( \frac{(64)^{2}a^{2}+G\left( b^{2}G+4\left(
r^{2}-\Delta^{2}\right) \left( -\Delta ^{2}G+Gr^{2}-8\right) \right) }{%
32G\left(r^{2}-\Delta ^{2}\right) }\right) ; \\
\tau _{1}=\sqrt{\frac{g_{0}g_{1}}{g_{2}g_{3}}}\,\tau \left( r\right) ,\qquad
\tau _{2}=\sqrt{\frac{g_{0}g_{2}}{g_{1}g_{3}}},\qquad \tau _{3}=\sqrt{\frac{%
g_{0}g_{3}}{g_{1}g_{3}}}, \\
\tau (r):=\frac{r-\Delta }{r+\Delta }; \\
F^{I}=(\pm )_{I}\frac{Gb}{64(r^{2}-\Delta ^{2})}
\begin{pmatrix}
\frac{(r+\Delta )}{g_{0}(r-\Delta )} \\
\frac{(r+\Delta )}{g_{1}(r-\Delta )} \\
\frac{(r-\Delta )}{g_{2}(r+\Delta )} \\
\frac{(r-\Delta )}{g_{3}(r+\Delta )}%
\end{pmatrix}
dt\wedge dr+(\pm )_{I}\frac{a}{g_{I}}\sinh {\theta }\,d\theta \wedge d\phi .
\end{gather}
This solution represents a non-extremal black hole in AdS$_{4}$ with
electric and magnetic charges
\begin{equation}
G=64\sqrt{g_{0}g_{1}g_{2}g_{3}},\qquad g_{I}>0,\qquad p^{I}=(\pm )_{I}\frac{a%
}{g_{I}},\qquad q_{I}=(\pm )_{I}b\,g_{I},\qquad (\text{no sum on }I)\text{,}
\end{equation}
where $(\pm )_{I}$ is a vector in which in each component one can choose
between the values $\pm =\{+1,-1\}$. When the extremality condition
\begin{equation}
(64)^{2}a^{2}G^{2}+b^{2}G^{4}\leqslant 64\,G^{2}
\end{equation}
is saturated, the unique event horizon is located at
\begin{equation}
r_{H}=\sqrt{\Delta ^{2}+\frac{4}{G}}.
\end{equation}
The Bekenstein-Hawking entropy reads
\begin{equation}
\frac{S}{\pi }=\frac{A}{4}=\left( \frac{\sqrt{-(64)^{2}a^{2}-b^{2}G^{2}+64}}{%
2G}+\frac{4}{G}\right) (g-1),
\end{equation}
where we compactified to a Riemann surface of genus $g$. The entropy in the
non-BPS (and non-extremal) case still does not depend on the values of the
scalars, but only on the values of the charges. This might seem quite
unexpected, since the attractor mechanism in the non-extremal case would not
work. In fact, there is no attractor mechanism, and the non-extremal
(non-BPS) BH entropy would depend also on the asymptotical values of scalar
fields, which however are stabilized in terms of the gauging parameters and
of the BH charges in the asymptotical background (as critical points,
actually local minima) of the gauge potential. Thus, the non-extremal BH
entropy may be recast in an explicit form depending only on the BH charges
and gauging parameters supporting the solution under consideration. The
potentials read
\begin{gather}
V_{BH}=\left( \frac{64a^{2}}{G}+\frac{Gb^{2}}{64}\right) \left( \frac{%
1+|\tau |^{2}}{\text{Re}\tau }\right) ,\quad \partial _{\tau }V_{BH}=\frac{1%
}{2}\left( \frac{64a^{2}}{G}+\frac{Gb^{2}}{64}\right) \left( \frac{\overline{%
\tau }^{2}-1}{\text{Re}^{2}\tau }\right) ;  \label{VBHP} \\
V=-\frac{G}{16}\left( 4+\frac{1+|\tau |^{2}}{\text{Re}\tau }\right) ,\qquad
\partial _{\tau }V=\frac{G}{32}\left( \frac{1-\overline{\tau }^{2}}{\text{Re}%
^{2}\tau }\right) ; \\
\partial _{\tau _{2}}V=0,\qquad \partial _{\tau _{3}}V=0,
\end{gather}
while the effettive potential is defined by (\ref{Veff}) with $\kappa =-1$.
\newline
Focusing on the extremal case, we can take the branch which allows the limit
$b=0$ which is supersymmetric, namely
\begin{equation}
a=\frac{1}{64}\sqrt{64-b^{2}G^{2}};
\end{equation}
the entropy density reduces to the supersymmetric value (cf. (\ref{S-k=-1}))
\begin{equation}
S=\frac{\mathbb{S}}{\mathbf{V}}=\frac{4}{G}=\frac{1}{16\sqrt{%
g_{0}g_{1}g_{2}g_{3}}}.  \label{Se}
\end{equation}
At the unique event horizon, we have the following values
\begin{gather}
B|_{H}=\frac{4}{G},\qquad \tau _{1}|_{H}=\sqrt{\frac{g_{0}g_{1}}{g_{2}g_{3}}}%
\,\frac{\sqrt{\Delta ^{2}G+4}-\Delta G}{\sqrt{\Delta ^{2}G+4}+\Delta G}, \\
V_{\text{eff}}|_{H}=\frac{4}{G},\qquad \partial _{\tau }{\ }V_{\text{eff}%
}|_{H}=0,
\end{gather}
and we see that these configurations are extremizing the effective
potential. It is here worth remarking that (\ref{Se}) yields
\begin{equation}
S^{2}=\frac{16}{G^{2}}=-\frac{\partial _{\tau _{1}}V_{BH}|_{H}}{%
\partial_{\tau _{1}}V|_{H}},
\end{equation}
consistent with the result (\ref{BPS-crit-2}) for BPS critical points of $V_{%
\text{eff}}$. Since $\partial _{\tau _{1}}V_{BH}|_{H}\neq 0$ and $%
\partial_{\tau _{1}}V|_{H}\neq 0$, one concludes that such BPS critical
points of $V_{\text{eff}}$ belong to \textbf{class II.15} , discussed in
Sec. \ref{Class-II}.

It is here worth remarking a curious fact : by varying the value of the
parameter $\Delta $, one can switch between \textbf{class II.15 }and class
\textbf{I.1 }of critical points of $V_{\text{eff}}$, respectively discussed
in Secs. \ref{Class-I} and \ref{Class-II}. In fact, a (continuous)
deformation of one into the other can be achieved by suitably choosing the
parametric dependence of the scalar fields $\tau _{I}$'s. By setting
\begin{equation}
\Delta =0,  \label{Delta=0}
\end{equation}
one obtains
\begin{gather}
\tau =1; \\
\partial _{I}V_{BH}=0, \\
\partial _{I}V=0,
\end{gather}
thus corresponding to the sub-class \textbf{1.I}$:=V_{BH}$.1$\otimes V$.1,
since for the extremal solution at the horizon holds that
\begin{equation}
V_{BH}=\left\vert \mathcal{Z}\right\vert ^{2}=\frac{2}{G},\qquad
V=-3\left\vert \mathcal{L}\right\vert ^{2}=-3\frac{G}{8}.
\end{equation}
By considering the entropy formula (\ref{1.1S}), one consistently obtains
the result (recall that $\kappa =-1$)
\begin{equation}
S=-2\kappa \left\vert \mathcal{Z}\right\vert ^{2}=\frac{4}{G},
\end{equation}
for any extremal black hole. This is a very interesting phenomenon, whose
investigation in detail is left to future work; here, we confine ourselves
to observe that the transition from $\Delta \neq 0$ to $\Delta =0$ as
specified by (\ref{Delta=0}) corresponds to a transition among \textit{%
different} classes of critical points of $V_{\text{eff}}$ which, in a
symmetric model like the $STU$ model, should correspond to a transition
among \textit{different} duality orbits in the representation spaces $%
\mathcal{Q}$ and $\mathcal{G}$. Namely, we have transited from class \textbf{%
II.15 }to class \textbf{I.1} by imposing (\ref{Delta=0}); this cannot be
achieved by a $U$-duality transformation, but rather through a symplectic
finite transformation belonging to the pseudo-Riemannian coset $Sp(8,\mathbb{%
R})/SL(2,\mathbb{R})^{3}$.

With the choices
\begin{equation}
p^{I}=-\frac{1}{8g_{I}},\qquad q_{I}=0,  \label{BPS-choice}
\end{equation}
the extremal critical point becomes ($\frac{1}{4}$-)BPS, and the
corresponding BPS entropy enjoys the expression (\ref{Ent}), thus belonging
to the noteworthy BPS sub-class discussed in Sec. \ref{note-BPS}. One can
thus conclude that the BPS extremal black hole supported by (\ref{Delta=0})
and (\ref{BPS-choice}), belonging to the \textbf{BPS sub-sector} of class
\textbf{I.1} of critical points of $V_{\text{eff}}$, is characterized by (%
\ref{BPSsub1}), and provides an example in which (\ref{pc1}), and thus (\ref%
{rrel}), is satisfied.\newline
Finally, by performing the further identifications
\begin{equation*}
g_{0},g_{1}\rightarrow g_{0}/2\qquad g_{2},g_{3}\rightarrow g_{2}/2,\qquad
\ell =\frac{8}{G}.
\end{equation*}
we get the static solution to the axion-dilaton model \cite{CK,K-Rot} $F=-%
\mathbf{i}X^{0}X^{1}$, presented in the next Section.


\subsection{$\overline{\mathbb{CP}}^{1}$\label{CP1}}

Starting from the $STU$ model, in order to obtain the \textit{minimally
coupled} model of $\mathcal{N}=2$, $D=4$ supergravity with $\overline{%
\mathbb{CP}}^{1}$ vector multiplet's scalar manifold in the symplectic frame
defined by
\begin{equation}
F_{\overline{\mathbb{CP}}^{1}}:=-\mathbf{i}X^{0}X^{1},
\end{equation}
one needs to identify the contravariant symplectic sections as follows :
\begin{align}
& X^{2}\Rightarrow X^{0}/\sqrt{2}, \\
& X^{0}\Rightarrow X^{0}/\sqrt{2}, \\
& X^{1}\Rightarrow X^{1}/\sqrt{2}, \\
& X^{3}\Rightarrow X^{1}/\sqrt{2},
\end{align}
thus getting that the quartic invariant boils down to be the square of a
quadratic invariant:
\begin{equation}
\mathbf{I}_{2}=(q_{0}q_{1}+p^{0}p^{1})^{2}.  \label{I2-square}
\end{equation}

We now present the investigation of the attractor dynamics of the complex
scalar field (axion-dilaton) within a subclass of extremal solutions
previously found in \cite{CK} and \cite{K-Rot}, in presence of $U(1)$ FI
gauging. These correspond to the choices ($A=0,1$)
\begin{equation}
\mathcal{G}=(0,g_{A})^{T},~\mathcal{Q}=(\kappa p^{A},0)^{T}.
\end{equation}
The symplectic section can be parametrised in terms of the complex scalar
field $\tau $ by choosing $X^{0}=1$, $X^{1}=\tau $, so that the holomorphic
symplectic section reads (cf. (\ref{hol-sec}))
\begin{equation}
\mathfrak{H}=
\begin{pmatrix}
1, & \tau , & -\mathbf{i}\tau , & -\mathbf{i}%
\end{pmatrix}%
^{T},
\end{equation}
where $\tau $ coordinatizes $M_{v}\equiv \overline{\mathbb{CP}}^{1}$. The K%
\"{a}hler potential and the non-vanishing components of the metric of the
scalar manifold are respectively
\begin{equation}
e^{-K}=4\text{Re}\tau \qquad g_{\tau \bar{\tau}}=g_{\bar{\tau}\tau}=\partial
_{\tau }\partial _{\bar{\tau}}K=(2\text{Re}\tau )^{-2}.
\end{equation}
By recalling (\ref{hol-sec}), the K\"{a}hler-covariantly holomorphic
symplectic section reads
\begin{equation}
\mathcal{V}=\frac{1}{2\sqrt{\text{Re}\tau }}\mathfrak{H}.
\end{equation}
Its derivative and the central charges respectively read
\begin{gather}
D_{\tau }\mathcal{V}\equiv \mathcal{V}_{\tau }=\frac{e^{K/2}}{2\tau }
\begin{pmatrix}
-1, & \tau , & \mathbf{i}\tau , & -\mathbf{i}%
\end{pmatrix}%
^{T}; \\
\mathcal{Z}:=\langle \mathcal{Q},\mathcal{V}\rangle =-\mathbf{i}
e^{K/2}\kappa (p^{0}\tau +p^{1}), \\
\mathcal{L}:=\langle \mathcal{G},\mathcal{V}\rangle
=-4g_{0}g_{1}e^{K/2}(p^{0}\tau +p^{1}),
\end{gather}
and the derivative of the central charges is
\begin{equation}
D_{\tau }\mathcal{Z}=\langle \mathcal{Q},\mathcal{V}_{\tau }\rangle =\mathbf{%
i}\,\kappa \frac{e^{K/2}}{2\tau }(p^{1}-p^{0}\tau ).
\end{equation}

Since also this model is symmetric, one can compute the 2-polarizations of
the quartic structure that, by virtue of (\ref{I2-square}), is \textit{%
non-primitive} (namely, the square of a quadratic invariant structure; see
Footnotes 5 and 11) :
\begin{align}
\mathbf{I}_{2}& =(p^{0}p^{1})^{2}, \\
\mathbf{I}_{1}& =0, \\
\mathbf{I}_{0}& =\frac{1}{3}p^{0}p^{1}g_{0}g_{1}, \\
\mathbf{I}_{-1}& =0, \\
\mathbf{I}_{-2}& =(g_{0}g_{1})^{2}.
\end{align}
In this noteworthy subclass, as discussed in Sec. \ref{note-BPS}, from (\ref%
{Ent}) the BPS extremal black hole entropy reads
\begin{equation}
S=\sqrt[4]{\frac{\mathbf{I}_{2}}{\mathbf{I}_{-2}}}\overset{\text{(\ref{rrel})%
}}{=}\sqrt[4]{9\frac{\mathbf{I}_{0}^{2}}{\mathbf{I}_{-2}^{2}}}=\sqrt{3\frac{%
\mathbf{I}_{0}}{\mathbf{I}_{-2}}}=\sqrt{\frac{p^{0}p^{1}}{g_{0}g_{1}}}.
\label{Ent2}
\end{equation}
We should recall that in \cite{CK} and \cite{K-Rot} the following
identifications were made:
\begin{equation}
p^{0}=-\frac{1}{4g_{0}},\qquad p^{1}=-\frac{1}{4g_{1}}.
\end{equation}
resulting into the entropy (\ref{Ent2}) to simplify down to
\begin{equation}
S=\frac{1}{4g_{0}g_{1}}.
\end{equation}
When the solution presents an hyperbolic horizon, $S$ denotes the entropy
density, and one can compactify to a Riemannian surface of genus $g$, and
the identification with the above formalism is trivial, since
\begin{equation}
\mathbf{V}=4\pi (g-1).  \label{V-g}
\end{equation}


\subsection{$T^{3}$}

Finally, we consider the solution in \cite{Hristov:2010ri} for the model
with prepotential
\begin{equation}
F=\frac{(X^{1})^{3}}{X^{0}},
\end{equation}
with non vanishing FI $U(1)$ gauging parameters $g_{0}=g\xi _{0}$ and $%
g^{1}=g\xi ^{1}$. The BH solution has one magnetic charge $p^{0}$ and one
electric charge $q_{1}$. The charges and the constants of the solution are
\begin{align}
p^{0}=& \mp \frac{1}{g\xi _{0}}\left( \frac{1}{8}+\frac{8(g\xi
^{1}\beta_{1})^{2}}{3}\right) ,\quad q_{1}=\pm \frac{1}{g\xi ^{1}}\left(
\frac{3}{8}-\frac{8(g\xi ^{1}\beta _{1})^{2}}{3}\right) , \\
\beta ^{0}=& \frac{\xi ^{1}\beta _{1}}{\xi _{0}},\qquad \alpha ^{0}=\pm\frac{%
1}{4\xi _{0}},\qquad \alpha _{1}=\mp \frac{3}{4\xi ^{1}},\qquad c=1-\frac{32%
}{3}(g\xi ^{1}\beta _{1})^{2}\ .
\end{align}
Then, the central charges and their derivatives read
\begin{align}
\mathcal{Z}=& -e^{K/2}\left( 3s\frac{\left( \frac{3}{8}-\frac{8}{3}%
(\beta_{1}g^{1})^{2}\right) }{g_{1}}-s^{2}\frac{\left( \frac{1}{8}+\frac{8}{3%
}(\beta _{1}g^{1})^{2}\right) }{g_{0}}\right) , \\
\mathcal{L}=& e^{K/2}\left( 3g^{1}s^{2}-g^{0}\right) , \\
D_{s}\mathcal{Z}=& -i\frac{1}{2Ims}\mathcal{Z}-e^{K/2}\left( \frac{\left(%
\frac{3}{8}-\frac{8}{3}(\beta _{1}g^{1})^{2}\right) }{g_{1}}-s^{2}\frac{%
\left( \frac{1}{8}+\frac{8}{3}(\beta _{1}g^{1})^{2}\right) }{g_{0}}\right) ,
\\
D_{s}\mathcal{L}=& e^{K/2}(3g^{1}s^{2}-g_{0}),
\end{align}
where $s$ is the only scalar field and $K$ is the K\"ahler potential as
usual. The $\mathbf{I}_{\pm 2}$ quartic invariants read
\begin{equation*}
\mathbf{I}_{2}=-4p^{0}q_{1}^{3},\qquad \mathbf{I}_{-2}=4g_{0}(g^{1})^{3}.
\end{equation*}
Thus, $\mathcal{L}$, $\mathbf{I}_{-2}$ and $\mathbf{I}_{2}$ are
non-vanishing, and exploiting the above results it is straightforward to
show that also Im$\left( \mathcal{L}N_{3}(\overline{\mathcal{L}},\overline{%
\mathcal{L}},\overline{\mathcal{L}})\right) $ and Im$\left( \mathcal{Z}N_{3}(%
\overline{\mathcal{Z}},\overline{\mathcal{Z}},\overline{\mathcal{Z}})\right)$
do not vanish, implying that this solution belongs to the class \textbf{II.15%
}$\mathbf{:}=\mathcal{Q}$.\textbf{3}$\otimes \mathcal{L}$.\textbf{5}.


\newpage

\section{\label{Conclusion}Conclusion}

In this paper, we have considered $\mathcal{N}=2$, $D=4$ supergravity
coupled to Abelian vector multiplets with U(1) Fayet-Iliopoulos gaugings.

By exploiting the identities determining the structure of projective special
K\"{a}hler geometry endowing the vector multiplets' scalar manifold in
presence of electric and magnetic BH charges as well as of (generally
dyonic) gauging parameters, we retrieved, extended and generalized various
results on the expression of Bekenstein-Hawking entropy of asymptotically AdS%
$_{4}$ BPS BHs in gauged supergravity. In doing this, we made use of the
quartic structure (and 2-polarizations thereof) characterizing the $U$%
-duality groups of type $E_{7}$ corresponding to symmetric scalar manifolds.
Then, we have presented a complete classification of the critical points of
the effective black hole potential $V_{\text{eff}}$ which governs the
attractor mechanism at the horizon of extremal BHs, relating - when possible
- the resulting attractors to the critical points of the gauge potential $V$
as well as of the effective black hole potential in the ungauged case, $%
V_{BH}$. In all cases, we have analyzed the existence of BPS sub-sectors and
studied their features. Finally, we have inserted explicit known examples of
asymptotically AdS$_{4}$ static extremal (BPS) BH in gauged supergravity in
the aforementioned classification, and, as a by-product of our treatment, we
also have provided a novel, static extremal BH solution to the STU model,
with the dilaton interpolating between a hyperbolic horizon and AdS$_{4}$ at
infinity. \newline

The classification of the critical points of $V_{\text{eff}}$ which we have
provided in the present work will hopefully be instrumental in order to
discover and explore new solutions of Maxwell-Einstein supergravity with
non-vanishing gauge potential. Some directions for possible further
developments also concern the extension to the planar case ($\kappa =0$),
the coupling of hypermultiplets (cf. e.g. \cite{HPZ, KPR}), and the
generalization to stationary solutions. It is finally worth remarking that
an almost uncharted territory is provided by non-Abelian gaugings of $%
\mathcal{N}=2$ $D=4$ supergravity, which just a few works (see e.g. \cite%
{Ortin-Vaula, Ortin-Santoli-1}) have hitherto dared to investigate; the
question whether in presence of non-Abelian gaugings an effective black hole
potential formalism for the (covariant) attractor mechanism can be
established, still remains unanswered. \newline

Finally, it is worth mentioning that the few examples discussed in
Sec. \ref{Ex} belong to two classes only. A quick procedure for the identification of a given solution into one class of our classification is not currently available; actually, a considerable deal of work and computations is needed in order to do so. While this is of course not an impossible task, it would nevertheless be helpful to develop some characterization theorems in order to simplify such an identification. Interestingly, such
a characterization would likely also provide a strategy for the construction of explicit solutions in any given class, or, possibly otherwise, prove the emptiness of some classes.

\pagebreak

\appendix

\section{\label{App-Explicit}Computation of the r.h.s. of Eq. (\protect\ref%
{rel2})}

In special K\"{a}hler geometry based on the \textit{cubic} holomorphic
prepotential (\ref{F}), named \textit{very special} geometry, the cubic form
is defined as (cfr. e.g. \cite{CFM, BMR1})
\begin{equation}
\mathbb{V}:=-\frac{1}{3!}d_{klm}\text{Im}\left( \frac{X^{k}}{X_{0}}\right)%
\text{Im}\left( \frac{X^{l}}{X_{0}}\right) \text{Im}\left( \frac{X^{m}}{X_{0}%
}\right) ,  \label{5Dconstr}
\end{equation}
and the scalar manifold of the corresponding minimal supergravity theory in $%
D=5$ is defined as the hypersurface at $\mathbb{V}=1$. In order to compute
the contractions in the r.h.s. of (\ref{rel2}), we have to recall some basic
formul\ae\ of very special K\"{a}hler geometry. From e.g. the treatment of
\cite{CFM}, choosing the so-called 4D/5D special coordinates' symplectic
frame and fixing the K\"{a}hler gauge such that $X^{0}=1$, with $\frac{X^{i}%
}{X^{0}}=:z^{i}=x^{i}-\mathbf{i}\mathbb{V}$ $^{1/3}\hat{\lambda}^{i}$ (such
that $\frac{1}{3!}d_{ijk}\hat{\lambda}^{i}\hat{\lambda}^{j}\hat{\lambda}%
^{k}=1$, which is a way to rewrite \ (\ref{5Dconstr})), one can define
\begin{eqnarray}
\hat{\kappa}_{ij} &:&=d_{ijk}\hat{\lambda}^{k},~\hat{\kappa}_{i}:=d_{ijk}%
\hat{\lambda}^{j}\hat{\lambda}^{k},~\hat{\kappa}:=d_{ijk}\hat{\lambda}^{i}%
\hat{\lambda}^{j}\hat{\lambda}^{k}=6; \\
h_{ij} &:&=d_{ijk}x^{k},~h_{i}:=d_{ijk}x^{j}x^{k},~h:=d_{ijk}x^{i}x^{j}x^{k}.
\end{eqnarray}
Then, the symplectic sections read
\begin{equation}
\mathcal{V}^{M}=:e^{K/2}\left(
\begin{array}{c}
X^{0} \\
X^{i} \\
F_{0} \\
F_{i}%
\end{array}
\right) =e^{K/2}\left(
\begin{array}{c}
1 \\
z^{i} \\
-F \\
\frac{\partial F}{\partial X^{i}}%
\end{array}
\right) ,
\end{equation}
where
\begin{equation}
e^{K/2}=\frac{1}{2\sqrt{2}}\mathbb{V}^{-1/2},
\end{equation}
and
\begin{eqnarray}
F &=&\frac{1}{3!}d_{ijk}z^{i}z^{j}z^{k}=\frac{1}{3!}d_{ijk}\left( x^{i}-%
\mathbf{i}\mathbb{V}^{1/3}\hat{\lambda}^{i}\right) \left( x^{j}-\mathbf{i}%
\mathbb{V}^{1/3}\hat{\lambda}^{j}\right) \left( x^{k}-\mathbf{i}\mathbb{V}%
^{1/3}\hat{\lambda}^{k}\right)  \notag \\
&=&\frac{1}{3!}d_{ijk}x^{i}x^{j}x^{k}-\frac{\mathbf{i}}{2}\mathbb{V}%
^{1/3}d_{ijk}x^{i}x^{j}\hat{\lambda}^{k}-\frac{1}{2}\mathbb{V}%
^{2/3}d_{ijk}x^{i}\hat{\lambda}^{j}\hat{\lambda}^{k}+\frac{\mathbf{i}}{3!}%
\mathbb{V} d_{ijk}\hat{\lambda}^{i}\hat{\lambda}^{j}\hat{\lambda}^{k}  \notag
\\
&=&\frac{h}{6}-\frac{1}{2}\mathbb{V}^{2/3}\hat{\kappa}_{i}x^{i}+\mathbf{i}%
\frac{\mathbb{V}^{1/3}}{2}\left( -h_{i}\hat{\lambda}^{i}+2\mathbb{V}%
^{2/3}\right)  \notag \\
&=&\frac{h}{6}-\frac{1}{2}\mathbb{V}^{2/3}h_{ij}\hat{\lambda}^{i}\hat{\lambda%
}^{j}+\mathbf{i}\frac{\mathbb{V}^{1/3}}{2}\left( -\hat{\kappa}%
_{ij}x^{i}x^{j}+2\mathbb{V}^{2/3}\right) ,
\end{eqnarray}
and
\begin{eqnarray}
F_{i} &=&\frac{\partial F}{\partial X^{i}}=\frac{1}{2}d_{ijk}z^{j}z^{k}=%
\frac{1}{2}d_{ijk}\left( x^{j}-\mathbf{i}\mathbb{V}^{1/3}\hat{\lambda}%
^{j}\right) \left( x^{k}-\mathbf{i}\mathbb{V}^{1/3}\hat{\lambda}^{k}\right)
\notag \\
&=&\frac{1}{2}d_{ijk}x^{j}x^{k}-\mathbf{i}\mathbb{V}^{1/3}d_{ijk}x^{j}\hat{%
\lambda}^{k}-\frac{1}{2}\mathbb{V}^{2/3}d_{ijk}\hat{\lambda}^{j}\hat{\lambda}%
^{k}=  \notag \\
&=&\frac{1}{2}h_{i}-\frac{1}{2}\mathbb{V}^{2/3}\hat{\kappa}_{i}-\mathbf{i}%
\mathbb{V}^{1/3}h_{ij}\hat{\lambda}^{j}=\frac{1}{2}h_{i}-\frac{1}{2}\mathbb{V%
}^{2/3}\hat{\kappa}_{i}-\mathbf{i}\mathbb{V}^{1/3}\hat{\kappa}_{ij}x^{j},
\end{eqnarray}
such that
\begin{equation}
\mathcal{Z}:=\left\langle \mathcal{Q},\mathcal{V}\right\rangle=e^{K/2}\left(
q_{0}+z^{i}q_{i}-p^{0}F_{0}-p^{i}F_{i}\right)
=e^{K/2}\left(q_{0}+z^{i}q_{i}+p^{0}F-p^{i}F_{i}\right) .
\end{equation}
On the other hand, the K\"{a}hler-covariant derivatives of the symplectic
sections read
\begin{gather}
\mathcal{V}_{i}^{M}\equiv D_{i}\mathcal{V}^{M}=e^{K/2}\left(
\begin{array}{c}
\tilde{D}_{i}X^{0} \\
\tilde{D}_{i}X^{j} \\
\tilde{D}_{i}F_{0} \\
\tilde{D}_{i}F_{j}%
\end{array}
\right) , \\
\tilde{D}_{i}\equiv \partial _{i}K+\partial _{i},
\end{gather}
where
\begin{eqnarray}
\tilde{D}_{i}X^{0} &=&-\frac{\mathbf{i}}{4}\mathbb{V}^{-1/3}\hat{\kappa}_{i};
\\
\tilde{D}_{i}X^{j} &=&\delta _{i}^{j}-\frac{\mathbf{i}}{4}\mathbb{V}^{-1/3}%
\hat{\kappa}_{i}\left( x^{j}-\mathbf{i}\mathbb{V}^{1/3}\hat{\lambda}%
^{j}\right) ; \\
\tilde{D}_{i}F_{0} &=&-\frac{h_{i}}{2}+\frac{1}{4}\mathbb{V}^{-1/3}\hat{%
\kappa}_{i}+\frac{1}{8}\hat{\kappa}_{i}\hat{\kappa}_{jk}x^{j}x^{k}  \notag \\
&&+\mathbf{i}\mathbb{V}^{1/3}\left( \frac{1}{24}\mathbb{V}^{-2/3}h\hat{\kappa%
}_{i}+\hat{\kappa}_{ij}x^{j}-\frac{1}{8}\hat{\kappa}_{i}\hat{\kappa}%
_{j}x^{j}\right) ; \\
\tilde{D}_{i}F_{j} &=&h_{ij}-\frac{1}{4}\hat{\kappa}_{i}\hat{\kappa}%
_{jk}x^{k}+\mathbf{i}\left( \frac{1}{8}\mathbb{V}^{1/3}\hat{\kappa}_{i}\hat{%
\kappa}_{j}-\mathbb{V}^{1/3}\hat{\kappa}_{ij}-\frac{1}{8}\mathbb{V}^{-1/3}%
\hat{\kappa}_{i}h_{j}\right) ,
\end{eqnarray}
such that
\begin{equation}
\mathcal{Z}_{i}\equiv D_{i}\mathcal{Z}=\left\langle \mathcal{Q},\mathcal{V}%
_{i}\right\rangle
=e^{K/2}%
\left(q_{0}D_{i}X^{0}+q_{j}D_{i}X^{j}-p^{0}D_{i}F_{0}-p^{j}D_{i}F_{j}\right)
.
\end{equation}

Thus, from (\ref{rel2}) and recalling the identity (\ref{adj-id}), one can
proceed and compute
\begin{eqnarray}
&&e^{-2K}\frac{1}{2}\Omega _{MNPQ}\overline{\mathcal{V}}^{M}\overline{%
\mathcal{V}}^{N}\overline{\mathcal{V}}^{P}\overline{\mathcal{V}}^{Q}  \notag
\\
&&\overset{\text{(\ref{I2})-(\ref{I2-1})}}{=}-\left( \overline{X}^{0}%
\overline{F}_{0}+\overline{X}^{i}\overline{F}_{i}\right) ^{2}+\frac{2}{3}%
\overline{F}_{0}d_{ijk}\overline{X}^{i}\overline{X}^{j}\overline{X}^{k}-
\frac{2}{3}\overline{X}^{0}d^{ijk}\overline{F}_{i}\overline{F}_{j}\overline{F%
}_{k}+d_{ijk}d^{ilm}\overline{X}^{j}\overline{X}^{k}\overline{F}_{l}%
\overline{F}_{m}  \notag \\
&=&-\left( -\overline{F}+\frac{1}{2}d_{ijk}\overline{z}^{i}\overline{z}^{j}%
\overline{z}^{k}\right) ^{2}-\frac{2}{3}\overline{F}d_{ijk}\overline{z}^{i}%
\overline{z}^{j}\overline{z}^{k}  \notag \\
&&-\frac{1}{12}d^{ijk}d_{ilm}d_{jnp}d_{krs}\overline{z}^{l}\overline{z}^{m}%
\overline{z}^{n}\overline{z}^{p}\overline{z}^{r}\overline{z}^{s}+\frac{1}{4}%
d_{ijk}d^{ilm}\overline{z}^{j}\overline{z}^{k}d_{lpq}d_{mrs}\overline{z}^{p}
\overline{z}^{q}\overline{z}^{r}\overline{z}^{s}  \notag \\
&=&-4\overline{F}^{2}-4\overline{F}^{2}-4\overline{F}^{2}+12\overline{F}%
^{2}=0.
\end{eqnarray}

\begin{eqnarray*}
&&e^{-2K}\frac{1}{2}\Omega _{MNPQ}\overline{\mathcal{V}}^{M}\overline{%
\mathcal{V}}^{N}\overline{\mathcal{V}}^{P}\overline{\mathcal{V}}_{\bar{\imath%
}}^{Q} \\
&&\overset{\text{(\ref{I1})-(\ref{I1-1})}}{=}-\frac{1}{2}\left[ \left(%
\overline{X}^{0}\right) ^{2}\left( D_{\bar{\imath}}\overline{F}_{0}\right)
\overline{F}_{0}+\overline{X}^{0}\left( D_{\bar{\imath}}\overline{X}%
^{0}\right) \overline{F}_{0}^{2}\right] -\frac{1}{2}\left[ \overline{X}^{k}
\overline{F}_{k}\overline{X}^{j}D_{\bar{\imath}}\overline{F}_{j}+\overline{X}%
^{k}\overline{F}_{k}\left( D_{\bar{\imath}}\overline{X}^{j}\right) \overline{%
F}_{j}\right] \\
&&-\frac{1}{2}\left[ \overline{X}^{0}\overline{F}_{0}\overline{X}^{j}D_{\bar{%
\imath}}\overline{F}_{j}+\overline{X}^{0}\overline{F}_{0}\left( D_{\bar{%
\imath}}\overline{X}^{j}\right) \overline{F}_{j}+\overline{X}^{0}\left( D_{%
\bar{\imath}}\overline{F}_{0}\right) \overline{X}^{j}\overline{F}%
_{j}+\left(D_{\bar{\imath}}\overline{X}^{0}\right) \overline{F}_{0}\overline{%
X}^{j}\overline{F}_{j}\right] \\
&&+\frac{1}{6}\left[ \left( D_{\bar{\imath}}\overline{F}_{0}\right) d_{jkl}%
\overline{X}^{j}\overline{X}^{k}\overline{X}^{l}+3\overline{F}_{0}d_{jkl}%
\overline{X}^{j}\overline{X}^{k}D_{\bar{\imath}}\overline{X}^{k}\right] -
\frac{1}{6}\left[ \left( D_{\bar{\imath}}\overline{X}^{0}\right) d^{jkl}%
\overline{F}_{j}\overline{F}_{k}\overline{F}_{l}+3\overline{X}^{0}d^{jkl}%
\overline{F}_{j}\overline{F}_{k}D_{\bar{\imath}}\overline{F}_{l}\right] \\
&&+\frac{1}{2}d_{njk}d^{nlm}\overline{X}^{j}\overline{X}^{k}\overline{F}%
_{l}D_{\bar{\imath}}\overline{F}_{m}+\frac{1}{2}d_{njk}d^{nlm}\overline{X}%
^{j}\left( D_{\bar{\imath}}\overline{X}^{k}\right) \overline{F}_{l}\overline{%
F}_{m} \\
&=&-\frac{1}{2}\left[ -\left( D_{\bar{\imath}}\overline{F}_{0}\right)%
\overline{F}+\left( D_{\bar{\imath}}\overline{X}^{0}\right) \overline{F}^{2} %
\right] -\frac{1}{2}\left[ \overline{z}^{k}\overline{F}_{k}\overline{z}%
^{j}D_{\bar{\imath}}\overline{F}_{j}+\overline{z}^{k}\overline{F}_{k}\left(
D_{\bar{\imath}}\overline{X}^{j}\right) \frac{1}{2}d_{jkl}\overline{z}^{k}%
\overline{z}^{l}\right] \\
&&-\frac{1}{2}\left[ -\overline{F}\overline{z}^{j}D_{\bar{\imath}}\overline{F%
}_{j}-\frac{1}{2}\overline{F}\left( D_{\bar{\imath}}\overline{X}^{j}\right)
d_{jmn}\overline{z}^{m}\overline{z}^{n}+\frac{1}{2}\left( D_{\bar{\imath}}%
\overline{F}_{0}\right) \overline{z}^{j}d_{jmn}\overline{z}^{m}\overline{z}%
^{n}-\frac{1}{2}\left( D_{\bar{\imath}}\overline{X}^{0}\right) \overline{F}
\overline{z}^{j}d_{jmn}\overline{z}^{m}\overline{z}^{n}\right] \\
&&+\frac{1}{6}\left[ \left( D_{\bar{\imath}}\overline{F}_{0}\right) d_{jkl}%
\overline{z}^{j}\overline{z}^{k}\overline{z}^{l}-3\overline{F}d_{jkl}%
\overline{z}^{j}\overline{z}^{k}D_{\bar{\imath}}\overline{X}^{k}\right] \\
&&-\frac{1}{6}\left[ \frac{1}{8}\left( D_{\bar{\imath}}\overline{X}%
^{0}\right) d^{jkl}d_{jmn}d_{kpq}d_{lrs}\overline{z}^{m}\overline{z}^{n}%
\overline{z}^{p}\overline{z}^{q}\overline{z}^{r}\overline{z}^{s}+\frac{3}{4}
d^{jkl}d_{jmn}d_{kpq}\overline{z}^{m}\overline{z}^{n}\overline{z}^{p}%
\overline{z}^{q}D_{\bar{\imath}}\overline{F}_{l}\right] \\
&&+\frac{1}{4}d_{njk}d^{nlm}\overline{z}^{j}\overline{z}^{k}d_{lrs}\overline{%
z}^{r}\overline{z}^{s}D_{\bar{\imath}}\overline{F}_{m}+\frac{1}{8}%
d_{njk}d^{nlm}\overline{z}^{j}\left( D_{\bar{\imath}}\overline{X}^{k}\right)
d_{lqs}d_{mrt}\overline{z}^{q}\overline{z}^{s}\overline{z}^{r}\overline{z}%
^{t}
\end{eqnarray*}
\begin{eqnarray}
&=&\frac{1}{2}\left( D_{\bar{\imath}}\overline{F}_{0}\right) \overline{F}-%
\frac{1}{2}\left( D_{\bar{\imath}}\overline{X}^{0}\right) \overline{F}^{2}-%
\frac{3}{2}\overline{F}\overline{z}^{j}D_{\bar{\imath}}\overline{F}_{j}-
\frac{3}{4}\overline{F}\left( D_{\bar{\imath}}\overline{X}^{j}\right) d_{jkl}%
\overline{z}^{k}\overline{\bar{z}}^{l}  \notag \\
&&+\frac{1}{2}\overline{F}\overline{z}^{j}D_{\bar{\imath}}\overline{F}_{j}+%
\frac{1}{4}\overline{F}\left( D_{\bar{\imath}}\overline{X}^{j}\right) d_{jmn}%
\overline{z}^{m}\overline{z}^{n}-\frac{3}{2}\overline{F}\left( D_{\bar{\imath%
}}\overline{F}_{0}\right) +\frac{3}{2}\left( D_{\bar{\imath}}\overline{X}%
^{0}\right) \overline{F}^{2}  \notag \\
&&+\overline{F}\left( D_{\bar{\imath}}\overline{F}_{0}\right) -\frac{1}{2}%
\overline{F}d_{jkl}\overline{z}^{j}\overline{z}^{k}D_{\bar{\imath}}\overline{%
X}^{k}  \notag \\
&&-\overline{F}^{2}\left( D_{\bar{\imath}}\overline{X}^{0}\right) -\overline{%
F}\overline{z}^{m}D_{\bar{\imath}}\overline{F}_{m}  \notag \\
&&+2\overline{F}\overline{z}^{m}D_{\bar{\imath}}\overline{F}_{m}+\overline{F}%
d_{njk}\overline{z}^{n}\overline{z}^{j}\left( D_{\bar{\imath}}\overline{X}%
^{k}\right)  \notag \\
&=&0.
\end{eqnarray}
Analogously, one can check that

\begin{equation}
e^{-2K}\frac{1}{2}\Omega _{MNPQ}\overline{\mathcal{V}}^{M}\overline{\mathcal{%
V}}^{N}\overline{\mathcal{V}}_{\bar{\imath}}^{P}\overline{\mathcal{V}}_{\bar{%
\jmath}}^{Q}=0.
\end{equation}

\begin{equation}
e^{-2K}\frac{1}{2}\Omega _{MNPQ}\overline{\mathcal{V}}^{M}\overline{\mathcal{%
V}}_{\bar{\imath}}^{N}\overline{\mathcal{V}}_{\bar{\jmath}}^{P}\overline{%
\mathcal{V}}_{\bar{k}}^{Q}=0.
\end{equation}

\begin{equation}
e^{-2K}\frac{1}{2}\Omega _{MNPQ}\overline{\mathcal{V}}_{\bar{\imath}}^{M}%
\overline{\mathcal{V}}_{\bar{\jmath}}^{N}\overline{\mathcal{V}}_{\bar{k}}^{P}%
\overline{\mathcal{V}}_{\bar{l}}^{Q}=0.
\end{equation}


\section{\label{App-comp}Proof of (\protect\ref{Rule-3-2})}

At the class $V$\textbf{.III} of critical points of $V$ (cf. Sec. \ref%
{Class-V}), it holds that
\begin{equation}
\mathcal{L}_{i}=\frac{\mathbf{i}}{2\overline{\mathcal{L}}}C_{ijk}\overline{%
\mathcal{L}}^{j}\overline{\mathcal{L}}^{k},
\end{equation}
which implies
\begin{equation}
2\overline{\mathcal{L}}\mathcal{L}_{i}=-\frac{\mathbf{i}}{4\mathcal{L}^{2}}%
C_{ijk}\overline{C}_{~(\overline{m}\overline{n}|}^{j}\overline{C}_{~|%
\overline{p}\overline{q})}^{k}\mathcal{L}^{\overline{m}}\mathcal{L}^{%
\overline{n}} \mathcal{L}^{\overline{p}}\mathcal{L}^{\overline{q}},
\end{equation}
and
\begin{equation}
\left\vert \mathcal{L}_{i}\right\vert ^{2}=\frac{\mathbf{i}}{2\overline{%
\mathcal{L}}}N_{3}(\overline{\mathcal{L}})=-\frac{\mathbf{i}}{2\mathcal{L}}%
\overline{N_{3}(\overline{\mathcal{L}})}\equiv -\frac{\mathbf{i}}{2\mathcal{L%
}}\overline{N}_{3}(\mathcal{L}),  \label{j1}
\end{equation}
where we have defined (cf. (\ref{NN3}))
\begin{equation}
N_{3}(\overline{\mathcal{L}})\equiv N_{3}(\overline{\mathcal{L}},\overline{%
\mathcal{L}},\overline{\mathcal{L}}):=C_{ijk}\overline{\mathcal{L}}^{i}%
\overline{\mathcal{L}}^{j}\overline{\mathcal{L}}^{k}.
\end{equation}
By using the special geometry identity (3.1.1.2.12) of \cite{K-rev}, one
obtains
\begin{equation}
2\overline{\mathcal{L}}\mathcal{L}_{i}=-\frac{\mathbf{i}}{3\mathcal{L}^{2}}%
\mathcal{L}_{i}\overline{N}_{3}\left( \mathcal{L}\right) -\frac{\mathbf{i}}{%
12\mathcal{L}^{2}}\left( D_{i}\overline{D}_{(\bar{\imath}}\overline{C}_{\bar{%
\jmath}\bar{k}\bar{l})}\right) \mathcal{L}^{\bar{\imath}}\mathcal{L}^{\bar{%
\jmath}}\mathcal{L}^{\bar{k}}\mathcal{L}^{\bar{l}}.  \label{j2}
\end{equation}
Therefore, (\ref{j1}) and (\ref{j2}) yield to
\begin{gather}
\left( \overline{\mathcal{L}}-\frac{1}{3}\frac{\left\vert \mathcal{L}%
_{i}\right\vert ^{2}}{\mathcal{L}}\right) \left\vert \mathcal{L}%
_{j}\right\vert ^{2}=-\frac{\mathbf{i}}{24\mathcal{L}^{2}}\left( D_{m}%
\overline{D}_{(\bar{\imath}}\overline{C}_{\bar{\jmath}\bar{k}\bar{l}%
)}\right) \overline{\mathcal{L}}^{m}\mathcal{L}^{\bar{\imath}}\mathcal{L}^{%
\bar{\jmath}}\mathcal{L}^{\bar{k}}\mathcal{L}^{\bar{l}}; \\
\Updownarrow  \notag \\
\left\vert \mathcal{L}\right\vert ^{2}-\frac{1}{3}\left\vert \mathcal{L}%
_{i}\right\vert ^{2}=-\frac{\mathbf{i}\left( D_{m}\overline{D}_{(\bar{\imath}%
}\overline{C}_{\bar{\jmath}\bar{k}\bar{l})}\right) \overline{\mathcal{L}}%
^{m} \mathcal{L}^{\bar{\imath}}\mathcal{L}^{\bar{\jmath}}\mathcal{L}^{\bar{k}%
}\mathcal{L}^{\bar{l}}}{24\mathcal{L}\left\vert \mathcal{L}%
_{j}\right\vert^{2}}\overset{\text{(\ref{j1})}}{=}\frac{\left( D_{m}%
\overline{D}_{(\bar{\imath}} \overline{C}_{\bar{\jmath}\bar{k}\bar{l}%
)}\right) \overline{\mathcal{L}}^{m}\mathcal{L}^{\bar{\imath}}\mathcal{L}^{%
\bar{\jmath}}\mathcal{L}^{\bar{k}}\mathcal{L}^{\bar{l}}}{12\overline{N}_{3}(%
\mathcal{L})}; \\
\Updownarrow  \notag \\
\left\vert \mathcal{L}_{i}\right\vert ^{2}=3\left\vert \mathcal{L}%
\right\vert ^{2}-\frac{\left( D_{m}\overline{D}_{(\bar{\imath}}\overline{C}_{%
\bar{\jmath}\bar{k}\bar{l})}\right) \overline{\mathcal{L}}^{m}\mathcal{L}^{%
\bar{\imath}} \mathcal{L}^{\bar{\jmath}}\mathcal{L}^{\bar{k}}\mathcal{L}^{%
\bar{l}}}{4\overline{N}_{3}(\mathcal{L})},
\end{gather}
which, by definition (\ref{DeltaL}), gives Eq. (\ref{Rule-3-2}) $%
\blacksquare $


\section{ \label{STU-Details}Details on the magnetic $STU$}

We use the conventions of \cite{Astesiano}. We consider $\mathcal{N}=2$, $%
D=4 $ gauged supergravity coupled to $n$ Abelian vector multiplets. Apart
from the vierbein $e_{\mu }^{a}$, the bosonic field content includes the
vectors $A_{\mu }^{I}$ enumerated by $I=0,\ldots ,n$, and the complex
scalars $z^{\alpha }$ where $\alpha =1,\ldots ,n$. These scalars parametrize
a special K\"{a}hler manifold $M_{v}$, i.e. an $n$-dimensional Hodge-K\"{a}%
hler manifold that is the base of a symplectic bundle, with the covariantly
holomorphic sections
\begin{equation}
{\mathcal{V}}=\left(
\begin{array}{c}
X^{I} \\
F_{I}%
\end{array}
\right) \,,\qquad D_{\bar{\alpha}}{\mathcal{V}}=\partial _{\bar{\alpha}}{%
\mathcal{V}}-\frac{1}{2}(\partial _{\bar{\alpha}}K){\mathcal{V}}=0\,,
\label{sympl-vec}
\end{equation}
where $K$ is the K\"{a}hler potential and $D$ denotes the K\"{a}%
hler-covariant derivative. ${\mathcal{V}}$ obeys the symplectic constraint
\begin{equation}
\langle {\mathcal{V}},\overline{\mathcal{V}}\rangle =X^{I}\overline{F}%
_{I}-F_{I}\overline{X}^{I}=\mathbf{i}\,.  \label{sympconst}
\end{equation}
To solve this condition, one defines
\begin{equation}
{\mathcal{V}}=:e^{{\mathcal{K}}(z,\bar{z})/2}\mathfrak{H}(z)\,,
\end{equation}
where $\mathfrak{H}(z)$ is a holomorphic symplectic vector,
\begin{equation}
\mathfrak{H}(z)=\left(
\begin{array}{c}
X^{I}(z) \\
\frac{\partial }{\partial X^{I}}F(X)\left( z\right)%
\end{array}
\right) \,.
\end{equation}
where $F$ is an holomorphic function homogeneous of degree two, called the
\textit{prepotential}, whose existence is assumed in order to obtain the
last expression. The K\"{a}hler potential is then (cf. (\ref{g1}))
\begin{equation}
e^{-K(z,\bar{z})}=\mathbf{i}\langle \mathfrak{H},\overline{\mathfrak{H}}
\rangle \,.
\end{equation}
The matrix ${\mathcal{N}}_{IJ}$ determining the coupling between the scalars
$z^{\alpha }$ and the vectors $A_{\mu }^{I}$ is defined by the relations
\begin{equation}
F_{I}={\mathcal{N}}_{IJ}X^{J}\,,\qquad D_{\bar{\alpha}}\overline{F}_{I}={%
\mathcal{N}}_{IJ}D_{\bar{\alpha}}\overline{X}^{J}\,.  \label{defN}
\end{equation}
The bosonic action reads
\begin{eqnarray}
e^{-1}{\mathcal{L}}_{\text{bos}} &=&\frac{1}{2}R+\frac{1}{4}(\text{Im}\,{%
\mathcal{N}})_{IJ}F_{\mu \nu }^{I}F^{J\mu \nu }-\frac{1}{8}(\text{Re}\,{%
\mathcal{N}})_{IJ}\,e^{-1}\epsilon ^{\mu \nu \rho \sigma }F_{\mu
\nu}^{I}F_{\rho \sigma }^{J}  \notag \\
&&-g_{\alpha \bar{\beta}}\partial _{\mu }z^{\alpha }\partial ^{\mu }%
\overline{z}^{\bar{\beta}}-V\,,  \label{action}
\end{eqnarray}
with the scalar potential
\begin{equation}
V=-2g^{2}\xi _{I}\xi _{J}[(\text{Im}\,{\mathcal{N}})^{-1|IJ}+8\overline{X}%
^{I}X^{J}],\,
\end{equation}
that results from $U(1)$ FI gauging. Here, $g$ denotes the gauge coupling
and the $\xi _{I}$ are FI gauging parameters. In what follows, we define $%
g_{I}\equiv g\xi _{I}$. The Einstein's equations of motion from %
\eqref{action} are given by
\begin{align}
G_{\mu \nu }=& T_{\mu \nu }=^{(0)}T_{\mu \nu }+^{(1)}T_{\mu \nu }-g_{\mu
\nu}V,  \label{var-einstein} \\
^{(0)}T_{\mu \nu }=& 2g_{\alpha \bar{\beta}}\partial _{(\mu
}z^{\alpha}\partial _{\nu )}\overline{z}^{\bar{\beta}}-g_{\mu \nu }g_{\alpha
\bar{\beta}}\partial _{\sigma }z^{\alpha }\partial ^{\sigma }\overline{z}^{%
\bar{\beta}},  \label{stress-tensorspin0} \\
^{(1)}T_{\mu \nu }=& -(Im\mathcal{N})_{IJ}F_{\mu \sigma
}^{I}F_{\nu}^{J\sigma }+g_{\mu \nu }\frac{1}{4}(Im\mathcal{N})_{IJ}F_{\sigma
\rho}^{I}F^{J\sigma \rho },  \label{stress-tensorspin1}
\end{align}
where we have made explicit the contribution form the spin $0$ and the spin $%
1$ part. We can rewrite the full system as

\begin{eqnarray}
R_{\mu \nu } &=&-(\text{Im}\,{\mathcal{N}})_{IJ}F_{\mu \lambda }^{I}F_{\ \nu
}^{J\ \lambda }+2g_{\alpha \bar{\beta}}\partial _{(\mu }z^{\alpha }\partial
_{\nu )}\overline{z}^{\bar{\beta}}+g_{\mu \nu }\left[ \frac{1}{4}(\text{Im}\,%
{\mathcal{N}})_{IJ}F_{\rho \sigma }^{I}F^{J\rho \sigma }+V\right] \,, \\
0 &=&\nabla _{\mu }\left[ (\text{Im}\,{\mathcal{N}})_{IJ}F^{J\mu \nu }-\frac{%
1}{2}(\text{Re}\,{\mathcal{N}})_{IJ}\,e^{-1}\epsilon ^{\mu \nu \rho \sigma
}F_{\rho \sigma }^{J}\right] ,  \label{var-maxwell} \\
0 &=&\frac{1}{4}\frac{\delta (\text{Im}\,{\mathcal{N}})_{IJ}}{\delta
z^{\alpha }}F_{\mu \nu }^{I}F^{J\mu \nu }-\frac{1}{8}\frac{\delta (\text{Re}%
\,{\mathcal{N}})_{IJ}}{\delta z^{\alpha }}\,e^{-1}\epsilon ^{\mu \nu \rho
\sigma }F_{\mu \nu }^{I}F_{\rho \sigma }^{J}+\frac{\delta g_{\alpha \bar{%
\beta}}}{\delta \overline{z}^{\bar{\gamma}}}\partial _{\lambda }\overline{z}%
^{\bar{\gamma}}\partial ^{\lambda }\overline{z}^{\bar{\beta}}  \notag \\
&&+g_{\alpha \bar{\beta}}\nabla _{\lambda }\nabla ^{\lambda }\overline{z}^{%
\bar{\beta}}-\frac{\delta V}{\delta z^{\alpha }},
\end{eqnarray}%
which hold independently from the existence and the choice of a prepotential
$F(X)$. \newline
Defining the tensor
\begin{equation}
G_{I\mu \nu }:=R_{IJ}F_{\mu \nu }^{I}+I_{IJ}\tilde{F}_{\mu \nu }\quad ,\quad
\tilde{F}_{\mu \nu }^{J}:=\frac{1}{2}\sqrt{-g}\epsilon _{\mu \nu \rho \sigma
}F^{J\rho \sigma }.
\end{equation}%
then eq.(\ref{var-maxwell}), the Bianchi identities and the charges can be
written as
\begin{equation}
\epsilon ^{\mu \nu \rho \sigma }\partial _{\mu }%
\begin{pmatrix}
F_{\rho \sigma }^{I} \\
G_{I\rho \sigma } \\
\end{pmatrix}%
=0\quad ,\quad \frac{1}{4\pi }\int_{\Sigma _{\infty }}%
\begin{pmatrix}
F^{I} \\
G_{I}%
\end{pmatrix}%
=:%
\begin{pmatrix}
p^{I} \\
q_{I} \\
\end{pmatrix}%
.  \label{MB}
\end{equation}%
\noindent For the magnetic $STU$ model, the prepotential is given by (\ref%
{MagneticSTU}), and the symplectic section can be parametrised in terms of
three complex scalar fields $\tau _{1}$, $\tau _{2}$ and $\tau _{3}$ by
choosing $X^{0}=1$, $X^{1}=\tau _{2}\tau _{3}$, $X^{2}=\tau _{1}\tau _{3}$, $%
X^{3}=\tau _{1}\tau _{2}$, so that
\begin{equation}
\mathfrak{H}=%
\begin{pmatrix}
1, & \tau _{2}\tau _{3}, & \tau _{1}\tau _{3}, & \tau _{1}\tau _{2}, & -%
\mathbf{i}\tau _{1}\tau _{2}\tau _{3}, & -\mathbf{i}\tau _{1}, & -\mathbf{i}%
\tau _{2}, & -\mathbf{i}\tau _{3}%
\end{pmatrix}%
^{T},
\end{equation}%
where $\tau _{\alpha }$'s coordinatize $M_{v}$. The K\"{a}hler potential and
the non-vanishing components of the metric on the scalar manifold are
respectively
\begin{equation}
e^{-K}=8\text{Re}\tau _{1}\text{Re}\tau _{2}\text{Re}\tau _{3},\qquad
g_{\alpha \bar{\alpha}}=g_{\bar{\alpha}\alpha }=\partial _{\alpha }\partial
_{\bar{\alpha}}K=(\tau _{\alpha }+\overline{\tau }_{\bar{\alpha}})^{-2}.
\end{equation}%
In particular, we notice the relations
\begin{equation}
F_{I}=\frac{F}{2X^{I}},\qquad \mathcal{N}_{IJ}=\frac{F}{2(X^{I})^{2}}\delta
_{IJ}
\end{equation}%
between the prepotential and the period matrix.\newline
We have
\begin{equation}
\tau _{1}=\sqrt{\frac{g_{0}g_{1}}{g_{2}g_{3}}}\tau (r)\equiv \sqrt{\frac{%
g_{0}g_{1}}{g_{2}g_{3}}}\left( f(r)+\mathbf{i}g(r)\right) ,\qquad \tau _{2}=%
\sqrt{\frac{g_{0}g_{2}}{g_{1}g_{3}}},\qquad \tau _{3}=\sqrt{\frac{g_{0}g_{3}%
}{g_{1}g_{2}}}.
\end{equation}%
The symplectic section $X^{I}$ and the period matrix $\mathcal{N}_{IJ}$ in
terms of these scalar fields boil down to
\begin{equation}
X^{I}=\frac{1}{8}\sqrt{\frac{G}{8}}\frac{1}{\sqrt{\text{Re}\tau }}\left(
\begin{array}{c}
\frac{1}{|g_{0}|} \\
\frac{1}{|g_{1}|} \\
\frac{\tau }{|g_{2}|} \\
\frac{\tau }{|g_{3}|}%
\end{array}%
\right) ,\qquad \mathcal{N}_{IJ}=-\mathbf{i}\frac{64}{G}%
\begin{pmatrix}
g_{0}^{2}\tau & 0 & 0 & 0 \\
0 & g_{1}^{2}\tau & 0 & 0 \\
0 & 0 & \frac{g_{2}^{2}}{\tau } & 0 \\
0 & 0 & 0 & \frac{g_{3}^{2}}{\tau } \\
&  &  &
\end{pmatrix}%
,
\end{equation}%
and Re$\tau >0$, in order to guarantee the positive definiteness of the
spin-1 kinetic terms of the action, namely the fact that Im$\mathcal{N}_{IJ}$
is negative definite. Explicitly, the real and imaginary parts read
\begin{align}
\text{Im}\mathcal{N}_{IJ}& =I_{IJ}=-\frac{64}{G}\text{diag}%
\begin{pmatrix}
g_{0}^{2}\text{Re}\tau \,, & g_{1}^{2}\text{Re}\tau \,, & g_{2}^{2}\frac{%
\text{Re}\tau }{|\tau |^{2}}\,, & g_{3}^{2}\frac{\text{Re}\tau }{|\tau |^{2}}%
\end{pmatrix}%
; \\
\text{Re}\mathcal{N}_{IJ}& =R_{IJ}=\frac{64}{G}\text{diag}%
\begin{pmatrix}
g_{0}^{2}\text{Im}\tau \,, & g_{1}^{2}\text{Im}\tau \,, & -g_{2}^{2}\frac{%
\text{Im}\tau }{|\tau |^{2}}\,, & -g_{3}^{2}\frac{\text{Im}\tau }{|\tau |^{2}%
}%
\end{pmatrix}%
.
\end{align}%
Now, we employ the two following \textit{Ans\"{a}tze} for the metric and the
electromagnetic field
\begin{gather}
ds^{2}=-A(r)dt^{2}+A(r)^{-1}dr^{2}+B(r)d\Omega _{k}^{2},  \label{GMA} \\
F_{tr}^{I}=\frac{(\text{Im}\,{\mathcal{N}})^{-1|IJ}}{B(r)}\left( (\text{Re}%
\mathcal{N})_{JS}\,p^{S}-q_{J}\right) ,\qquad F_{\theta \phi
}^{I}=p^{I}f_{\kappa }(\theta ),  \label{EMF}
\end{gather}%
where $d\Omega _{k}^{2}=d\theta ^{2}+f_{\kappa }^{2}(\theta )d\phi ^{2}$ is
the metric on the two-dimensional surfaces $\Sigma =\{S^{2},H^{2}\}$ of
constant scalar curvature $R=2\kappa $, with $\kappa =\pm 1$, and (cf. e.g.
(5.10) of \cite{Chimento} and Refs. therein)
\begin{equation}
f_{\kappa }(\theta )=\frac{1}{\sqrt{\kappa }}\sin (\sqrt{\kappa }\theta
)=\left\{
\begin{array}{c@{\quad}l}
\sin \theta \, & \kappa =1\,, \\
\sinh \theta \, & \kappa =-1\,.%
\end{array}%
\right.  \label{fk}
\end{equation}%
The stress tensor for the spin-1 part $T_{\mu \nu }^{(1)}$ can be computed
as
\begin{equation}
^{(1)}T_{\,0}^{0}=^{(1)}T_{\,1}^{1}=-^{(1)}T_{\,2}^{2}=-^{(1)}T_{\,3}^{3}=-%
\frac{1}{B^{2}}V_{BH};
\end{equation}%
one can also check that
\begin{equation}
\frac{1}{4}\frac{\partial I_{IJ}}{\partial z^{a}}F_{\mu \nu }^{I}F^{J|\,\mu
\nu }-\frac{1}{8}\frac{\partial R_{IJ}}{\partial z^{a}}F_{\mu \nu }^{I}{%
^{\ast }F}^{J|\,\mu \nu }=-\frac{1}{B^{2}}\frac{\partial V_{BH}}{\partial
z^{a}},
\end{equation}%
where we defined the so-called black hole (BH) potential \cite{CDF, AM2}
\begin{equation}
V_{BH}:=-\frac{1}{2}\mathcal{Q}^{T}\mathcal{M}(\mathcal{N})\mathcal{Q}.
\end{equation}%
One also have for $g_{I}>0$
\begin{equation}
V=-\frac{G}{16}\left( 4+\frac{1+|\tau |^{2}}{\text{Re}\tau }\right) ,\qquad
\partial _{\tau }V=\frac{G}{32}\left( \frac{1-\overline{\tau }^{2}}{\text{Re}%
^{2}\tau }\right) .
\end{equation}%
The field Maxwell and Bianchi field equations (\ref{MB}) are satisfied,
while the Einsteins equations of motion and the scalar field equation read%
\footnote{%
The priming denotes differentiation with respect to the radial coordinate.}
\begin{align}
R_{tt}& =\frac{A}{2B}\left( A^{\prime \prime }B+A^{\prime }B^{\prime
}\right) =\frac{A}{B^{2}}V_{BH}-AV;  \label{SRtt1} \\
R_{rr}& =-\frac{1}{2AB^{2}}\left( A^{\prime \prime 2}B^{2}+A^{\prime
}B^{\prime }B+2AB^{\prime \prime }B-AB^{\prime 2}\right) =-\frac{1}{AB^{2}}%
V_{BH}+\frac{1}{A}V+\frac{\tau ^{\prime }\overline{{\tau }}{^{\prime }}}{2%
\text{Re}^{2}\tau };  \label{SRrr1} \\
R_{\theta \theta }& =-\frac{1}{2}\left( A^{\prime }B^{\prime }+AB^{\prime
\prime }\right) +\kappa =\frac{1}{B}V_{BH}+BV;  \label{SRthth1} \\
0& =\frac{1}{B^{2}}\partial _{\tau }V_{BH}+\partial _{\tau }V-\frac{1}{B}%
\frac{\left( BA\overline{\tau }^{\prime }\right) ^{\prime }}{4\text{Re}%
^{2}\tau }+A\frac{\overline{{\tau }}^{\prime 2}}{4\text{Re}^{3}\tau }.
\label{SS1}
\end{align}%
The system can be rewritten as
\begin{gather}
\frac{\left( 2B^{\prime \prime }B-B^{\prime 2}\right) }{B^{2}}=-\frac{\tau
^{\prime }\overline{{\tau }}{^{\prime }}}{\text{Re}^{2}\text{ }\tau };
\label{EQ1} \\
A^{\prime \prime }B-AB^{\prime \prime }=4\frac{V_{BH}}{B}-2\kappa ;
\label{EQ2} \\
(AB)^{\prime \prime }=-4BV+2\kappa ;  \label{EQ3} \\
0=\frac{1}{B^{2}}\partial _{\tau }V_{BH}+\partial _{\tau }V-\frac{1}{B}\frac{%
\left( BA\overline{{\tau }}{^{\prime }}\right) ^{\prime }}{4\text{Re}%
^{2}\tau }+A\frac{\overline{{\tau }}{^{\prime 2}}}{4\text{Re}^{3}\tau }.
\label{EQ4}
\end{gather}%
Assuming $B:=r^{2}-\Delta ^{2}$ and hyperbolic symmetry $\kappa =-1$,
equation (\ref{EQ1}) is solved by
\begin{equation}
f(r)=\frac{r-\Delta }{r+\Delta },\qquad g(r)=0.
\end{equation}%
With such a position, the equation (\ref{EQ2}) reads
\begin{equation}
\left( r^{2}-\Delta ^{2}\right) A^{\prime \prime }(r)-2A(r)-\frac{\left(
\Delta ^{2}+r^{2}\right) \left( (64)^{2}a^{2}+b^{2}G^{2}\right) }{%
8G(r^{2}-\Delta ^{2})^{2}}-2=0,
\end{equation}%
and the part of solution which is consistent with the remaining equations of
motion is
\begin{equation}
A(r)=\frac{(64)^{2}a^{2}+b^{2}G^{2}}{32G\left( r^{2}-\Delta ^{2}\right) }%
+c_{1}\left( r^{2}-\Delta ^{2}\right) -\frac{r^{2}}{\Delta ^{2}},
\end{equation}%
where $c_{1}$ is a constant, that can be fixed using the remaining equations
(\ref{EQ3}) and (\ref{EQ4}). In fact, those equations are satisfied if we
require the following condition,
\begin{equation}
-8c_{1}\Delta ^{2}+\Delta ^{2}G+8=0.
\end{equation}


\section{\label{App-Eff}Near-horizon limit and $V_{\text{eff}}$}

By considering the treatment of the magnetic $STU$ model given in Sec. \ref%
{STU} as well as in App. \ref{STU-Details}, we recall here the near-horizon
limit of the equations of motion, which yields to the definition of the
effective black hole potential $V_{eff}$ in gauged supergravity \label%
{BFMY-FI, Chimento, KPR}. We start with the metric \textit{Ansatz}
\begin{equation}
ds^{2}=-A(r)dt^{2}+\frac{1}{A(r)}dr^{2}+B(r)d\Omega _{\kappa }^{2},
\end{equation}
where
\begin{equation}
d\Omega _{\kappa }^{2}=d\theta ^{2}+f_{\kappa }^{2}\left( \theta \right)
d\varphi ^{2}
\end{equation}
is the metric on the two-dimensional surface $S^{2}$ for $\kappa =1$ or $%
H^{2}$ for $\kappa =-1$. Such a surface has constant scalar curvature $%
R=2\kappa $, and $f_{\kappa }\left( \theta \right) $ is defined by (\ref{fk}%
). In the near-horizon limit, it must hold that
\begin{eqnarray}
A(r) &\rightarrow &\frac{r_{H}^{2}}{r_{A}^{2}}\Rightarrow
A^{\prime}(r)\rightarrow \frac{2r_{H}}{r_{A}^{2}},~A^{\prime \prime
}(r)\rightarrow \frac{2}{r_{A}^{2}}; \\
B(r) &\rightarrow &r_{H}^{2}\Rightarrow B^{\prime }(r)\rightarrow
0,~B^{\prime \prime }(r)\rightarrow 0.
\end{eqnarray}
Thus :

\begin{enumerate}
\item the near-horizon limit of Eq. (\ref{SRtt1}) reads
\begin{gather}
\frac{A}{2B}\left( A^{\prime \prime }B+A^{\prime }B^{\prime }\right) =\frac{A%
}{B^{2}}V_{BH}-AV; \\
\Downarrow  \notag \\
\frac{r_{H}^{2}}{2r_{A}^{2}r_{H}^{2}}\frac{2}{r_{A}^{2}}r_{H}^{2}=\frac{%
r_{H}^{2}}{r_{A}^{2}r_{H}^{4}}V_{BH}-\frac{r_{H}^{2}}{r_{A}^{2}}%
V\Leftrightarrow \frac{1}{r_{A}^{2}}=\frac{1}{r_{H}^{4}}V_{BH}-V,  \label{1}
\end{gather}
which matches Eq. (2.17) of \cite{BFMY-FI}, or Eq. (5.33) of \cite{Chimento}
(by setting $V_{BH~\text{them}}\rightarrow \frac{V_{BH~\text{us}}}{%
\left(8\pi \right) ^{2}}$ and $V_{\text{them}}\rightarrow \frac{V_{\text{us}}%
}{2}$).

\item the near-horizon limit of Eq. (\ref{SRrr1}) reads
\begin{gather}
-\frac{1}{2AB^{2}}\left( A^{\prime \prime }B^{2}+A^{\prime
}B^{\prime}B+2AB^{\prime \prime }B-AB^{\prime 2}\right) =-\frac{1}{AB^{2}}%
V_{BH}+\frac{1}{A}V+\frac{\tau ^{\prime }\overline{\tau }^{\prime }}{2\left(
\text{Re}\tau \right) ^{2}}; \\
\Downarrow  \notag \\
-\frac{1}{2}\frac{r_{A}^{2}}{r_{H}^{2}}\frac{1}{r_{H}^{4}}\frac{2r_{H}^{4}}{%
r_{A}^{2}}=-\frac{r_{A}^{2}}{r_{H}^{6}}V_{BH}+\frac{r_{A}^{2}}{r_{H}^{2}}%
V\Leftrightarrow \frac{1}{r_{A}^{2}}=\frac{1}{r_{H}^{4}}V_{BH}-V,  \label{2}
\end{gather}
again matching Eq. (2.17) of \cite{BFMY-FI}, or Eq. (5.33) of \cite{Chimento}%
.

\item the near-horizon limit of Eq. (\ref{SRthth1}) reads
\begin{gather}
-\frac{1}{2}\left( A^{\prime }B^{\prime }+AB^{\prime \prime }\right) +\kappa=%
\frac{1}{B}V_{BH}+BV; \\
\Downarrow  \notag \\
\kappa =\frac{1}{r_{H}^{2}}V_{BH}+r_{H}^{2}V\Leftrightarrow \frac{\kappa }{%
r_{H}^{2}}=\frac{1}{r_{H}^{4}}V_{BH}+V,  \label{3}
\end{gather}
which generalizes to $\kappa =\pm 1$ Eq. (2.16) of \cite{BFMY-FI} (to which
it reduces by setting $\kappa =1$), or Eq. (5.34)\footnote{%
Let us remark that (\ref{3}) fixes a typo in Eq. (5.34) of \cite{Chimento}.}
of \cite{Chimento}.

\item the near-horizon limit of Eq. (\ref{SS1}) reads
\begin{gather}
0=\frac{1}{B^{2}}\partial _{\tau }V_{BH}+\partial _{\tau }V-\frac{1}{B}\frac{%
\left( BA\overline{\tau }^{\prime }\right) ^{\prime }}{4\left( \text{Re}%
\tau\right) ^{2}}+A\frac{\left( \overline{\tau }^{\prime }\right) ^{2}}{%
4\left(\text{Re}\tau \right) ^{3}}; \\
\Downarrow  \notag \\
0=\frac{1}{B^{2}}\partial _{\tau }V_{BH}+\partial _{\tau }V\Leftrightarrow 0=%
\frac{1}{r_{H}^{4}}\partial _{\tau }V_{BH}+\partial _{\tau }V,  \label{4}
\end{gather}
which - by trivial extension to the generic case with many scalars - matches
Eq. (2.18) of \cite{BFMY-FI}, or Eq. (5.35) of \cite{Chimento}.
\end{enumerate}

Thus, by solving (\ref{3}), one obtains
\begin{equation}
\frac{\kappa }{r_{H}^{2}}=\frac{1}{r_{H}^{4}}V_{BH}+V\Leftrightarrow
Vr_{H}^{4}-\kappa r_{H}^{2}+V_{BH}=0\Leftrightarrow r_{H,\pm }^{2}=\frac{%
\kappa }{2V}\pm \frac{\sqrt{\kappa^2-4VV_{BH}}}{2V}.  \label{5}
\end{equation}
On the other hand, from (\ref{1}) or (\ref{2}) one obtains
\begin{equation}
\frac{1}{r_{A}^{2}}=\frac{1}{r_{H,\pm }^{4}}V_{BH}-V\overset{\text{(\ref{5})}%
}{\Leftrightarrow }\frac{1}{r_{A}^{2}}=-2V+\frac{\kappa }{r_{H,\pm }^{2}}%
\Leftrightarrow r_{A,\pm }^{2}=\frac{1}{\frac{\kappa }{r_{H,\pm }^{2}}-2V}=
\frac{r_{H,\pm }^{2}}{\kappa -2Vr_{H,\pm }^{2}},
\end{equation}
such that
\begin{equation}
r_{A,\pm }^{2}=\frac{r_{H,\pm }^{2}}{\kappa -2Vr_{H,\pm }^{2}}=\mp \frac{%
r_{H,\pm }^{2}}{\sqrt{\kappa^2-4VV_{BH}}},  \label{6}
\end{equation}
where in the last step the result
\begin{equation}
\text{Eq.~(\ref{5})}\Leftrightarrow \kappa -2Vr_{H,\pm }^{2}=\mp \sqrt{%
\kappa^2-4VV_{BH}}
\end{equation}
has been used. Since $r_{A}^{2}>0$, only $r_{A,-}^{2}$ and $r_{H,-}^{2}$
make sense:
\begin{eqnarray}
r_{H}^{2} &=&\frac{\kappa -\sqrt{\kappa^2-4VV_{BH}}}{2V}; \\
r_{A}^{2} &=&\frac{r_{H}^{2}}{\sqrt{\kappa^2-4VV_{BH}}}.
\end{eqnarray}

Thus, one can define\footnote{%
It should be noted that, apart from the renaming $V_{BH~\text{them}%
}\rightarrow \frac{V_{BH~\text{us}}}{\left( 8\pi \right) ^{2}}$ and $V_{%
\text{them}}\rightarrow \frac{V_{\text{us}}}{2}$, the effective potential $%
V_{\text{eff}}$ defined by (\ref{Veff}) is $\kappa $ times the effective
potential defined by Eq. (5.39) of \cite{Chimento}.}
\begin{equation}
V_{\text{eff}}:=\frac{1-\kappa \sqrt{\kappa ^{2}-4VV_{BH}}}{2V},
\label{Veff}
\end{equation}
such that for $\kappa =\pm 1$ it holds that
\begin{equation}
S=\kappa \left. V_{\text{eff}}\right\vert _{\partial V_{\text{eff}}=0}.
\label{S}
\end{equation}
In fact,
\begin{gather}
\partial _{i}V_{\text{eff}}=0 \\
\Updownarrow  \notag \\
\kappa ^{2}\partial _{i}V_{BH}+\frac{\left( \kappa
^{2}-2\kappa^{2}V_{BH}V-\kappa \sqrt{\kappa ^{2}-4VV_{BH}}\right) }{2V^{2}}%
\partial _{i}V  \notag \\
\overset{\text{(\ref{Veff})}}{=}\partial _{i}V_{BH}+\kappa ^{2}V_{\text{eff}%
}^{2}\partial _{i}V=\partial _{i}V_{BH}+r_{H}^{4}\partial _{i}V=0,
\end{gather}
which is satisfied by virtue of the trivial generalization of (\ref{4}) to
the generic case of many scalars, namely%
\begin{equation}
\partial _{\tau }V_{BH}+r_{H}^{4}\partial _{\tau }V=0\Rightarrow \partial
_{i}V_{BH}+r_{H}^{4}\partial _{i}V=0,~\forall i.
\end{equation}


\end{document}